\documentclass{jfm}
\usepackage{graphicx}
\usepackage{latexsym}
\usepackage[dvipsnames]{xcolor}
\usepackage{epstopdf, epsfig}
\usepackage{caption}
\usepackage{subcaption}
\usepackage{hyperref}
\usepackage{natbib}
\usepackage{graphicx}
\usepackage{dcolumn}
\usepackage{bm}
\usepackage{amsmath}
\usepackage{amssymb}
\usepackage{placeins}
\usepackage{tikz}
\usepackage{pgfplots}
\usepackage{color}

\title{Orientation instability of settling spheroids in a linearly density stratified fluid}

\author{Rishabh V. More\aff{1},
  Mehdi N. Ardekani\aff{2},
  Luca Brandt\aff{3}
  Arezoo M. Ardekani\aff{1},
   \corresp{\email{ardekani@purdue.edu}}
  }

\affiliation{\aff{1}School of Mechanical Engineering, Purdue University, West Lafayette, IN 47907, USA
\aff{2}Chemical Engineering, Stanford University, Stanford, CA 94305, USA
\aff{3} Flow and SeRC (Swedish e-Science Research Centre), Department of Engineering Mechanics, KTH, SE-100 44 Stockholm, Sweden
}

\begin{document}

\maketitle

\begin{abstract}
Much work has been done to understand the settling dynamics of spherical particles in a homogeneous and a stratified fluid. However, the effects of shape anisotropy on the settling dynamics of a particle in a stratified fluid are not completely understood. To this end, we perform numerical simulations for settling oblate and prolate spheroids in a stratified fluid. We present the results for the Reynolds number, $Re$, in the range $80-250$ and the Richardson number, $Ri$, in the range $0-10$. We find that both the oblate and prolate spheroids reorient to the edge-wise and partially edge-wise orientations, respectively, as they settle in a stratified fluid completely different from the steady-state broad-side on orientation observed in a homogeneous fluid. We observe that reorientation instabilities emerge when the velocity magnitude of the spheroids fall below a particular threshold. We also report the enhancement of the drag on the particle from stratification. The torque due to buoyancy effects tries to orient the spheroid in an edge-wise orientation while the hydrodynamic torque tries to orient it to a broad-side on orientation. Below the velocity threshold, the buoyancy torque dominates; resulting in the onset of reorientation instability. Finally, the asymmetry in the distribution of the baroclinic vorticity generation term around the spheroids explains the onset of the reorientation instability.  
\end{abstract}


\section{Introduction}
Particles settling in a fluid medium under the influence of gravity has historically been a widely investigated research problem \citep{boussinesq1985resistance, basset1888treatise, gatignol1983faxen, maxey1983equation, magnaudet1997forces}. In the past few decades, researchers have devoted many efforts to understand the effects of fluid density stratification on the settling dynamics of spherical particles, mainly motivated by geophysical applications \citep{ardekani2017transport,Doostmohammadi2014b, Magnaudet2019}. The most notable effect of density stratification on the motion of a spherical particle is drag enhancement. This observation has been confirmed by experiments \citep{lofquist1984drag, srdic1999gravitational, yick2009enhanced}, theory \citep{mehaddi2018inertial} and computations \citep{torres2000flow, hanazaki2009schmidt}. The immediate effect of this drag enhancement is to reduce the settling velocity of a sphere falling through a stratified fluid under the influence of gravity, an effect which should therefore be considered in large-scale transport models of environmental interest \citep{Doostmohammadi2014b}.  

Fluid stratification also modifies the flow structures around spherical particles in interesting ways. Depending on the Reynolds number of the moving particle, $Re=UD/\nu$, and the Froude number of the flow, $Fr=U/ND$, a variety of jet structures can be observed \citep{hanazaki2009jets} behind a sphere with diameter $D$ moving vertically with a velocity $U$ in a stratified fluid with kinetic viscosity $\nu$ and Brunt–Väisälä frequency $N$. The formation of the jet influences a variety of phenomena in the oceans, such as the vertical movement of zooplankton and buoys used for ocean observation. Owing to the ubiquity of the density stratification due to salinity and/or temperature gradients in nature, e.g., in the atmosphere, lakes, and oceans \citep{jacobson2005fundamentals,macintyre1995accumulation}, it is obvious that studying how the density stratification influences the dynamics of settling/moving particles is crucial to understand a plethora of natural phenomenon.

In oceans, the top layer, $O \approx (1-1000)m$ deep, is associated with intense biological and ecological activities which are strongly influenced by the density stratification. The formation of algal blooms has been known to be a direct consequence of marine organisms' interactions with density stratification \citep{macintyre1995accumulation}. Stratification significantly alters the stability, interaction, and nutrient uptake of organisms \citep{ardekani2010stratlets, doostmohammadi2012low,  more2020interaction, more2020motion}. Stratification impacts carbon fluxes into the ocean by inhibiting the descent of marine snow particles (aggregates $>$ 0.5mm in diameter) \citep{alldredge1989direct}. Furthermore, the vertical density stratification promotes accumulation of marine snow \citep{alldredge1989direct} and of phytoplankton \citep{cloern1984temporal}. The bio-convection in the oceans is an important step in the carbon cycle and is responsible for transferring about 300 million tons of carbon from the atmosphere to the oceans every year \citep{henson2011reduced, stone2010invisible}. These observations make it imperative to investigate the role of density stratification on the dynamics of settling particles. However, the particles/organisms which are influenced by stratification are not exactly spherical. They come in a variety of shapes \citep{morris1980physiological}. The most common shapes that can be imagined are plate-like flat \citep{gibson2007inherent} or rod-like elongated \citep{bainbridge1957size}. The extra degree of freedom introduced by the anisotropy of the particle shape leads to interesting settling dynamics. 

In a homogeneous fluid, the anisotropy of the settling particle shape leads to more convoluted phenomena like breaking of the flow axial symmetry, oscillatory settling path, and wake instability not observed for a spherical particle \citep{ern2012wake}. The influence of the body degrees of freedom on the wake dynamics along with the vorticity production at the body surface can explain the wake instabilities and their consequences on the body path. The influence of the degree of freedom introduced by the particle shape anisotropy is clear if we look at the critical Reynolds number, $Re_c$, above which the unsteady motion of settling particle sets in. The critical value for the spheroidal particles is found to be lower than the corresponding value of $Re_c$ for a spherical particle \citep{fernandes2007oscillatory, Ardekani2016}. Furthermore, it has been shown that, for spherical particles, the critical Reynolds number $Re_c$ is larger than that at which the wake loses its symmetry for a fixed sphere \citep{jenny2004instabilities} with $Re_c$ depending on the ratio, $\rho_r = \rho_p/\rho_f$, between the particle density, $\rho_p$ and the fluid density, $\rho_f$. 

The path oscillations for infinitely long circular cylinders falling freely in the direction perpendicular to their axis can also be explained with the help of wake instabilities \citep{namkoong2008numerical}. The role of particle shape anisotropy is to make the generalized inertia tensor anisotropic resulting in an added torque that essentially links the particle translational and rotational velocities. Specifically, for oblate spheroids, four different states for the particle motion are observed for Galileo number, $Ga=\sqrt{(|\rho_r-1|gD^3)/\nu^2}$, between 50 to 250 \citep{chrust2012etude, Ardekani2016} for aspect ratio, $\mathcal{AR}=1/3$. Here $g$ is the acceleration due to gravity and $D$ is the diameter of a sphere with the same volume as the spheroidal particle. The transition between the four states takes place at $Ga \approx$ 120, 210, and 240 for $\rho_r=1.14$. These findings are similar to a disk or a circular plate type particles \citep{ern2012wake}. These observations for an oblate spheroid are consistent with the buoyancy-driven motion of a disk in a homogeneous fluid and have been confirmed by experiments \citep{willmarth1964steady, field1997chaotic, fernandes2007oscillatory}, numerical calculations \citep{auguste2010bifurcations, chrust2013numerical} and theory \citep{fabre2008bifurcations, tchoufag2014global}.

On the other hand, the motion of freely falling prolate spheroids has been subjected to much fewer investigations than the disk-shaped or oblate spheroid. The onset of secondary motions for prolate spheroids occurs at a considerably lower $Ga$ than for an oblate spheroid. The peculiar feature of settling prolate spheroids is that they attain a terminal rotational velocity about the axis parallel to the vertical direction in which it is falling freely for $Ga>70$ in the case of aspect ratio, $\mathcal{AR}=3$. This behavior is again related to the wake instability and can be explained by the four thread-like quasi-axial vortices appearing in the wake of a prolate spheroid \citep{Ardekani2016}. 

Although particle shape anisotropy leads to path and wake instabilities in the settling motion of particles in a homogeneous fluid, it does not change the steady-state settling orientations of the particles. The spheroidal particles have been observed to settle such that their long axis is always perpendicular to the settling direction \citep{feng1994direct, fernandes2007oscillatory, fernandes2008dynamics, Ardekani2016} for $Re>0.1$. In addition, the particles reach a constant terminal velocity when falling freely in a homogeneous fluid. The terminal velocity depends on the $Ga$ and the aspect ratio of the particles.  

Introducing a linear density stratification in the fluid significantly alters the settling dynamics of spherical as well as non-spherical particles. A first significant departure from the settling in homogeneous fluids is the absence of a terminal velocity. This is due to the fact that stratification increases the drag experienced by the settling particles which therefore reduces their settling speeds. In addition, increasing buoyancy leads to the deceleration of the particle as it approaches the neutrally buoyant position and can cause oscillations in the particle velocity depending on the strength of stratification \citep{Doostmohammadi2014b}.

Recently, researchers have started exploring the effects of stratification on the settling dynamics of anisotropically shaped particles in a stratified fluid. Most of the investigations are limited to disks. Experiments of a disk settling encountering a stratified two-layer fluid show that the disk reorients itself such that the long axis is perpendicular to the vertical direction while it moves through the transition layer between the two fluids \citep{mrokowska2018stratification, mrokowska2020dynamics, mrokowska2020influence}. Further, a disk settling in a linearly stratified fluid has been observed to go through three regimes as it settles. First, there is a quasi-steady state with the disk long axis perpendicular to the vertical direction. Then, there is a change in the stability for the disk orientation when it changes its orientation from long axis normal to the vertical direction (broad-side on) to long axis parallel to the vertical axis (edge-wise). Finally, the disk settles edge-wise at its neutrally buoyant position \citep{mercier2020settling}. As concerns prolate spheroids, we can only mention the numerical study by \cite{doostmohammadi2014reorientation}, on the settling across a density interface. Hence, we are still far from completely understanding the settling and orientation dynamics of spheroidal shaped particles in a stratified fluid.  

To gain some new understanding of the problem, we numerically simulate the free-falling motion of spheroidal particles, an oblate spheroid with an aspect ratio, $\mathcal{AR}=1/3$ and a prolate spheroid with $\mathcal{AR}=2$, in a linearly stratified fluid for different $Ga$ and $Fr$ values. The aim of this effort is to investigate the possible mechanism which leads to the orientational instability of a freely falling spheroidal object in a linearly stratified fluid. 
 
\section{Governing equations}

We present the governing equations and the solution methodology implemented to solve them in this section. To calculate the flow field $\mathbf{u}$, we solve the Navier-Stokes equations and the continuity equation. We assume the fluid to be Newtonian and incompressible which results in the following equations, written in the reference frame translating with the particle velocity $\mathbf{U}_p$:

\begin{equation}
 \rho_f\left( \frac{\partial\mathbf{u}}{\partial{t}} + \left( \mathbf{u}-\mathbf{U}_p \right) \cdot \nabla{\mathbf{u}} \right) = -\nabla{p} + \mu{\nabla}^{2}\mathbf{u}+\rho_f \left( {\mathbf{g}}+\mathbf{f} \right),  
  \label{eq:NS}
\end{equation}
\begin{equation}
  \nabla \cdot \bf{u} = 0,
  \label{eq:cont}
\end{equation}
where $\mathbf{u}$ is the fluid velocity, $\rho_f$ is the density field, $\mathbf{U}_p$ is the instantaneous particle translational velocity, $p$ is the pressure, $\mu$ is the fluid dynamic viscosity, $\mathbf{g}$ is the acceleration due to gravity. The additional term $\mathbf{f}$ on the right-hand-side of (\ref{eq:NS}) accounts for the presence of particle, modelled with the immersed boundary method (IBM). This IBM force is active in the immediate vicinity of a particle to impose the no-slip and no-penetration boundary conditions indirectly. In other words, the force distribution $\mathbf{f}$ ensures that the fluid velocity at the surface is equal to the particle surface velocity ($ \mathbf{U}_p + \boldsymbol{\omega}_p \times \mathbf{r}$ ).

The particle motion is solution of the following Newton-Euler Lagrangian equation of particle motion:
\begin{equation}
  {\rho}_p {V_{p}} \frac{\textrm{d}\mathbf{U}_p}{\textrm{d}t} =   \oint_{\partial{V}_p} \boldsymbol{\tau} \cdot \mathbf{n} \, \textrm{d}A, 
  \label{eq:tran}
\end{equation}
\begin{equation}
  \frac{ \textrm{d} \left( \mathbf{I}_p \boldsymbol{\omega}_p \right) }{\textrm{d}t} = 
  \oint_{\partial{V}_p} {\mathbf{r} \times \left( \boldsymbol{\tau} \cdot \mathbf{n} \right)  \, \textrm{d}A},
  \label{eq:ang}
\end{equation}
here $\mathbf{U}_p$ and $\boldsymbol{\omega}_p$ are the particle translational and angular velocities. $\rho_p$, $V_p$ and $\mathbf{I}_p$ represent the particle density, particle volume and the particle moment of inertia matrix. $\mathbf{n}$ is the unit normal vector pointing outwards on the particle surface, while $\mathbf{r}$ is the position vector from the particle's center. $\boldsymbol{\tau} = -p\mathbf{I}+\mu \left( \nabla\mathbf{u} + \nabla\mathbf{u}^{T} \right)$ is the stress tensor and its integration on the particle surface accounts for the fluid-particle interaction.

Accounting for the inertia and Buoyancy forces of the fictitious fluid phase inside the particle volume and using IBM, Eqs.~\ref{eq:tran} and \ref{eq:ang} are rewritten as below:
\begin{equation}
\label{eq:NewtonEulerHeat}
\rho_p V_p \frac{ \mathrm{d} \textbf{U}_{p}}{\mathrm{d} t}   \, \approx \, -\rho_0  \sum\limits_{l=1}^{N_L} \textbf{F}_l \Delta V_l + \rho_0  \frac{ \mathrm{d}}{\mathrm{d} t} \left( \int_{V_p} \textbf{u} \mathrm{d} V  \right)
- \int_{V_p}  \rho_f  \textbf{g} \mathrm{d} V +  \rho_p V_p \textbf{g}  \, ,
 \end{equation}
\begin{equation}
\label{eq:NewtonEulerHeat2}
\begin{split}
 \frac{ \mathrm{d} \left( \textbf{I}_p \, \pmb{\omega}_{p} \right) }{\mathrm{d} t}  \, \approx \, -\rho_0 \sum\limits_{l=1}^{N_L} \textbf{r}_l \times \textbf{F}_l \Delta V_l + \rho_0  \frac{ \mathrm{d}}{\mathrm{d} t}  \left( \int_{V_p} \textbf{r}  \times \textbf{u} \mathrm{d} V  \right) - \int_{V_p} \textbf{r} \times \rho_f  \textbf{g} \mathrm{d} V \, , 
 \end{split}
 \end{equation}
where the first two terms on the right-hand-side of the equations denote the hydrodynamic force and torque $F_h$ and $T_h$, while the third terms indicate the buoyancy force and torque $F_b$ and $T_b$. More details on the numerical model can be found in \cite{niazi_ardekani2018,majlesara2020numerical}.

The vertical variation in the fluid density can either be due to the vertical variation in the fluid temperature or salinity or both. For this study we consider the density stratification to arise from the fluid temperature variation. Thus, the particle sediments in a linearly density stratified fluid with the initial vertical density stratification given by $\bar{\rho}(z) = \rho_0 - \gamma{z}$. $\rho_0$ is the reference density, $\gamma$ is the vertical density gradient and $z$ is the vertical coordinate. The fluid density increases linearly in the downward $z$ direction (gravity direction). The density variation across thermocline occurs due to the vertical variation in the temperature, since $\rho = \rho_0 \left( 1 - \beta\left( T - T_0\right) \right)$, where $\beta$ is the coefficient of thermal expansion, $T$ is the temperature field and $T_0$ is the reference temperature corresponding to the reference density, $\rho_0$. Thus, the initial temperature of the background fluid is given by, $\bar{T}(z) = T_0 + (\gamma/\beta\rho_0)z$. The energy equation for an incompressible fluid flow in the frame of reference moving with the particle can be simplified to, 
\begin{equation}
   \frac{\partial{T}}{\partial{t}} + \left( \mathbf{u}-\mathbf{U}_p \right) \cdot \nabla{T}  = \nabla \cdot \left( \alpha T \right). 
  \label{eq:temp}
\end{equation}
$\alpha$ is the thermal diffusivity. We split the temperature field in the linear component and the perturbation ($T^{'}$) as $T = \bar{T}(z) + T^{'}$. We solve for the temperature perturbation, $T^{'}$, and add it to the linear component to get the temperature field at any instance of time. Eq. (\ref{eq:temp}) can be rewritten in term of the temperature perturbation field, $T^{'}$ as follow:
\begin{equation}
   \frac{\partial{T^{'}}}{\partial{t}} + \left( \mathbf{u}-\mathbf{U}_p \right) \cdot \nabla({\bar{T}(z) + T^{'}}) =  \nabla \cdot \left( \alpha T^{'} \right). 
  \label{eq:temp_per}
\end{equation}
We set $\alpha = 0$ for the particle \citep{Doostmohammadi2014b} and $\alpha=\nu/Pr$ for the fluid phase. $\nu$ is the fluid kinematic viscosity and $Pr$ is the Prandtl number. This is equivalent to the insulating/impermeable/no-flux boundary condition on the surface of the particle \citep{hanazaki2009schmidt, Doostmohammadi2014b} which is also true if the stratifying agent is salt or having an adiabatic particle. We also investigate the effects of relaxing the no-flux boundary condition on the particle surface by varying $\alpha$ for the particle by changing the particle heat conductivity, $k$, in Sec.~\ref{sec:k}. 

We use $Pr=0.7$ for this study, which is the $Pr$ value for temperature stratified atmosphere. In a stratified fluid, a density boundary layer is present in addition to the velocity boundary layer near the particle surface. The thickness of this density boundary layer scales as $\approx O(D/\sqrt{RePr})$. For accurate resolution of the flow within this boundary layer, it is necessary to have at least a few grid points in it. This imposes limitations on the maximum mesh size that can be used for the simulations. Owing to large size of the domain, using such a fine grid becomes computationally expensive. Hence, we use a smaller value for the $Pr$ which enables us to resolve the fluid flow as well as the density field in both the boundary layer and the outside. It has been shown in previous studies that, changing the value of $Pr$ merely changes the magnitudes of the velocities of the objects \citep{Doostmohammadi2014b} moving in a stratified fluid conserving the overall qualitative trends and behaviors. The details on the numerical algorithm to solve the governing equations and validations of the numerical tool are provided elsewhere \citep{Ardekani2016, Ardekani2018, niazi2019numerical} and hence not discussed here.

\subsection{Dimensionless parameters \& simulation conditions}

Re-writing the equations in the non-dimensional form results in the following equations: 

\begin{equation}
\frac{\partial \mathbf{u}}{\partial{t}} + \left( \left( \mathbf{u}-\mathbf{U}_p \right) \cdot \nabla \right) \mathbf{u}  = -\nabla{P} +  \frac{1}{Re}{\nabla}^{2}\mathbf{u}+ \frac{Ri}{Re} T + \mathbf{f},  
  \label{eq:NSn}
\end{equation}
\begin{equation}
\frac{\partial T}{\partial{t}} + \left( \mathbf{u}-\mathbf{U}_p \right) \cdot \nabla{T} + \mathbf{u} \cdot \hat{e}_g  =  \frac{1}{\rho^* C^*_p} \nabla \cdot \left( \frac{k^*}{Re Pr} \nabla T \right),  
  \label{eq:Tn}
\end{equation}
\begin{equation}
  \nabla \cdot \bf{u} = 0,
  \label{eq:contn}
\end{equation}
 where, $\mathbf{u}$, $T$ and $P$ now denote perturbations in velocity, temperature and pressure field. Temperature is normalized with the temperature difference of $1$ equivalent particle diameter in the gravity direction. $\rho^*$, $C^*_p$ and $k^*$ indicate particle density, heat capacity and heat conductivity ratio ($\rho_r$, ${C_p}_r$ and $k_r$) inside the particles and are equal to $1$ in the fluid region. We investigate the sedimentation of spheroidal particles in a quiescent but linearly density stratified fluid with finite inertia. The non-dimensional parameters defining the problem are described below.

\begin{figure*}
    \centering
        \includegraphics[width=0.7\textwidth]{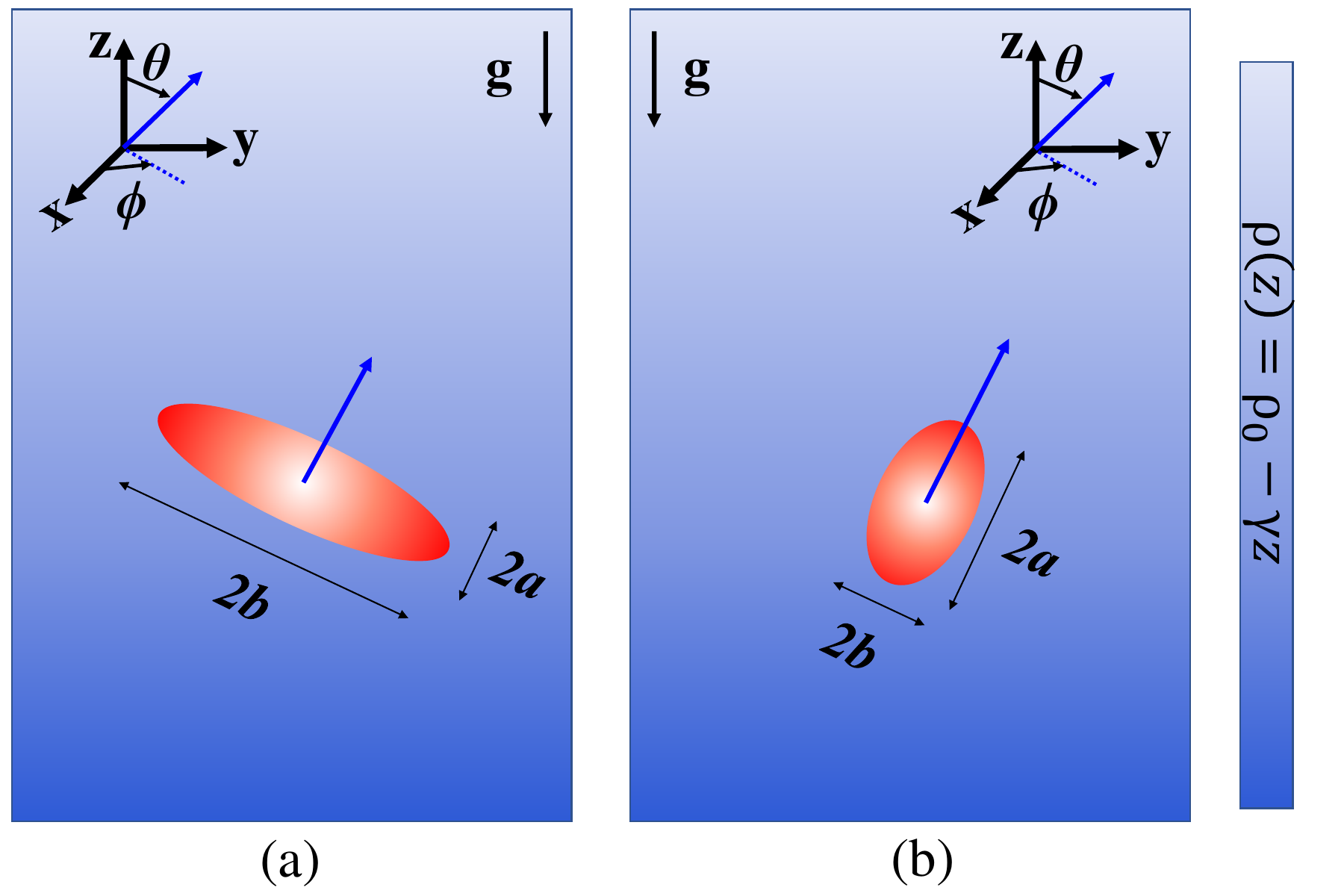}
    \caption{\label{fig:setup_all}Schematic of the settling spheroidal objects in a linearly density stratified fluid. a) Oblate spheroid ($\mathcal{AR}<1$) and b) prolate spheroid ($\mathcal{AR}>1$). Here $a$ and $b$ are the semi-major and the semi-minor axis. The aspect ratio $\mathcal{AR}$ is given by $a/b$. For spherical particles $\mathcal{AR}=1$. The orientation of the particle is quantified in terms of the polar angle $\theta$ and the azimuthal angle $\phi$ for a vector directed along the major axis of the spheroids. The coordinate system used is shown at the top of the figures.}
\end{figure*}

\begin{enumerate}

    \vspace{5pt}

    \item The Reynolds number, $Re= \mathcal{U} D/\nu$, which quantifies the relative importance of the inertial and the viscous forces. Here $Re$ is equivalent to Galileo number, $Ga$, with the reference velocity $\mathcal{U}$ defined as $\mathcal{U}=\sqrt{D|\rho_r-1|g}$.  $D$ is the length scale corresponding to the particle size, set as the diameter of a sphere with the same volume as that of the spheroidal particle ($D=(b^2a)^{(1/3)}$). $a$ and $b$ denote the polar and the equatorial radius of the spheroidal particle.
    
    \vspace{5pt}
    
    \item The Richardson number, $Ri=\gamma g D^3 / (\mathcal{U} \rho_0 \nu) = D^3N^2/ (\mathcal{U} \nu)$, which quantifies the relative importance of buoyancy and the viscous forces. $N= {(\gamma{g}/\rho_{0})}^{1/2}$ is the Brunt–Väisälä frequency. It is the natural frequency of oscillation of a vertically displaced fluid parcel in a stratified fluid. 
    \vspace{5pt}
    
    \item The Prandtl number, $Pr = C_p \mu / k$, defined as the ratio of momentum diffusivity to thermal diffusivity inside the fluid region.
    
    \vspace{5pt}
    
    \item The particle density ratio, {indicating the ratio between the particle density and the reference density of the fluid.} $\rho_r = \rho_p/\rho_0$. 
    
    \vspace{5pt}
    
    \item The particle heat conductivity ratio, $k_r = k_p/k_f$, with subscripts ${}_p$ and ${}_f$ denoting the particle phase and the fluid phase.
    
    \vspace{5pt}
    
    \item The particle heat capacity ratio, ${C_p}_r = {C_p}_p / {C_p}_f$. 
    
    \vspace{5pt}
    
    \item the particle aspect ratio, $\mathcal{AR}=a/b$.
    
\end{enumerate}
    \vspace{5pt}
 Finally, we note that $Ga$ and $Re$ are defined in the same way and used in this study interchangeably. This is different to the effective Reynolds number defined in terms of the particle velocity ($Re(t) = U_p(t)D/\nu$) that changes in time. The characteristic time scale, $\tau$, used to make $t$ dimensionless is chosen to be $\tau=D/\mathcal{U}$.


We simulate the sedimenting motion of a spheroidal shaped particle in a linearly density stratified fluid using a 3D rectangular domain of size $20D \times 20D \times 80D$ ($10D \times 10D \times 40D$) for an oblate (prolate) spheroid with grid size equal to $D/32$ ($D/48$), resulting in $\approx O(10^9)$ ($\approx O(5 \times 10^8)$) grid points. We use periodic boundary conditions for the velocity field and the temperature perturbations on all the sides of the domain. We consider an oblate particle with aspect ratio, $\mathcal{AR} = a/b = 1/3$ (fig.~\ref{fig:setup_all}a) and a prolate particle with $\mathcal{AR} = a/b = 2$ (fig.~\ref{fig:setup_all}b). Since we solve the flow field in the frame translating with the particle, the particle stays at its initial position, i.e., ([10D, 10D, 20D] for an oblate and [5D, 5D, 10D] for a prolate spheroid). Depending on the hydrodynamic torque it experiences, the particle can rotate freely. The orientation of the spheroid is measured in terms of the polar angle $\theta$, which is the angle made by the major axis of the spheroid with the $z-$axis as shown in fig.~\ref{fig:setup_all}. In the atmosphere, the typical values of $N$ is $10^{-2} s^{-1}$ while in the ocean $N$ is around $10^{-4}-0.3 s^{-1}$ depending on the strength of density stratification \citep{wust2017variability, geyer2008quantifying}. We perform simulations for $Re = 80-250$, while we vary $Ri$ from $0-10$ which are consistent with the typical value of $N$ mentioned above. $Ri=0$ represents a particle settling in a homogeneous fluid with a constant density. We fix the density ratio, $\rho_r=1.14$ in all the cases. {The temperature inside the particle is set similar to the surrounding fluid initially, resulting in a domain with zero temperature fluctuations at the start of the simulations.} Table~\ref{tab:tab1} summarizes the values of all the relevant parameters investigated. 

\begin{table}

 \begin{center}
\begin{tabular}{c c c c c c c}
  $\mathcal{AR}$ \, & ${C_p}_r$  & $k_r$  & $\rho_r$  & $Re$ ($Ga$)  & $Ri$  & $Pr$ \\[3pt]
\hline
 $1/3$  & $1$  & $(0, 0.001, 1)$  & $1.14$  & $(80, 170, 210, 250)$  & $0 - 5$  & $0.7$  \\
 $2$  & $1$  & $(0, 0.001, 1)$  & $1.14$  & $(80, 170)$  & $0-10$  & $0.7$  \\
\end{tabular}
\caption{\label{tab:tab1}Values of relevant parameters investigated in this study}
\end{center}
\end{table}

\section{Results and discussion}
The following subsections present the simulation results for settling spheroids in a stratified fluid. We present the settling velocities and orientations of the spheroids for the range of $Re$ and $Ri$ investigated. We first present and discuss the results for an oblate spheroid followed by the results for the prolate spheroid. We compare the data from the stratified fluid case with the data from the homogeneous fluid case for better understanding the results. We use ``broad-side on" to indicate an orientation of the spheroidal particles such that their broader side is horizontal, i.e., $\theta=0^{\circ}$ for an oblate spheroid and $\theta = 90^{\circ}$ for a prolate spheroid. On the other hand, ``edge-wise" indicates the orientation of the particles in which their broader side is perpendicular to the horizontal direction, i.e., $\theta = 90^{\circ}$ for an oblate spheroid and $\theta = 0^{\circ}$ for a prolate spheroid. 

\subsection{Settling dynamics of an oblate spheroid in a stratified fluid}

\subsubsection{Fluid stratification slows down and reorients a settling oblate spheroid}\label{sec:oblmain}

\begin{figure*}
    \centering
     \begin{subfigure}[t]{0.45\textwidth}
        \centering
        \includegraphics[width=\textwidth]{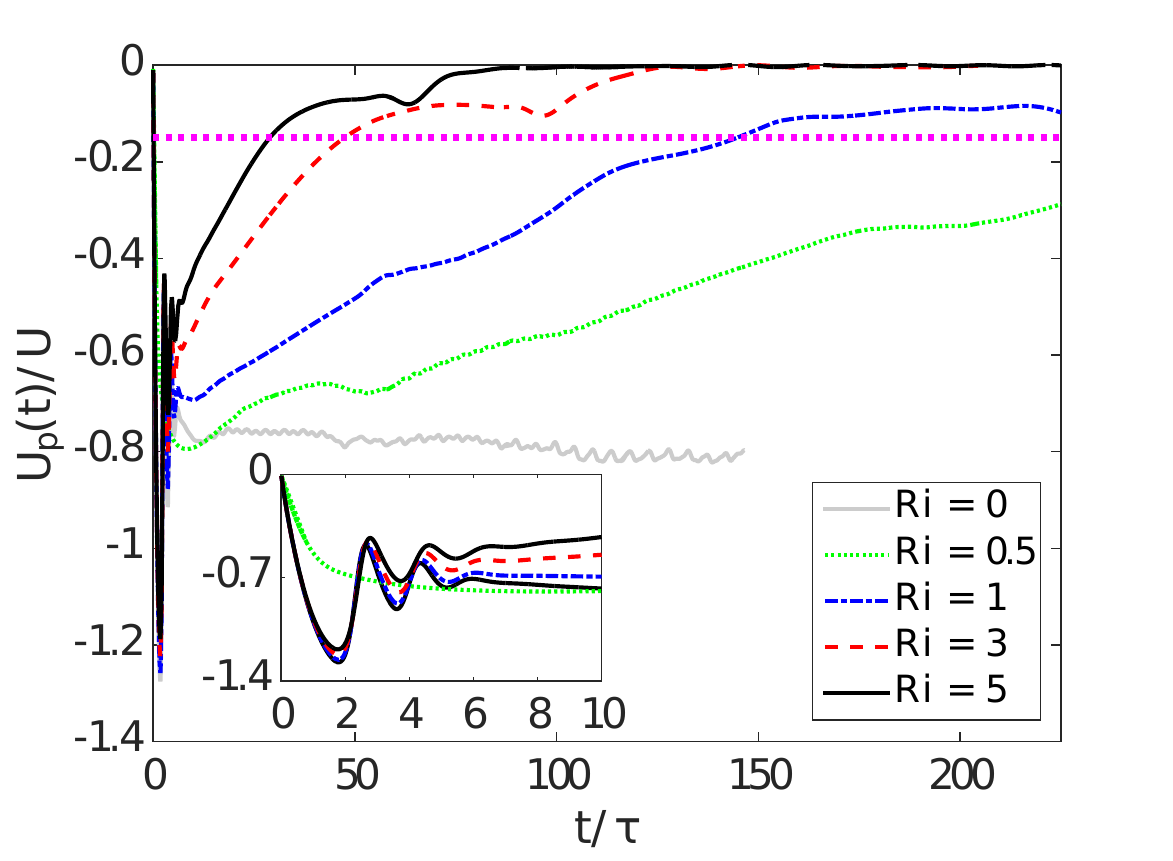}
    	\caption{\label{fig:210_vel}}
       
    \end{subfigure}
    ~ 
    \begin{subfigure}[t]{0.45\textwidth}
        \centering
        \includegraphics[width=\textwidth]{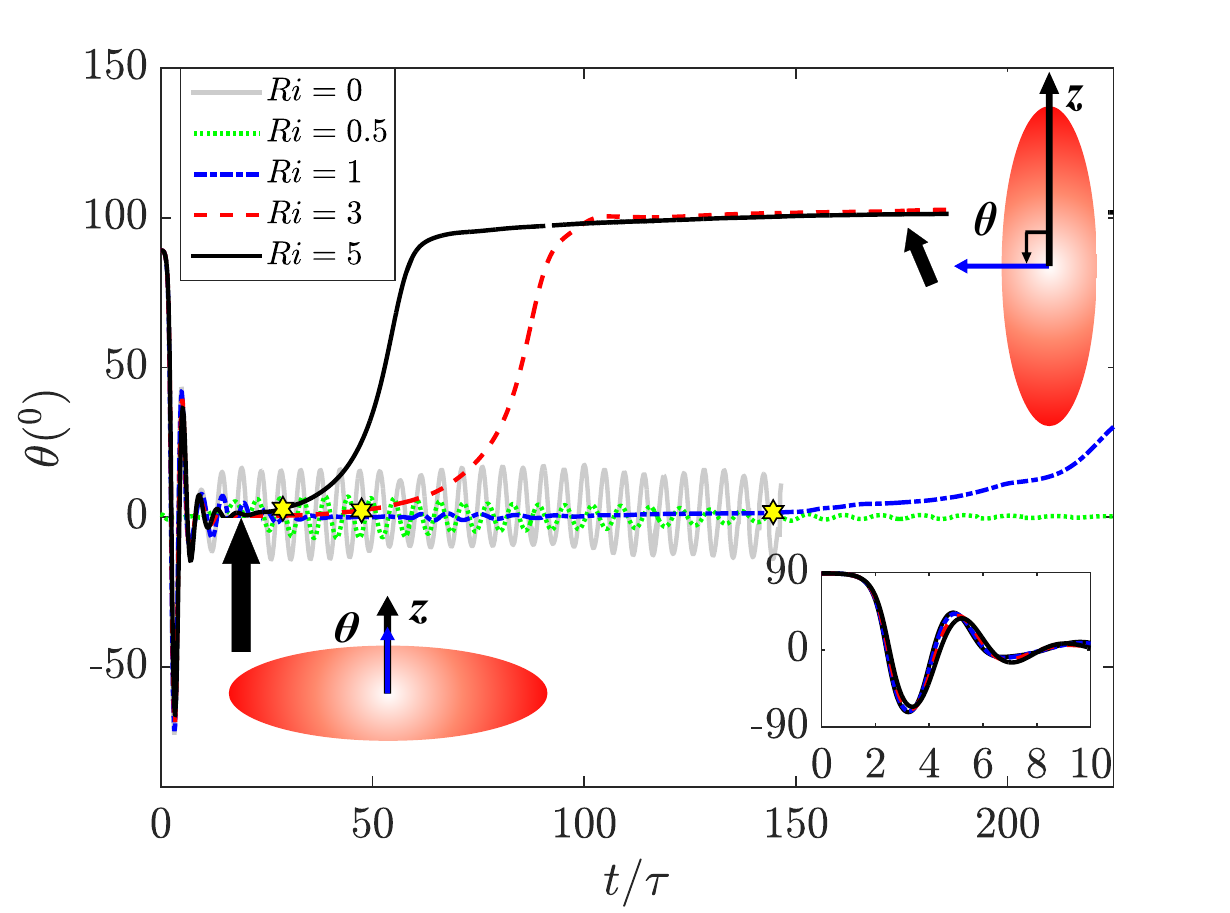}
    	\caption{\label{fig:210_or}}
      
    \end{subfigure}

    \caption{\label{fig:Ga210}Settling dynamics of an oblate spheroid ($\mathcal{AR}=1/3$) with $Re=210$ in a homogeneous fluid ($Ri=0$) and a stratified fluid with different $Ri$ values: a) Settling velocity evolution, b) spheroid orientation evolution versus time. The insets in both the figures show the initial oscillations with decreasing amplitudes in the velocity and orientation of the spheroid. The oblate spheroid attains a steady state terminal velocity and oscillates about broad-side on orientation in a homogeneous fluid after the initial transients. Stratification leads to a reduction in the spheroid velocity and a continuous deceleration of the spheroid velocity until it stops. The magnitude of the deceleration increases with stratification. In addition, the steady state orientation of the oblate spheroid changes from broad-side on (i.e., $\theta = 0^{\circ}$) in a homogeneous fluid to edge-wise (i.e., $\theta \approx 90^{\circ}$) in a stratified fluid. The transition in the orientation starts once the magnitude of the dimensionless spheroid velocity drops below a particular threshold. Here for $|U_p/U| < 0.15$. The onset of transition in the spheroid orientation is denoted by dotted horizontal line in (a) and yellow stars in (b).}
\end{figure*}

This subsection presents the simulation results for an oblate spheroid with $\mathcal{AR}=1/3$ settling in a stratified fluid. The oblate spheroid starts from rest in an initially quiescent fluid. The spheroid velocity then evolves depending on the hydrodynamic and buoyancy forces acting on it as the flow evolves. We initialize the orientation of the oblate spheroid such that $\theta = 90^{\circ}$ or in edge-wise orientation. In a homogeneous fluid, the oblate spheroid accelerates and attains a terminal velocity after the initial transients (which are due to the oscillations in the spheroid orientation) as shown in fig.~\ref{fig:210_vel}. In addition, as the oblate spheroid accelerates, it topples from its initial edge-wise to a broad-side on orientation. However, due to its inertia and periodic shading of hair-pin like vortex structures from alternate edges \citep{Ardekani2016}, it oscillates around the broad-side on ($\theta=0^{\circ}$) orientation. So, for $Re=210$, an oblate spheroid settles in an oscillatory orientation about $\theta=0^{\circ}$ as shown in fig.~\ref{fig:210_or} for $Ri=0$. The oscillations are not present at lower $Re$ ($<120$) \citep{mehaddi2018inertial}. 

\begin{figure*}
    \centering
     \begin{subfigure}[t]{0.45\textwidth}
        \centering
        \includegraphics[width=\textwidth]{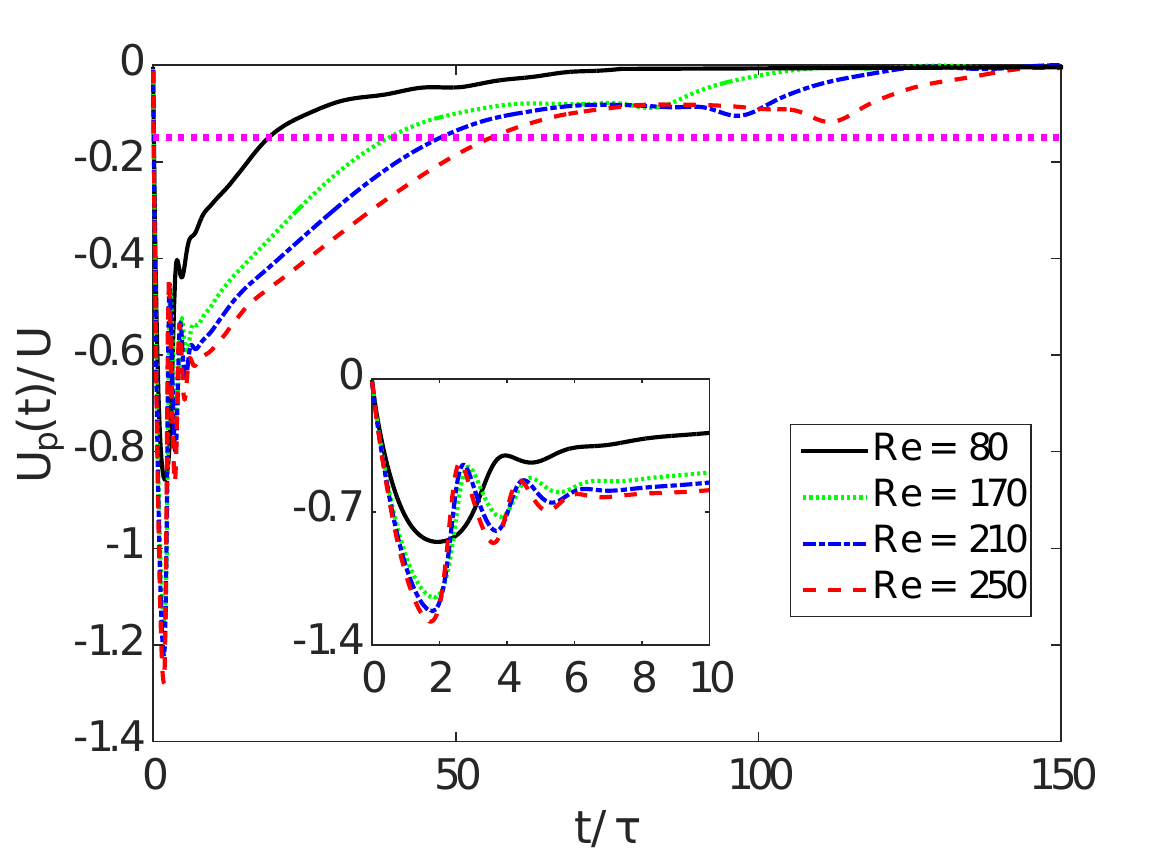}
    	\caption{\label{fig:Re_vel}}
       
    \end{subfigure}
    ~ 
    \begin{subfigure}[t]{0.45\textwidth}
        \centering
        \includegraphics[width=\textwidth]{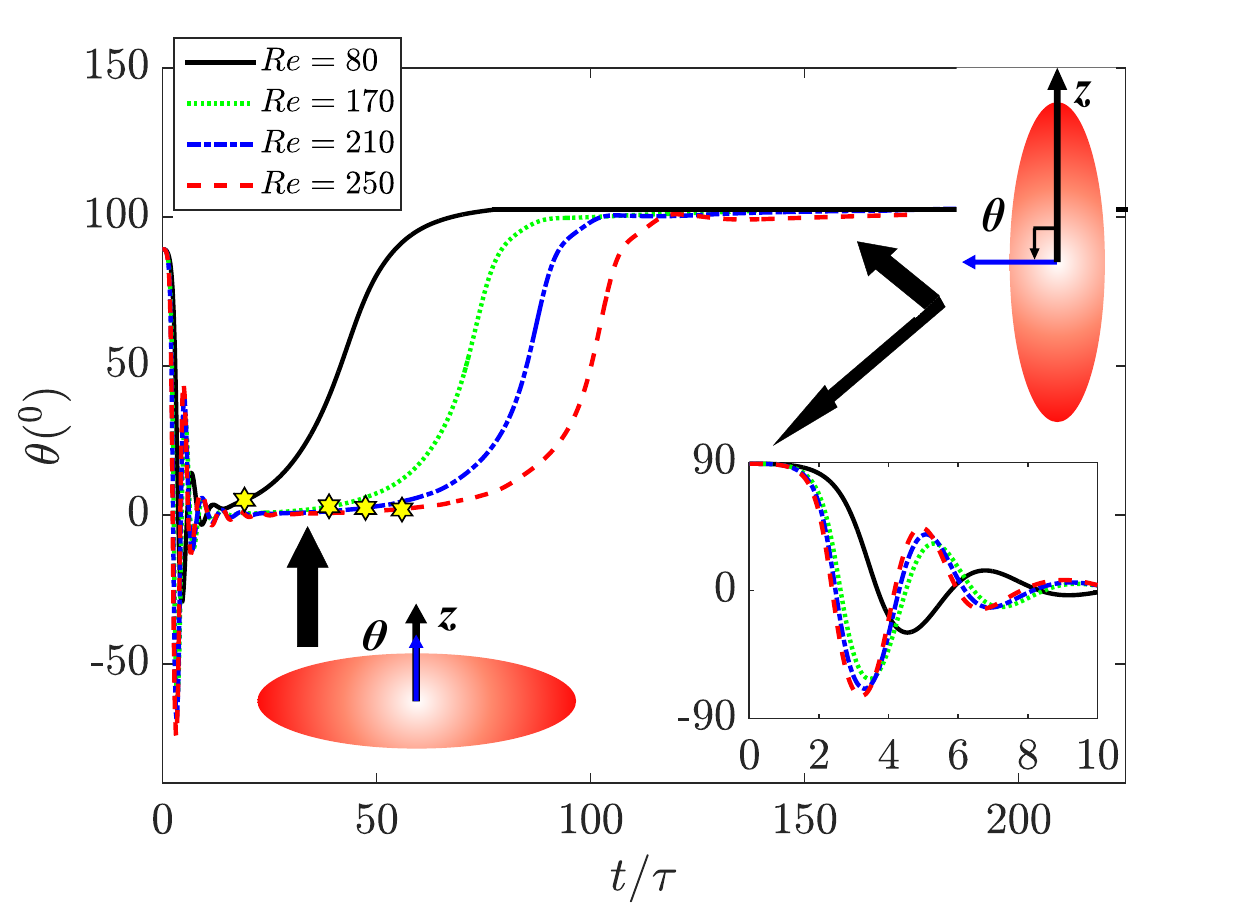}
    	\caption{\label{fig:Re_or}}
  
    \end{subfigure}
    \caption{\label{fig:Ri3}Settling dynamics of an oblate spheroid in a stratified fluid and $Ri=3$ with different $Re$ values: a) Settling velocity, b) spheroid orientation evolution versus time. The insets show the initial oscillations with decreasing amplitude. The oblate spheroid attains a steady state terminal velocity and orientation (broad-side on, $\theta= 0 ^{\circ}$) in a homogeneous fluid. Stratification leads to a reduction in the spheroid velocity and a continuous deceleration of the spheroid velocity until it stops. The magnitude of the deceleration decreases with increasing the particle inertia. In addition, the steady state orientation of the oblate spheroid changes from broad-side on (i.e., $\theta = 0^{\circ}$) in a homogeneous fluid to broad-side perpendicular (i.e., $\theta \approx 90^{\circ}$) in a stratified fluid. The transition in the orientation starts once the magnitude of the dimensionless spheroid velocity drops below a threshold. Here for $|U_p/U| < 0.15$. The onset of transition in the spheroid orientation is denoted by the dotted horizontal line in (a) and the yellow stars in (b).}
\end{figure*}

Introducing density stratification in the fluid significantly changes the settling dynamics of an oblate spheroid. This is shown in fig.~\ref{fig:Ga210} for $Re=210$ and various $Ri$ as well as in fig.~\ref{fig:Ri3} for $Ri=3$ and various $Re$ values. As the oblate spheroid sediments in a stratified fluid, it moves from a region with lighter fluid into a region with heavier fluid. As a result, it experiences an increasing buoyancy force which essentially opposes its settling motion. Hence, the particle cannot attain a steady state terminal velocity. This phenomenon is clearly depicted in fig.~\ref{fig:210_vel} and~\ref{fig:Re_vel} where the particle velocity decreases continuously after the initial transients. The suppression of the fluid flow due to the tendency of the displaced iso-density difference surfaces (isopycnals) to return to their original locations is another reason for the reduction in the particle velocity (see the detailed discussion in Sec.~\ref{sec:oblres} and fig.~\ref{fig:obl_wg}).

An increase in the stratification strength of the background fluid increases the magnitude of the particle deceleration. This is expected as the magnitude of the buoyancy force experienced by the particle increases with the fluid stratification. As a result, the particle stops at earlier times for increasing $Ri$ values as shown in fig.~\ref{fig:210_vel}. Another consequence of this increased opposition to the settling motion is the reduction in its peak velocity when increasing the stratification as shown in fig.~\ref{fig:210_vel}. In addition, as the $Re$ of the particle increases for a fixed $Ri$, the magnitude of deceleration decreases as shown in fig.~\ref{fig:Re_vel}. This is because of the increase in the inertia of the particle with $Re$. 

\begin{figure*}
    \centering
     \begin{subfigure}[t]{0.45\textwidth}
        \centering
        \includegraphics[width=\textwidth]{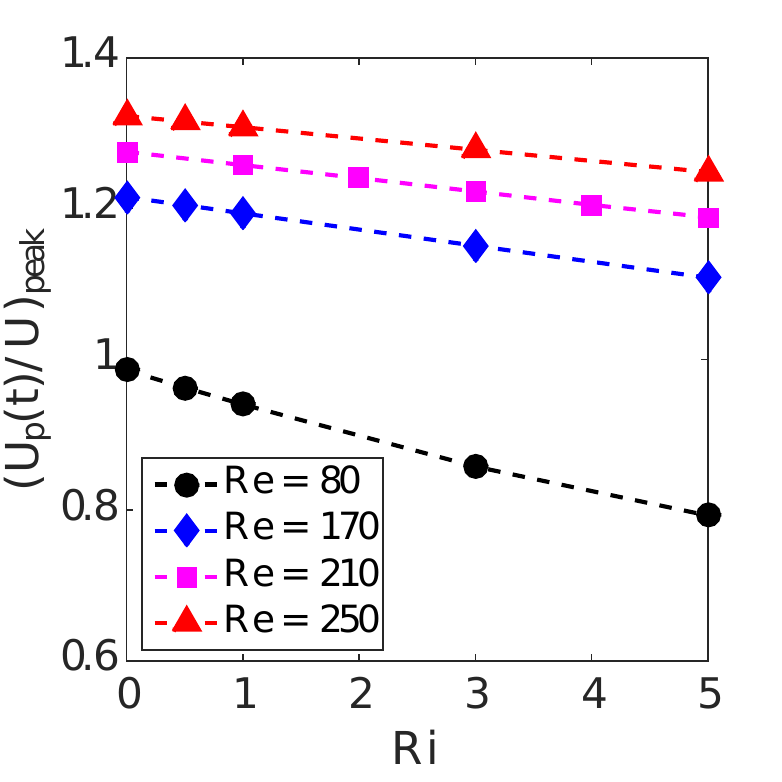}
    	\caption{\label{fig:AR3_peak}}

    \end{subfigure}
    ~ 
    \begin{subfigure}[t]{0.45\textwidth}
        \centering
        \includegraphics[width=\textwidth]{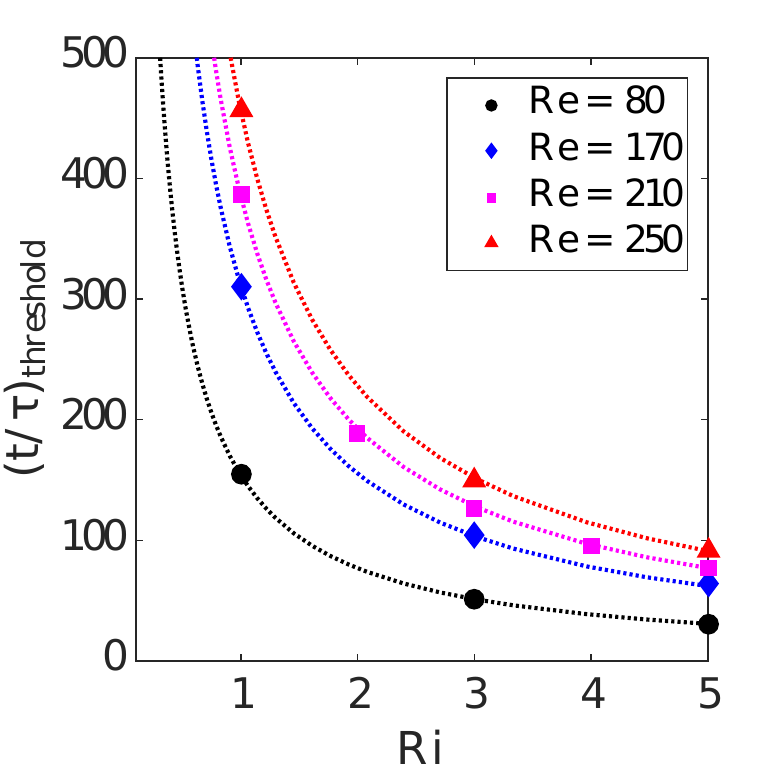}
    	\caption{\label{fig:AR3_th}}
    
    \end{subfigure}
    \caption{\label{fig:AR3}Effect of inertia and stratification strength on a) the peak velocity, $(U_p(t)/U)_{peak}$, of a settling oblate spheroid with $\mathcal{AR}=1/3$. The peak velocity attained by the particle decreases stratification and increases with increase in particle inertia, and b) the time ($(t/\tau)_{threshold}$) at which $|U_p(t)/U| < 0.15$. The dashed line in (a) is a guide to the eye. The dotted line in (b) is the $(t/\tau)_{threshold} = A*Ri^{-1}$ fit with A = 153.7, 310.5, 384.8 and 455.5 for $Re=80$, $170$, $210$ and $250$, respectively. }
\end{figure*}

A closer comparison between the time histories of the velocity and orientation reveals that, the onset of reorientation of the oblate spheroid is connected to the reduction of the settling velocity below a certain threshold. From the simulation data, we observe that, the reorientation starts once the magnitude of the dimensionless velocity of the particle falls below $\approx 0.15$. This is indicated by a horizontal dashed line in the velocity evolution plots and a star in the spheroid orientation evolution plots (see fig.~\ref{fig:Ga210} and~\ref{fig:Ri3}). This observation is consistent with the experimental and numerical study on the orientation of a settling disk in a stratified fluid by \cite{mercier2020settling}. Since stratification leads to a reduction in the particle velocity, an oblate spheroid eventually settles in an edge-wise orientation. This is because after a long enough time, the particle velocity goes below the threshold velocity for the onset of reorientation in a stratified fluid. 

We quantify the effects of fluid density stratification on the peak velocity of the particles in fig.~\ref{fig:AR3_peak}. We define the peak velocity as the maximum velocity achieved by the particles as it settles. We observe that the peak velocity decreases monotonically with the fluid stratification strength and increases with increasing $Re$. Also, the relative decrease in the peak velocity for the lowest to the highest stratification strengths explored reduces with the Reynolds number. For $Re=80$ it decreases by $\approx 20 \%$ while for $Re=250$ it decreases by $\approx 6 \%$. This is due the increase in the strength of the inertial effects as compared to the stratification effects with increasing $Re$ at fixed $Ri$. As concluded from fig.~\ref{fig:210_vel} and~\ref{fig:Re_vel}, increasing the stratification strength or reducing the inertia of the particle moves the onset of the reorientation instability to an earlier time. Fig.~\ref{fig:AR3_th} shows the effect of changing particle $Re$ and $Ri$ on the time for the onset of reorientation instability. We observe that, the time ($(t/\tau)_{threshold}$) at which particle velocity falls below the threshold velocity for the onset of reorientation instability decreases as $O(Ri^{-1})$. 

\subsubsection{Drag enhancement due to stratification}\label{sec:obl_drag}

{In the previous section, we observed that the density stratification results in a reduction in the settling velocity of the oblate and the reduction is higher for higher stratification strengths. The reason behind this decrease in the settling velocity is the entrainment of the lighter fluid by the oblate as it settles into a heavier fluid and the increase in the buoyancy forces acting on it. We can quantify the effect of buoyancy by calculating the added drag due to stratification as the oblate sediments. It has been shown in previous studies on spheres and disks that the stratification results in a significant additional drag on the settling particle \citep{yick2009enhanced, Doostmohammadi2014b, mercier2020settling}.}

{To calculate the stratification drag, we assume that the oblate undergoes a quasi-steady settling. This means that the buoyancy force acting on the oblate and the drag are instantaneously balanced \citep{mercier2020settling}. The quasi-steady assumption allows us to neglect the added mass and the history effects which are anyway smaller than the buoyancy force and the drag force \citep{Doostmohammadi2014b, mercier2020settling}. Consequently, the stratified drag coefficient can be defined as \citep{yick2009enhanced}}

\begin{equation}
C_D^S = \frac{2\left(\rho_P/\rho(z)-1 \right)gD}{U_P^2(z)}.
\end{equation}
Here $\rho({z})$ and $U_P(z)$ are the unperturbed background density and the particle velocity at the instantaneous particle location $z$. Hence, various dimensionless parameters also vary with $z$ and can be written as a function of the instantaneous particle location as
\begin{equation}
\rho_r(z) = \rho_P/\rho(z), 
\end{equation}
\begin{equation}
Re(z) = \frac{|U_P(z)|D}{\nu},
\end{equation}
\begin{equation}
Fr(z) = \frac{|U_P(z)|}{ND}. 
\end{equation}
Here, $Fr(z)$ is the instantaneous Froude number which can also be written as $Fr(z) = \sqrt{Re(z)/Ri}$. Please note that $Ri$ remains constant irrespective of the particle speed and location. 

For spheroids in a homogeneous fluid, we use the following correlation for the drag coefficient ($C_D^H$) which is valid in the range $1 \le Re \le 200$ and $0.25 \le \mathcal{AR} \le 2.5$ \citep{kishore2011momentum}

\begin{equation}
C_D^H = \frac{24\mathcal{AR}^{0.49}}{Re(z)}\left( 1.05 + 0.152Re(z)^{0.687}\mathcal{AR}^{0.671}\right).
\end{equation}

\begin{figure*}
    \centering
     \begin{subfigure}[t]{0.45\textwidth}
        \centering
        \includegraphics[width=\textwidth]{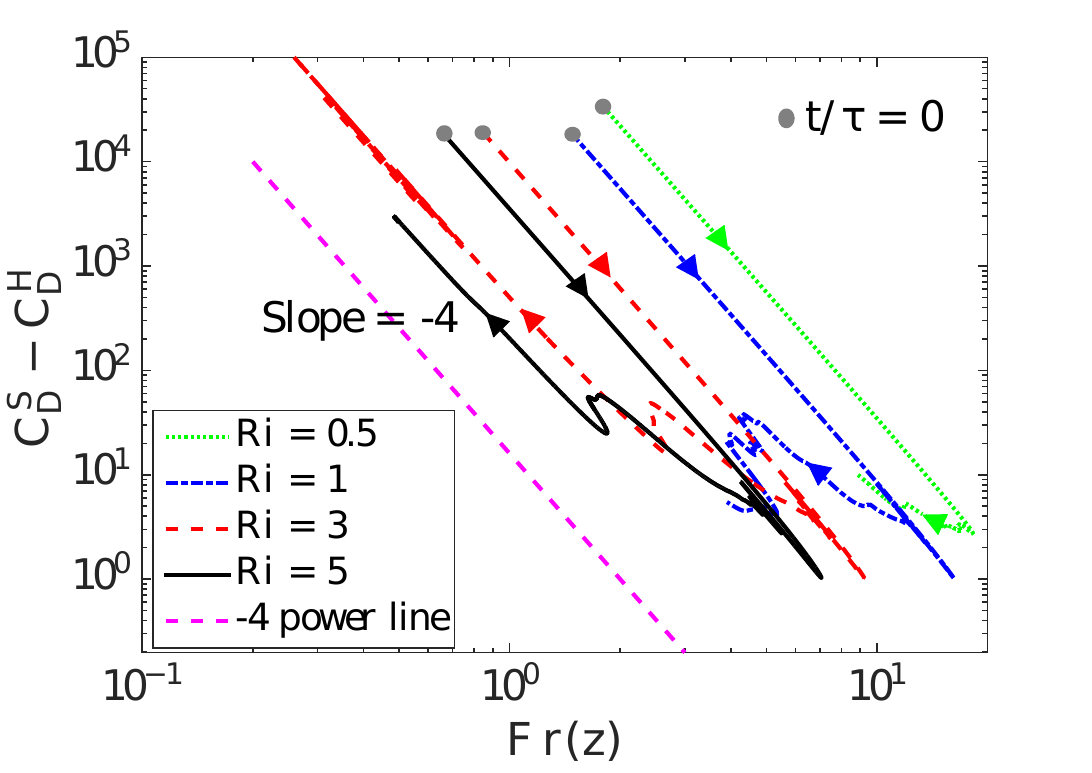}
    	\caption{\label{fig:drag_210}}
       
    \end{subfigure}
    ~ 
    \begin{subfigure}[t]{0.45\textwidth}
        \centering
        \includegraphics[width=\textwidth]{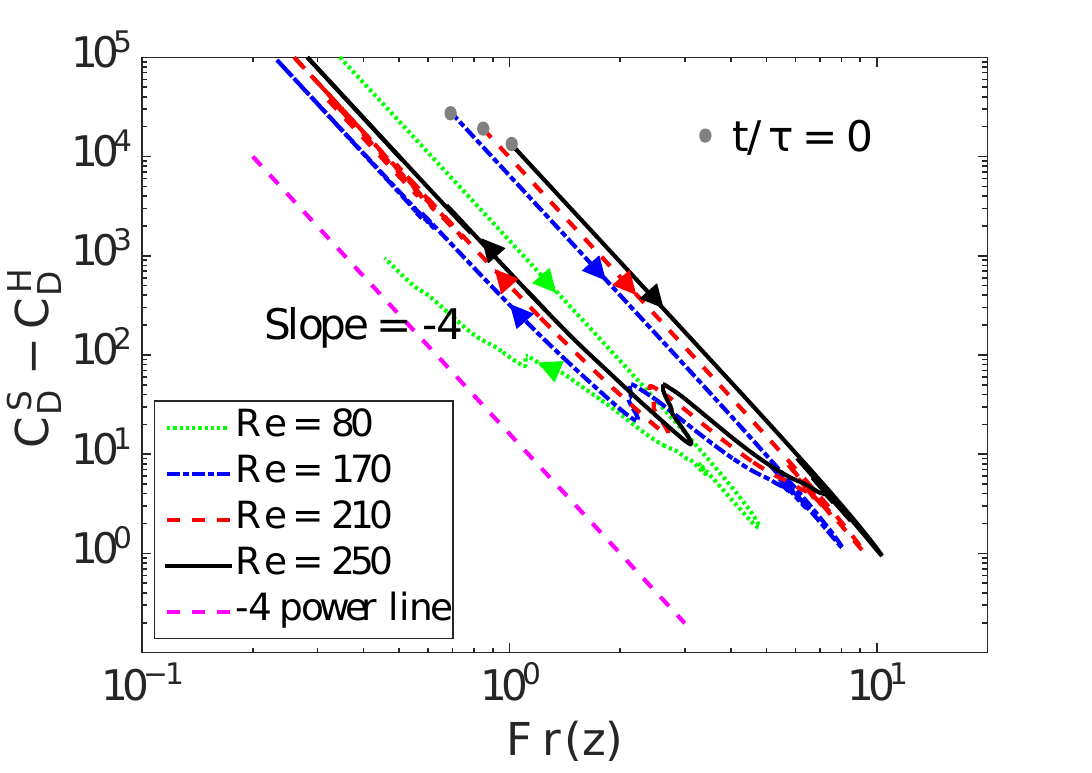}
    	\caption{\label{fig:drag_3}}
      
    \end{subfigure}

    \caption{\label{fig:drag_obl}{Added drag due to stratification, $C_D^S-C_D^H$, for an oblate spheroid with $\mathcal{AR}=1/3$ as a function of the instantaneous particle Froude number, $Fr(z)$, for a) $Re=210$ and different stratification strengths. b) Added drag for $Ri=3$ for different $Re$. The arrows show the direction of increasing time and the filled dots show the simulation start time. The dashed pink line shows the $-4$ power line to indicate a $Fr(z)^{-4}$ scaling of $C_D^S-S_D^H$.}}
\end{figure*}

{Fig.~\ref{fig:drag_obl} presents the variation in the added drag due to stratification ($C_D^S-C_D^H$) for different stratification strengths (fig.~\ref{fig:drag_210}) and different $Re$ (fig.~\ref{fig:drag_3}). As the particle starts from rest, it accelerates initially and $Fr(z)$ increases. As the particle accelerates, the stratification drag acting on it decreases and hence $C_D^S-C_D^H$ decreases. This is expected as the inertial effects dominate in the initial phase of the settling until the particle attains a peak velocity. Hence, $C_D^S-C_D^H$ reaches a minimum when the particle attains its peak velocity. }

{Once the particle reaches its peak velocity, it starts to decelerate as the buoyancy and stratification effects start to dominate over the inertial effects. As a result, the stratification drag starts to increase again. The difference, $C_D^S-C_D^H$ scales as $Fr(z)^{-4}$ as shown in fig.~\ref{fig:drag_obl} and increases with increasing $Re$ (Fig.~\ref{fig:drag_3}).}

\subsubsection{Disappearance of oscillatory paths of settling oblate spheroid}

An oblate spheroid settling in a homogeneous fluid exhibits four distinct trajectories depending on its $Re$ \citep{Ardekani2016}. An oblate spheroid with $\mathcal{AR} = 1/3$ falls in a straight line with an axisymmetric wake for $Re \lessapprox 120$. Increasing $Re$ further eliminates the axisymmetry and introduces oscillations in the settling path. The path is fully vertical with periodic oscillations for $Re \lessapprox 210$. A weakly oblique oscillatory state motion is observed in the range $210 \lessapprox Ga \lessapprox 240$ whereas for $Ga \gtrapprox 240$ the particle path becomes chaotic with patterns of quasi-periodicity. These four states of motion can be explained by the wake instabilities behind a settling oblate spheroid \citep{Ardekani2016} similar to the wake instabilities behind a settling disk \citep{ern2012wake, magnaudet2007wake, yang2007linear}.

\begin{figure*}
    \centering
    \begin{tabular}[t]{ccc}
     \begin{tabular}{c}
            \begin{subfigure}[t]{0.27\textwidth}
                \centering
                \includegraphics[width=\textwidth]{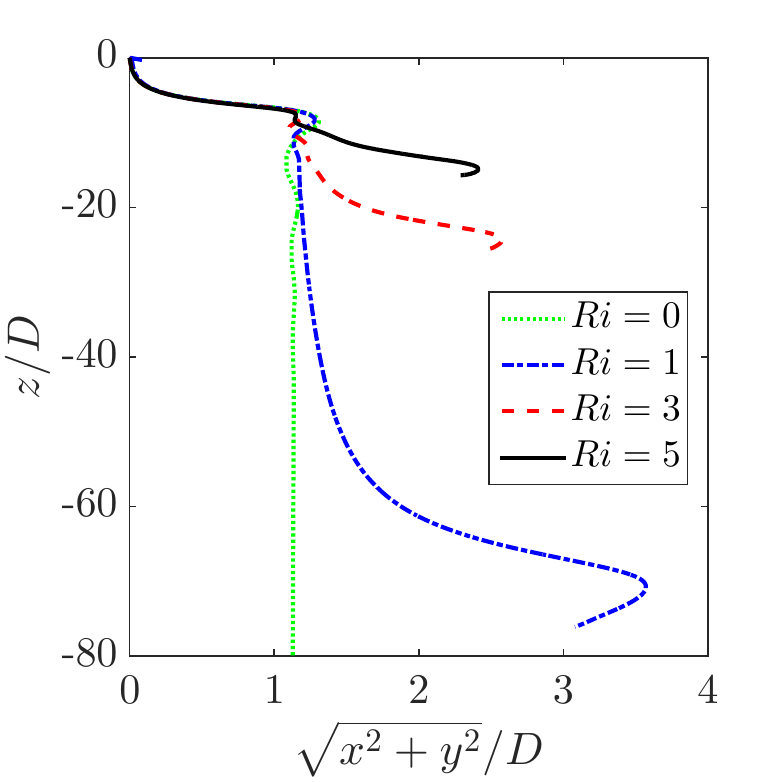}
                \caption{$Re=80$\label{fig:traj80}}
            \end{subfigure}\\
            \begin{subfigure}[t]{0.27\textwidth}
                \centering
                \includegraphics[width=\textwidth]{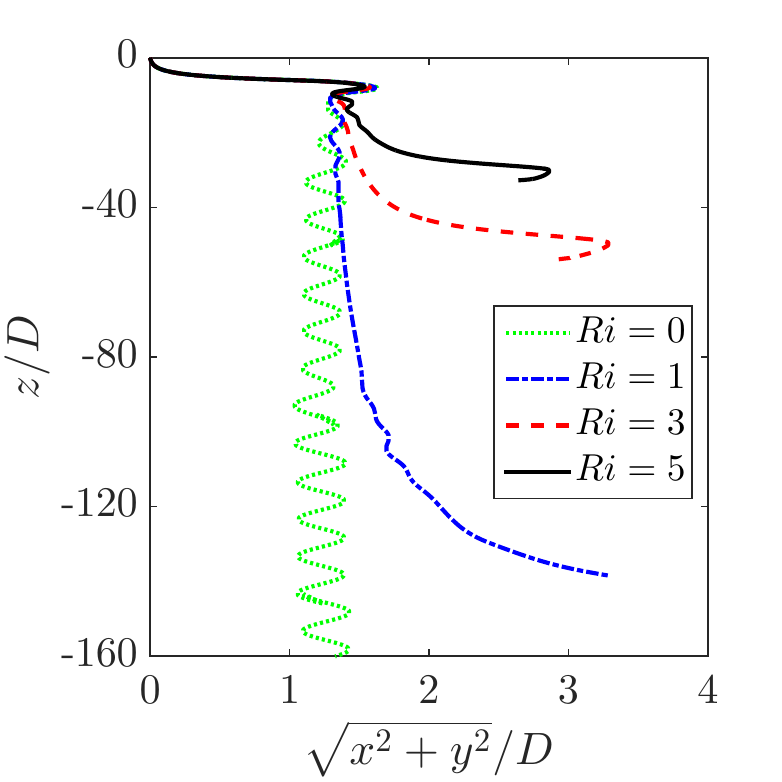}
                \caption{$Re=170$\label{fig:traj170}}
            \end{subfigure}
        \end{tabular} 
    &
        \begin{tabular}{c}
            \begin{subfigure}[t]{0.27\textwidth}
                \centering
                \includegraphics[width=\textwidth]{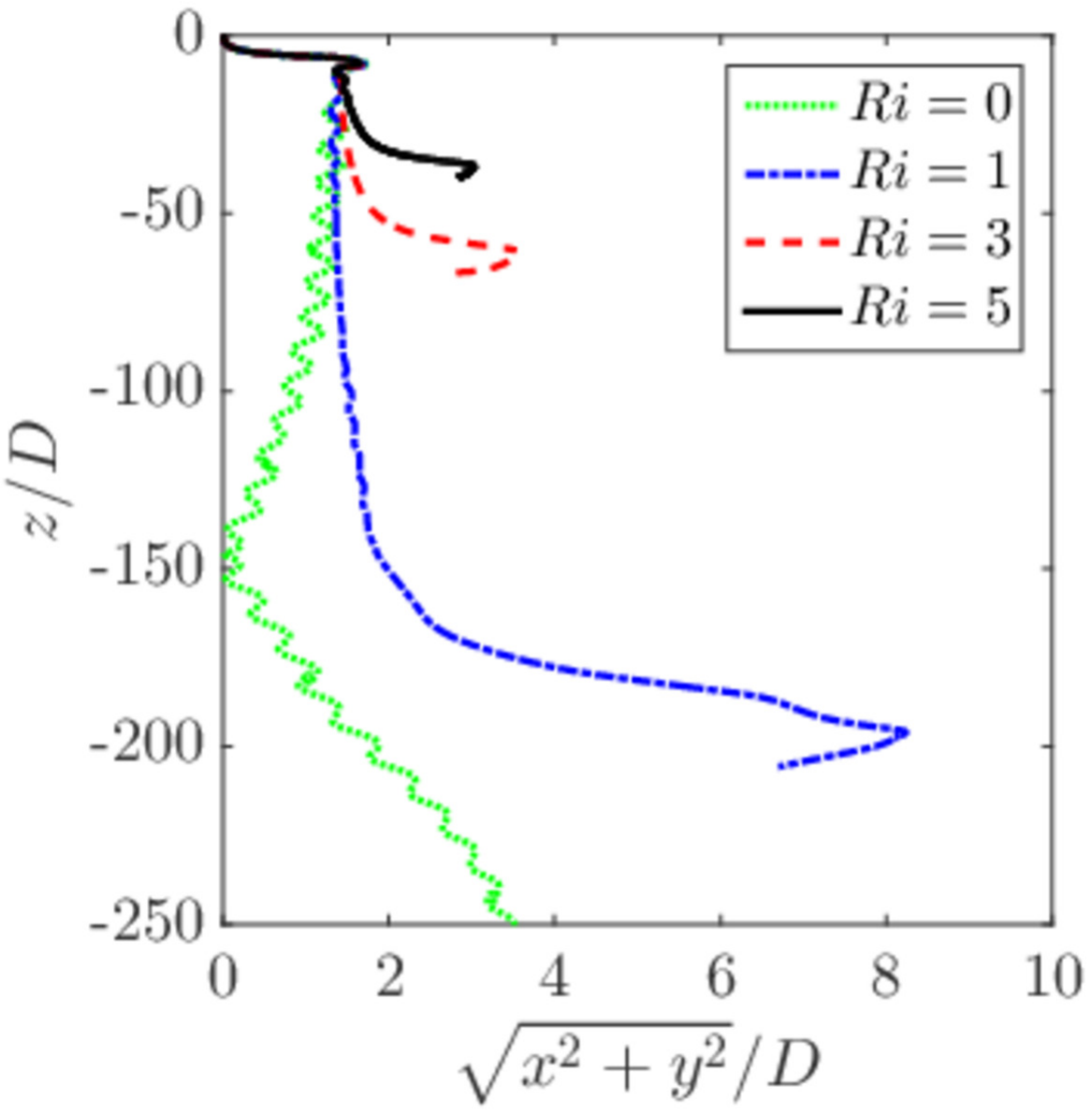}
                \caption{$Re=210$\label{fig:traj210}}
            \end{subfigure}\\
            \begin{subfigure}[t]{0.27\textwidth}
                \centering
                \includegraphics[width=\textwidth]{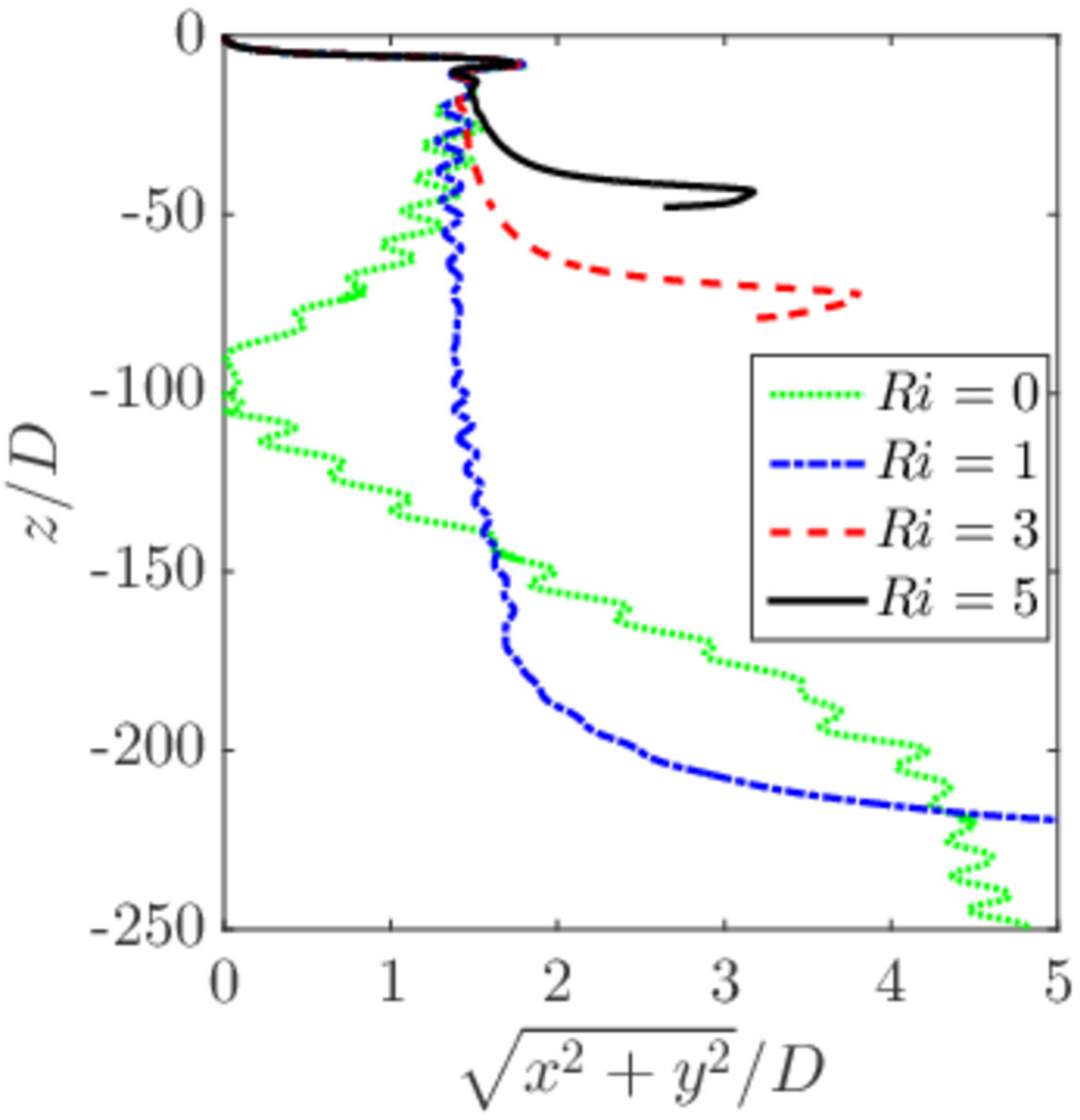}
                \caption{$Re=250$\label{fig:traj250}}
            \end{subfigure}
        \end{tabular} 
  %
    &
    \begin{subfigure}{0.4\textwidth}
        \centering
        \includegraphics[width=\textwidth]{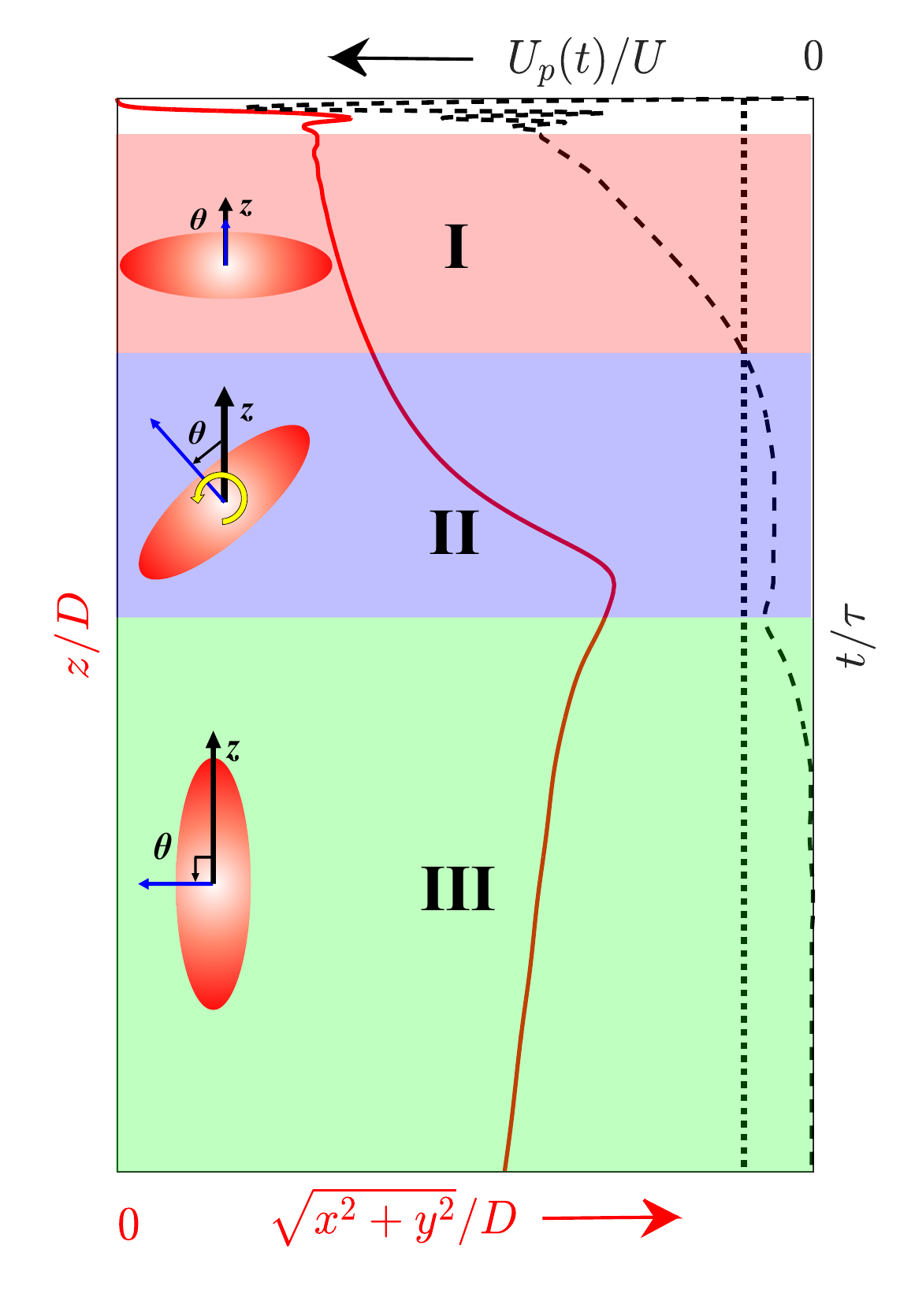}
        \caption{Schematic\label{fig:vel_traj_cartoon}}
    \end{subfigure}
\end{tabular}
    \caption{\label{fig:trajRe}Trajectories of an oblate spheroid with $\mathcal{AR}=1/3$ in a homogeneous and a stratified fluid for different $Re$ and $Ri$. a) $Re=80$, b) $Re=170$, c) $Re=210$, d) $Re=250$, and e) a schematic summarizing the settling velocity, particle trajectory and the orientation in the three zones identified in the settling motion of an oblate spheroid in a stratified fluid. Left vertical axis and bottom horizontal axis indicate spheroid position (solid line is the settling trajectory). Right vertical axis and top horizontal axis are for particle settling velocity vs time (dashed line is the settling velocity).}
\end{figure*}

Stratification significantly alters the settling paths of an oblate spheroid. In particular, it completely annihilates the oscillatory trajectories experienced by a settling oblate spheroid at $Re \gtrapprox 120$ as shown in fig.~\ref{fig:traj170},~\ref{fig:traj210}, and~\ref{fig:traj250}. Comparing the trajectories at different non-zero $Ri$ for various $Re$ in fig.~\ref{fig:trajRe} shows that an oblate spheroid experiences a qualitatively similar trajectory (after the initial transients which will be absent if we initialize the oblate spheroid with the broad-side on orientation) irrespective of its $Re$ and $Ri$. The settling path can be divided into three regions. 


Initially, as the spheroid accelerates from rest, it sediments approximately in a straight line until its velocity approaches the threshold for the reorientation onset. We call this region $I$. In region $II$, the oblate spheroid starts reorienting due to the onset of the reorientation instability. This induces a non-zero horizontal velocity component in the settling of an oblate spheroid. As a result, the particle moves in the horizontal direction, breaking the straight line motion and getting deflected in the transverse direction. This region can also be identified in the settling velocity of the oblate spheroid. The settling velocity attains a temporary plateau after it falls below the threshold for reorientation. During this time, the oblate spheroid experiences reorientation from broad-side on to edge-wise and gets deflected in the horizontal direction. This horizontal deflection has previously been observed for disks \citep{mrokowska2018stratification, mercier2020settling, mrokowska2020dynamics}. This region ends when the reorientation is over and the settling velocity increases momentarily as can be seen in fig.~\ref{fig:210_vel}. Finally, in region $III$, as the particle comes close to its neutrally buoyant position, its velocity quickly decelerates and stops which is indicated by the reversal of the horizontal trajectory at the end of the settling path in fig.~\ref{fig:trajRe}. These settling trajectories and regions are similar to those observed for a disk in a stratified fluid \citep{mercier2020settling}. However, we do not observe any change in the orientation of an oblate spheroid from edge-wise at the end of region $III$ as observed for a disk \citep{mercier2020settling}. This is most likely because of the ideal conditions in simulations as opposed to experiments. Fig.~\ref{fig:vel_traj_cartoon} summarizes the three regions of the settling path of an oblate spheroid in a stratified fluid along with their onset conditions on the settling velocity evolution plot. 

\subsection{What causes deceleration and reorientation of an oblate spheroid in a stratified fluid?}\label{sec:oblres}

\begin{figure*}
    \centering
    \begin{tabular}[t]{ccccc}
    
    \begin{subfigure}[t]{0.19\textwidth}
        \centering
        \includegraphics[width=\textwidth]{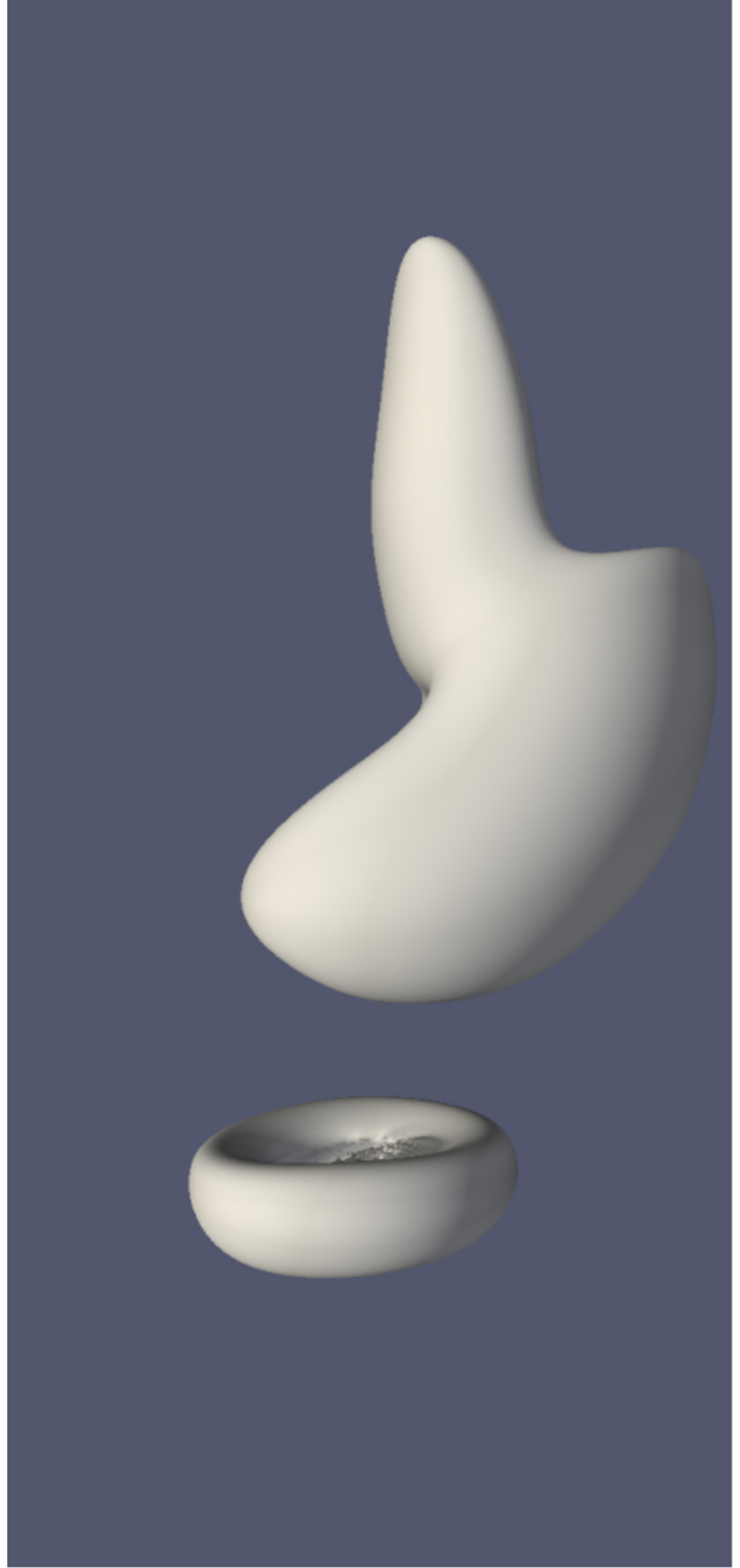}

    \end{subfigure}
   &
    \begin{subfigure}[t]{0.19\textwidth}
        \centering
        \includegraphics[width=\textwidth]{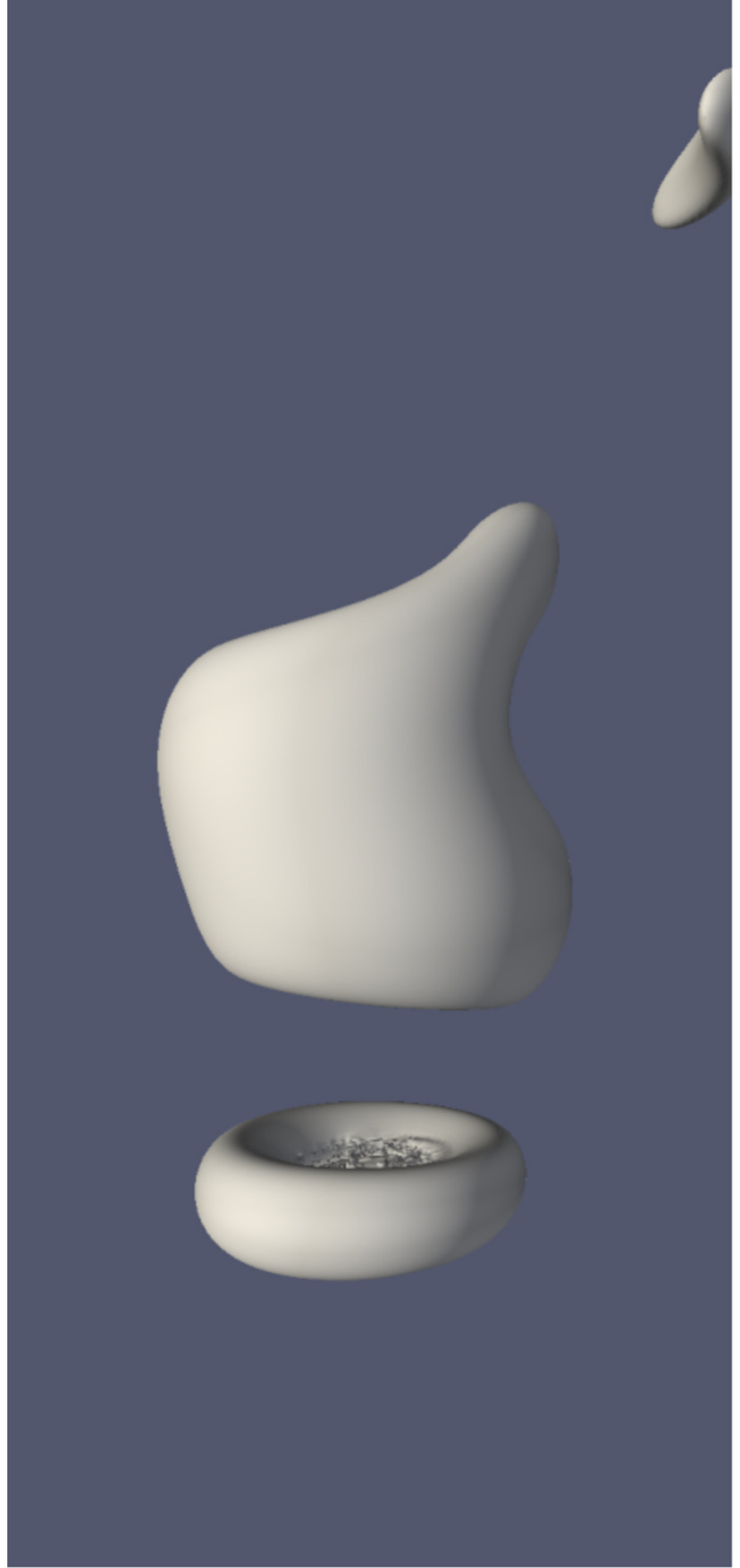}
 
    \end{subfigure}
    & 
    \begin{subfigure}[t]{0.19\textwidth}
        \centering
        \includegraphics[width=\textwidth]{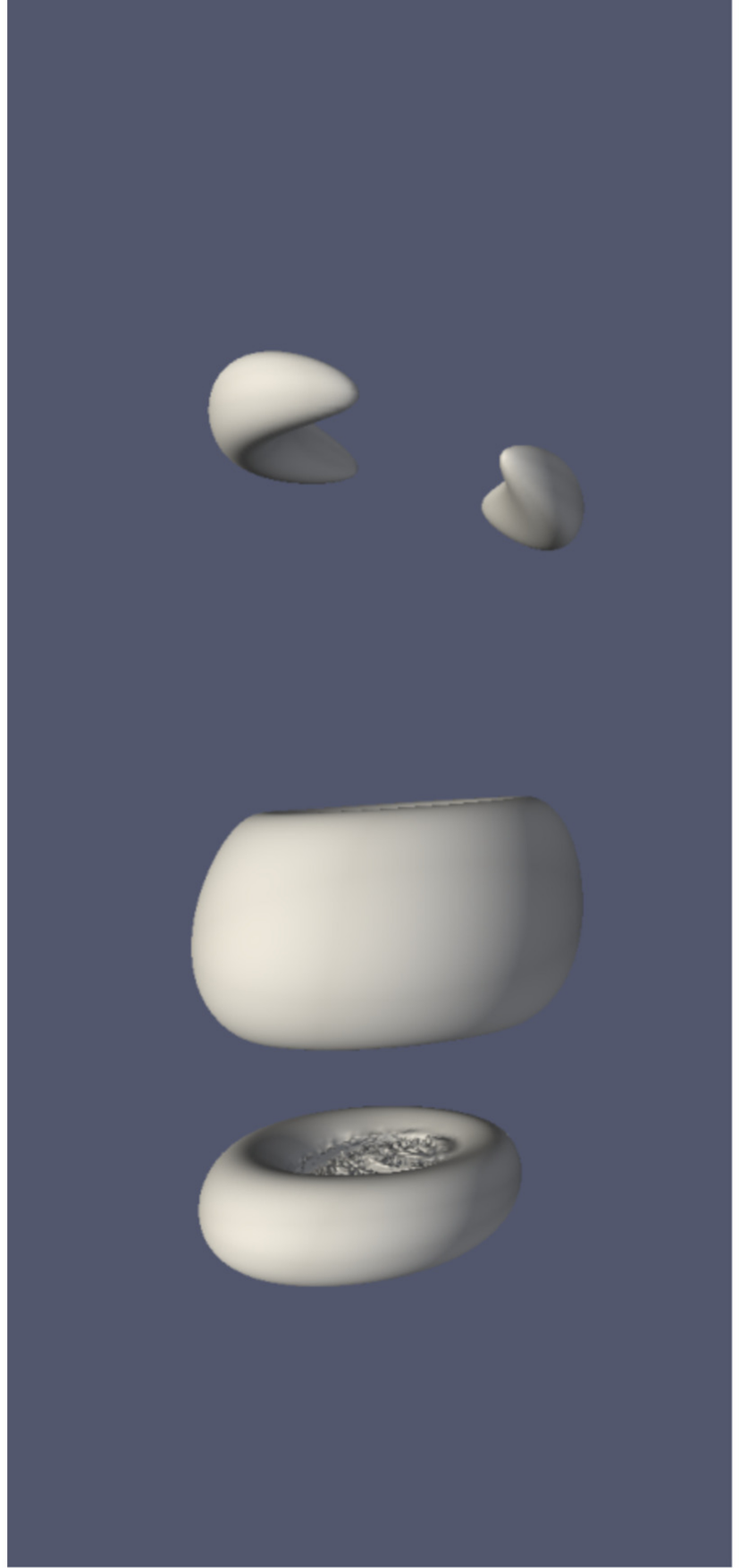}
 
    \end{subfigure}
    & 
    \begin{subfigure}[t]{0.19\textwidth}
        \centering
        \includegraphics[width=\textwidth]{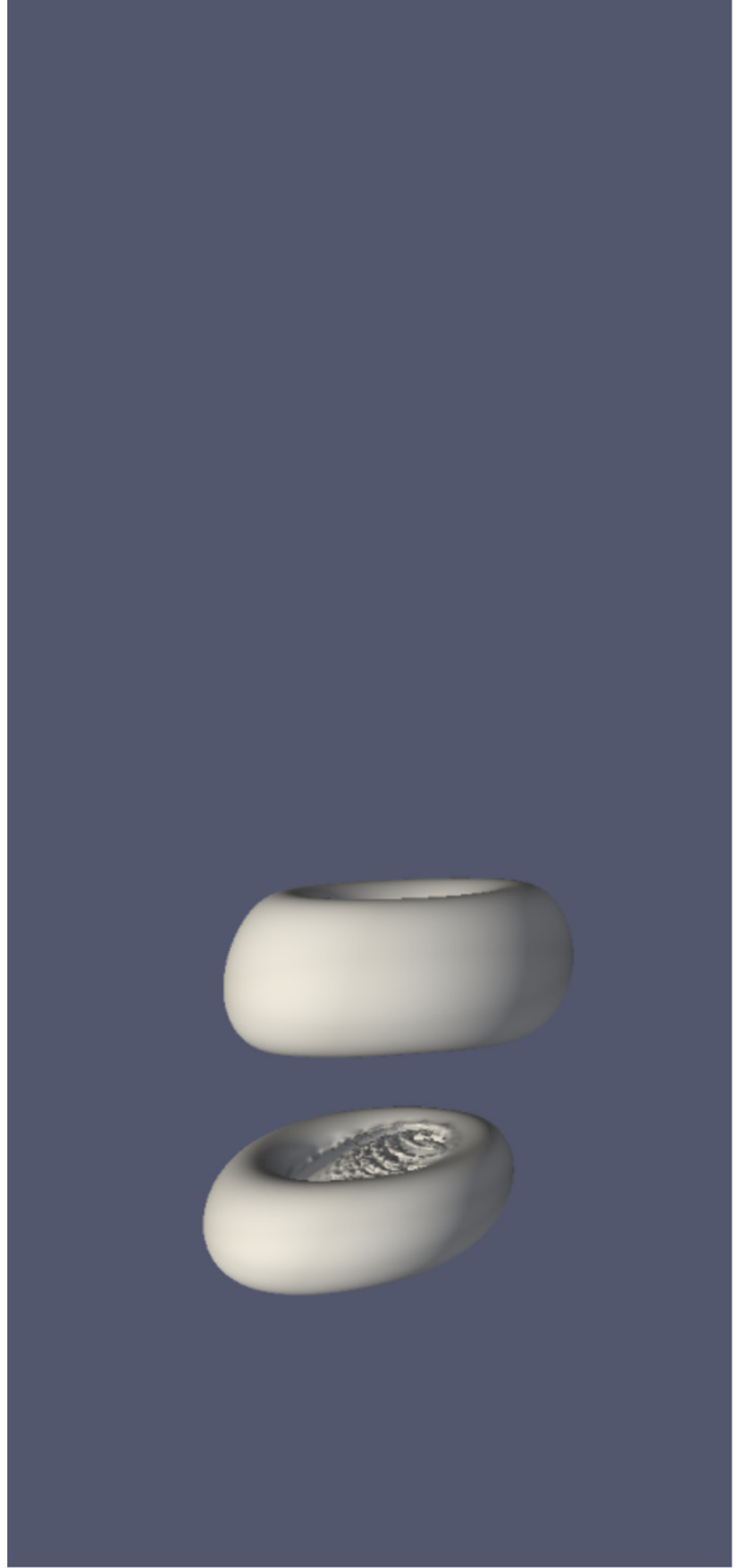}
  
    \end{subfigure}
    &
    \begin{subfigure}[t]{0.19\textwidth}
        \centering
        \includegraphics[width=\textwidth]{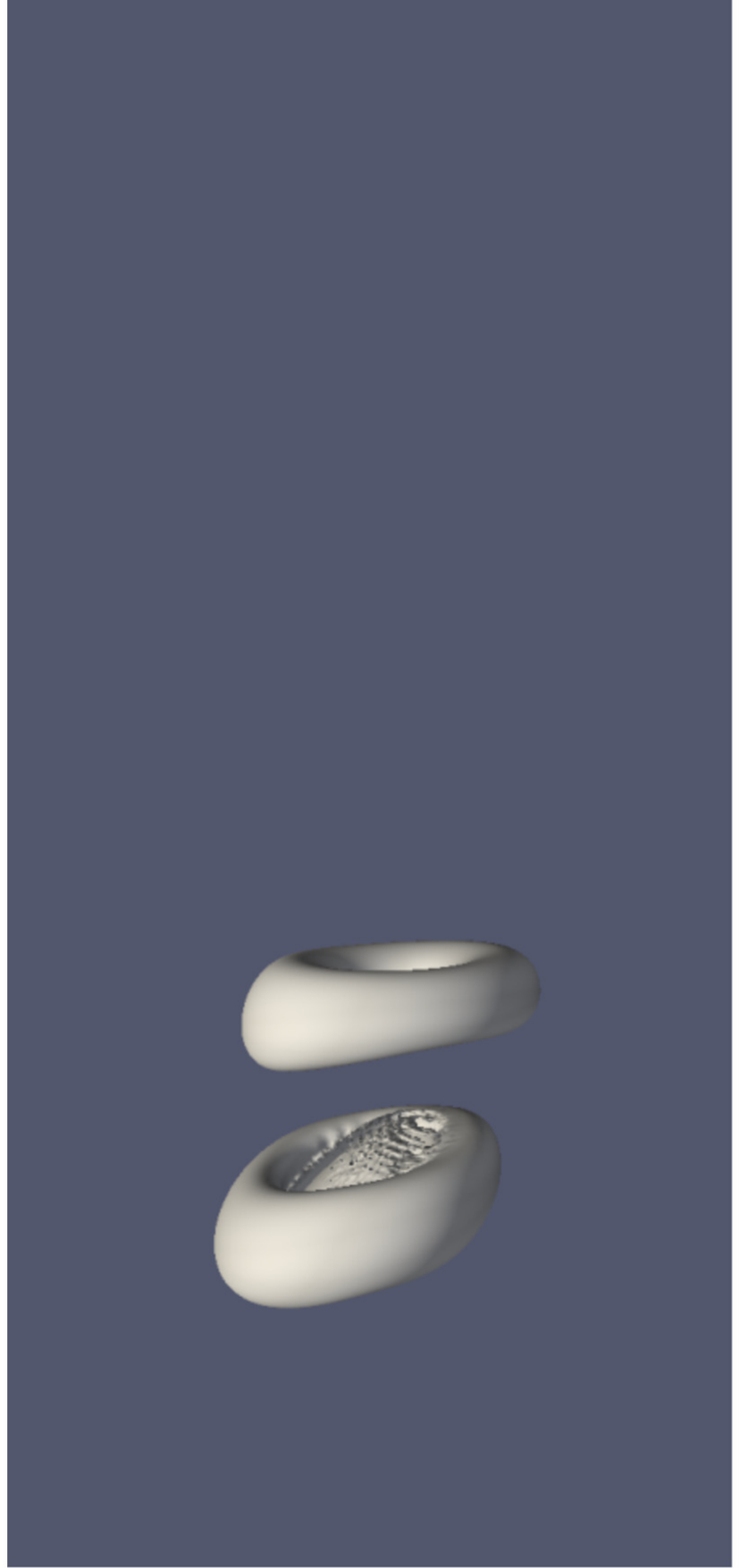}
    \end{subfigure}

    \end{tabular}
    \caption{\label{fig:obl_q}Dimensionless iso-surfaces of Q-criterion equal to $5 \times 10^{-4}$ for an oblate spheroid with $\mathcal{AR}=1/3$, $Re=80$ and $Ri=5$ at equal time intervals of $t/\tau=10.74$ starting from $t/\tau = 18.78$. These contours show the evolution of vortices. The vortical structures identified by the positive Q-criterion are associated with a lower pressure region behind the particle.}
\end{figure*}

In the case of disk-like bodies settling in a homogeneous fluid, the path instabilities as described in the last subsection can be explained by the wake instabilities \citep{magnaudet2007wake, yang2007linear, ern2012wake}. Therefore, analysing the wake vortices can provide insight into the mechanisms leading to a particular type of motion in either a homogeneous or a stratified fluid. For an oblate spheroid settling in a homogeneous fluid, a single toroidal vortex attached to the particle is initially formed. This is similar to a spherical particle moving with a steady velocity in a homogeneous fluid. As time passes, instabilities develop and the particle starts rotating around one of its major axes, normal to the direction of gravity as shown in fig.~\ref{fig:210_or}. As the angle of the oblate spheroid with respect to the horizontal axis increases, a part of this toroidal vortex  detaches from the particle in a hairpin like structure \citep{Ardekani2016}. Vortices are associated with low pressure regions than the ambient. So, as a result of the detachment of the toroidal vortex, the oblate spheroid experiences a torque due to the formation of this low pressure region behind it which directly opposes the rotation of the particle in the other direction. Owing to inertia, the particle then rotates in the other direction. New hairpin vortices keep detaching from the oblate spheroid alternatively from either sides as it settles, leading to periodic changes in the orientation and oscillatory paths \citep{Ardekani2016}.

\begin{figure*}
    \centering
     \begin{subfigure}[t]{0.45\textwidth}
        \centering
        \includegraphics[width=\textwidth]{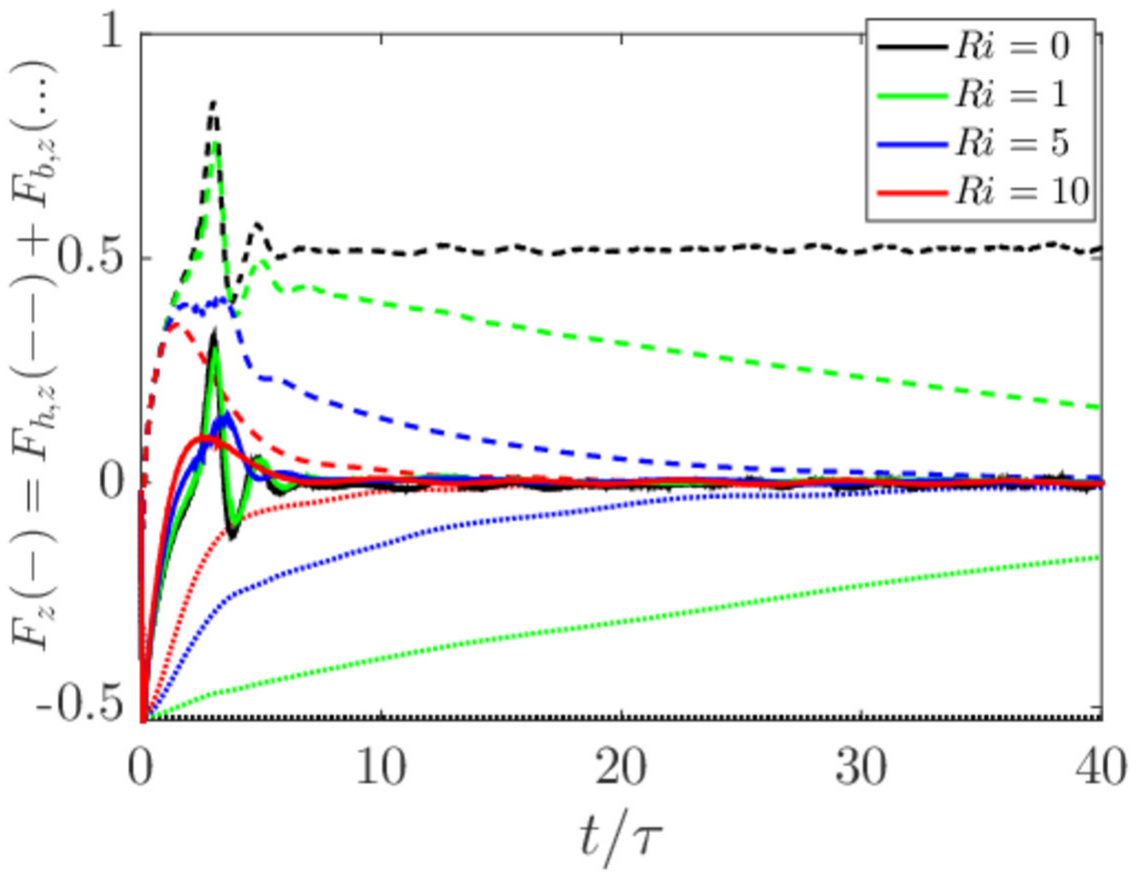}
    	\caption{\label{fig:force}}
    
    \end{subfigure}
    ~ 
    \begin{subfigure}[t]{0.45\textwidth}
        \centering
        \includegraphics[width=\textwidth]{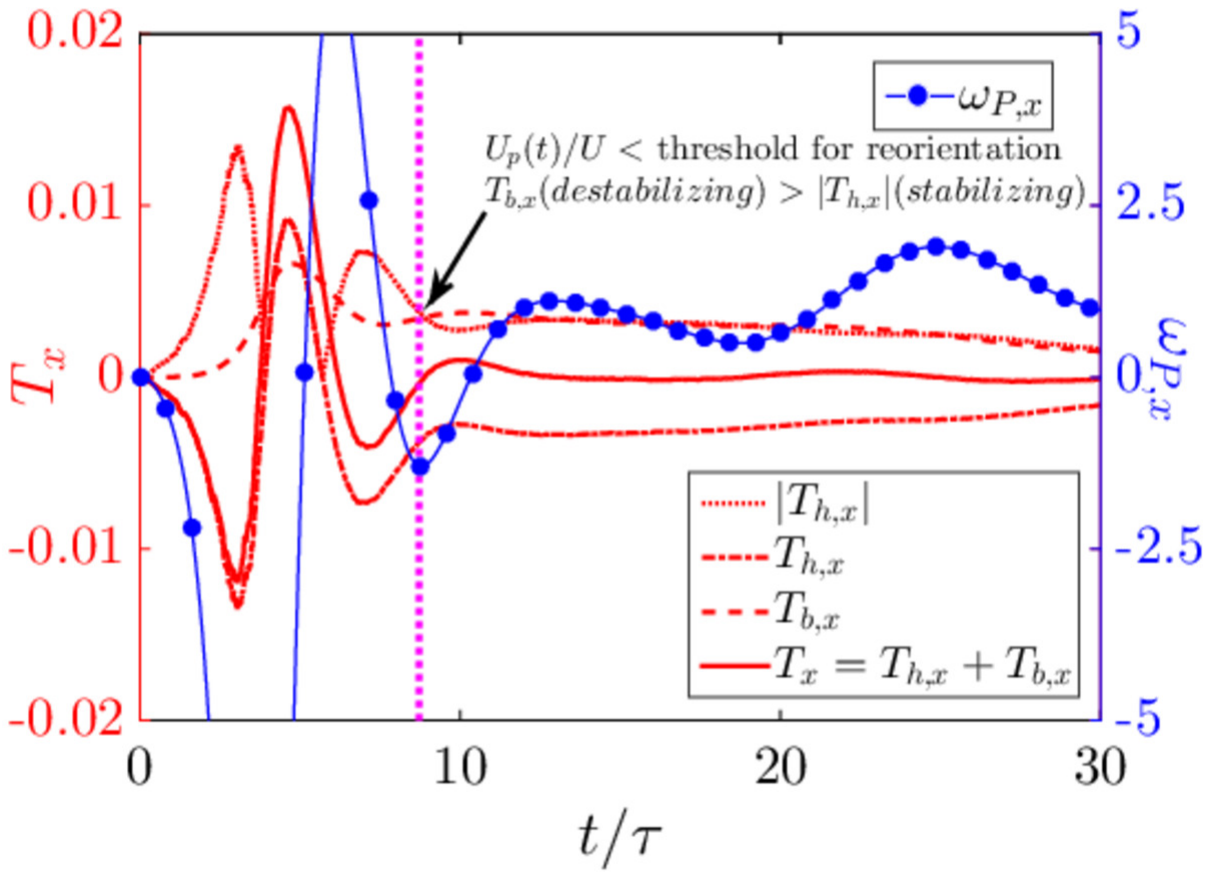}
    	\caption{\label{fig:torq}}
      
    \end{subfigure}
    \caption{\label{fig:frc} {a) Forces acting on the oblate spheroid with $Re=80$ as it settles in a stratified fluid with varying $Ri$ shown with different colors. The total force (solid line) can be split into two components, the hydrodynamic component (dashed line) and the buoyancy component (dotted line). b) x-component of the torque acting on the oblate spheroid with $Re=80$ as it sediments in a stratified fluid with $Ri=5$ along with the x-component of the angular velocity. The net torque (solid line) is split into two components, the hydrodynamic torque (dotted line) which tries to orient the prolate in a broadside on orientation (hence stabilizing) and the buoyancy component (dashed-dotted line) which is destabilizing and tries to reorient the prolate in a edgewise orientation. The reorientation starts once the magnitude of hydrodynamic torque falls below the buoyancy torque which happens when the prolate velocity falls below the threshold for reorientation as discussed in Sec.~\ref{sec:oblmain}. }}
\end{figure*}

\begin{figure*}
    \centering
    \begin{tabular}[t]{ccccc}
    \hspace{-4mm}
     \begin{tabular}{c}
    \begin{subfigure}[t]{0.19\textwidth}
        \centering
        \includegraphics[width=\textwidth]{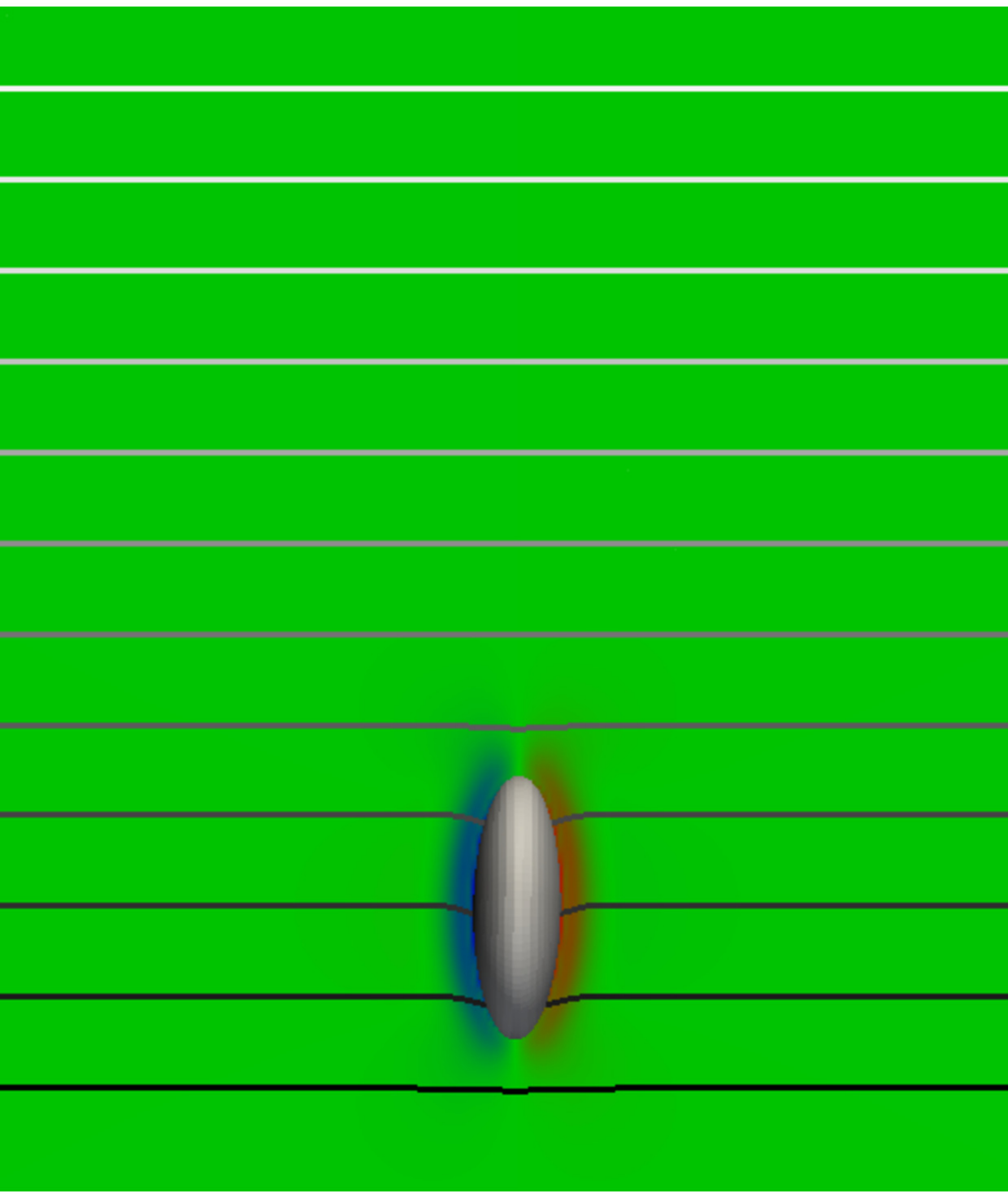}
    
    \end{subfigure} \\
     \begin{subfigure}[t]{0.19\textwidth}
        \centering
        \includegraphics[width=\textwidth]{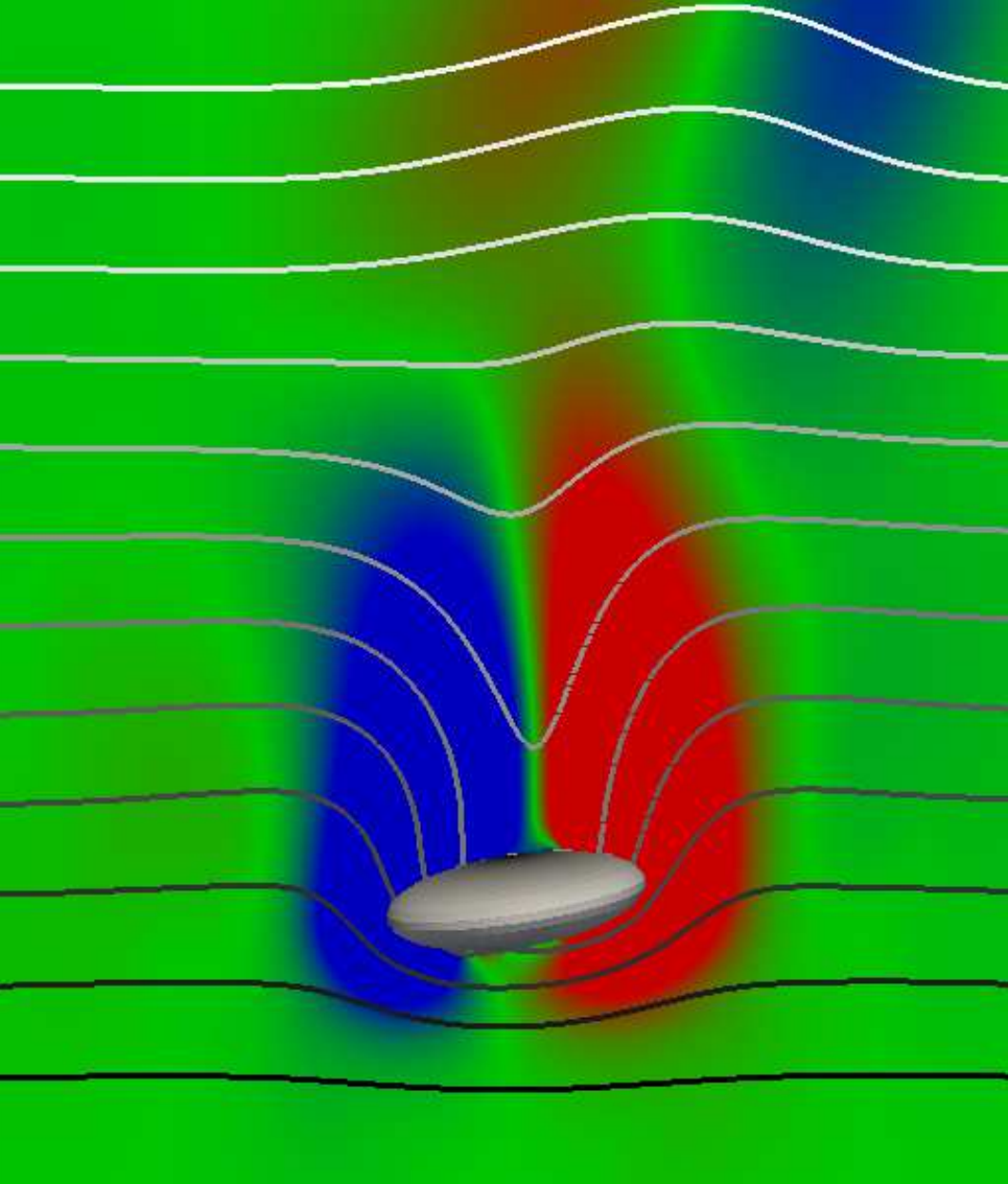}
    
    \end{subfigure} \\
     \begin{subfigure}[t]{0.19\textwidth}
        \centering
        \includegraphics[width=\textwidth]{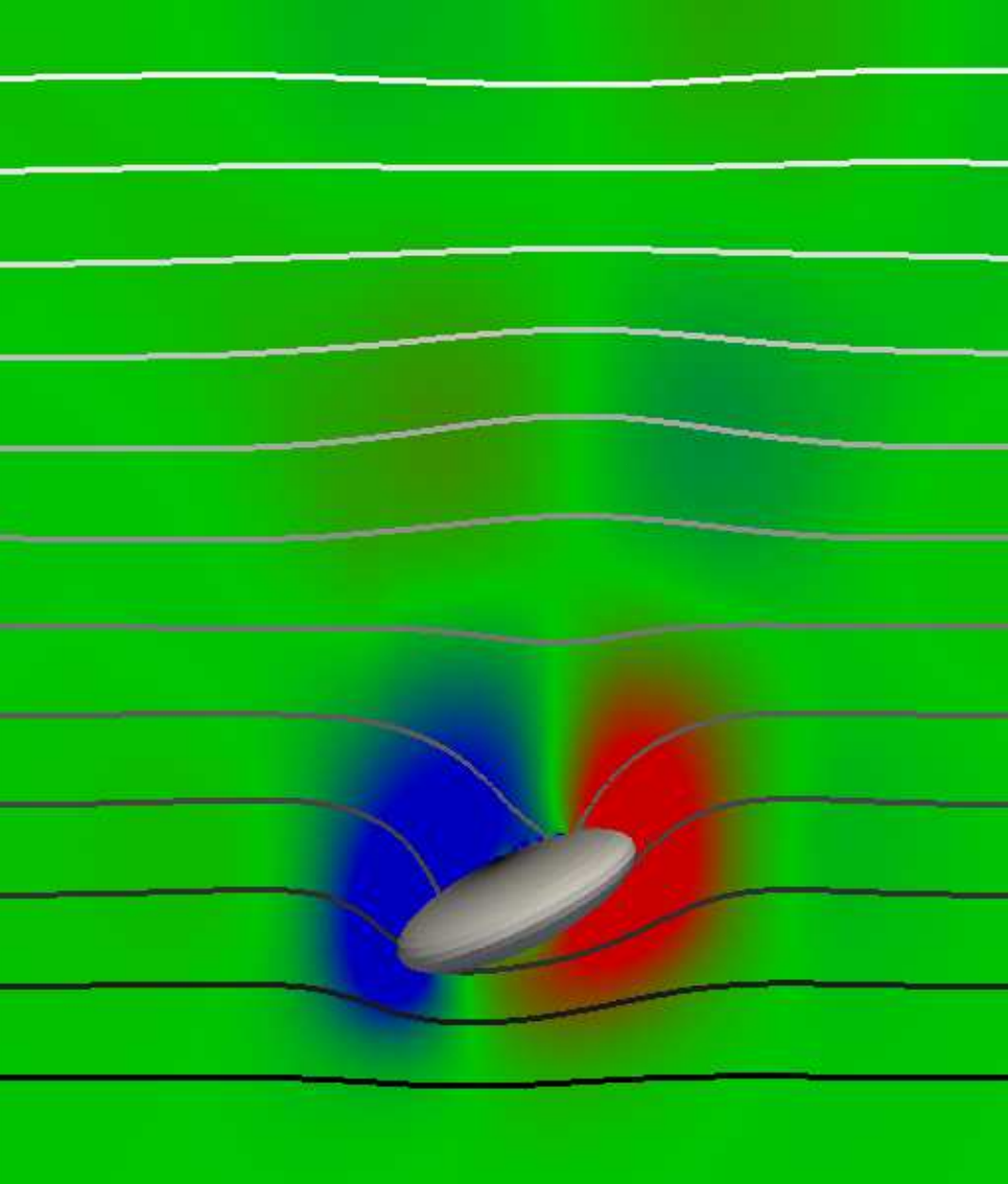}
  
    \end{subfigure} 
    \end{tabular}
    \hspace{-4mm}
   &
   \begin{tabular}{c}
     \begin{subfigure}[t]{0.19\textwidth}
        \centering
        \includegraphics[width=\textwidth]{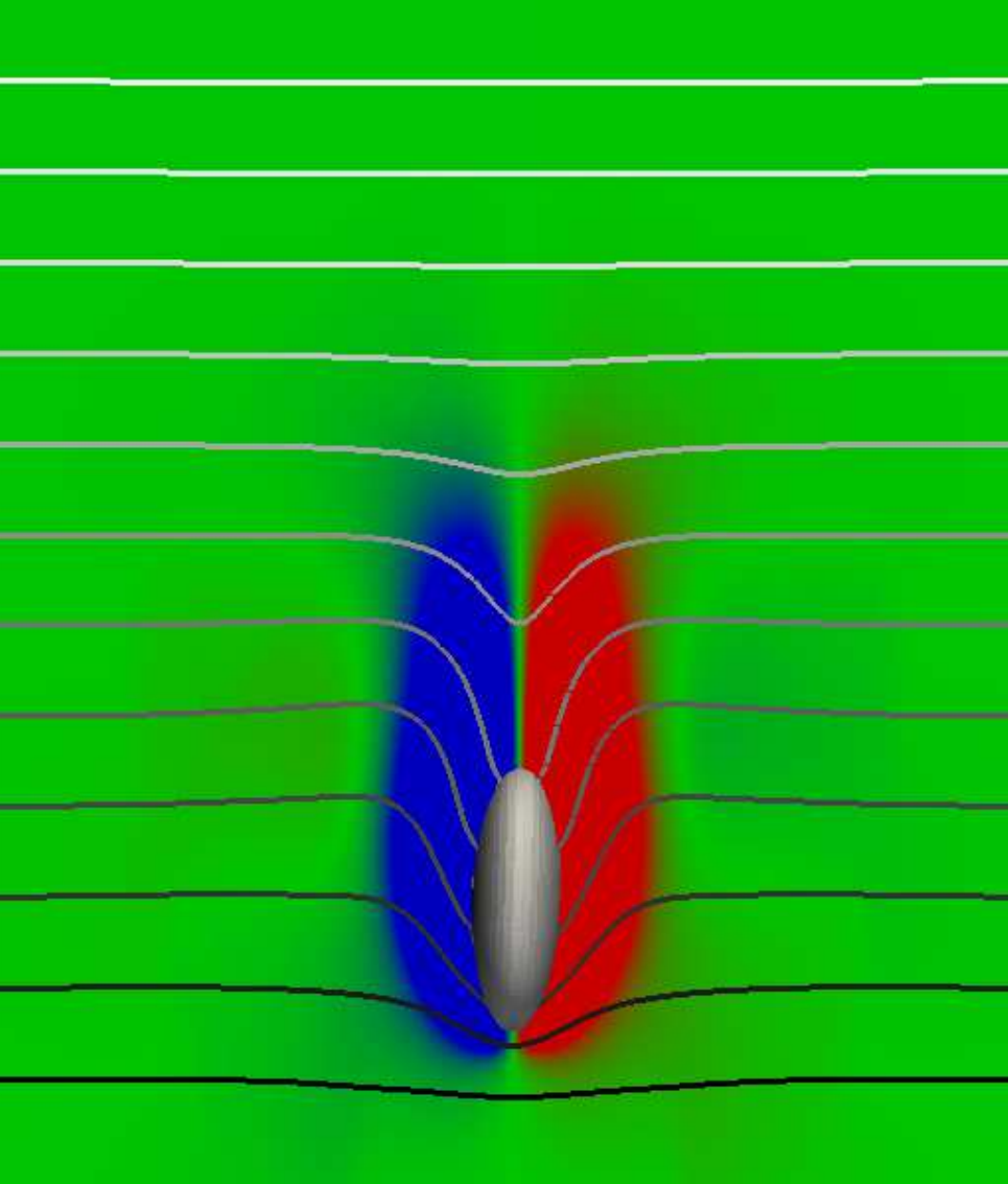}
   
    \end{subfigure} \\
    
     \begin{subfigure}[t]{0.19\textwidth}
        \centering
        \includegraphics[width=\textwidth]{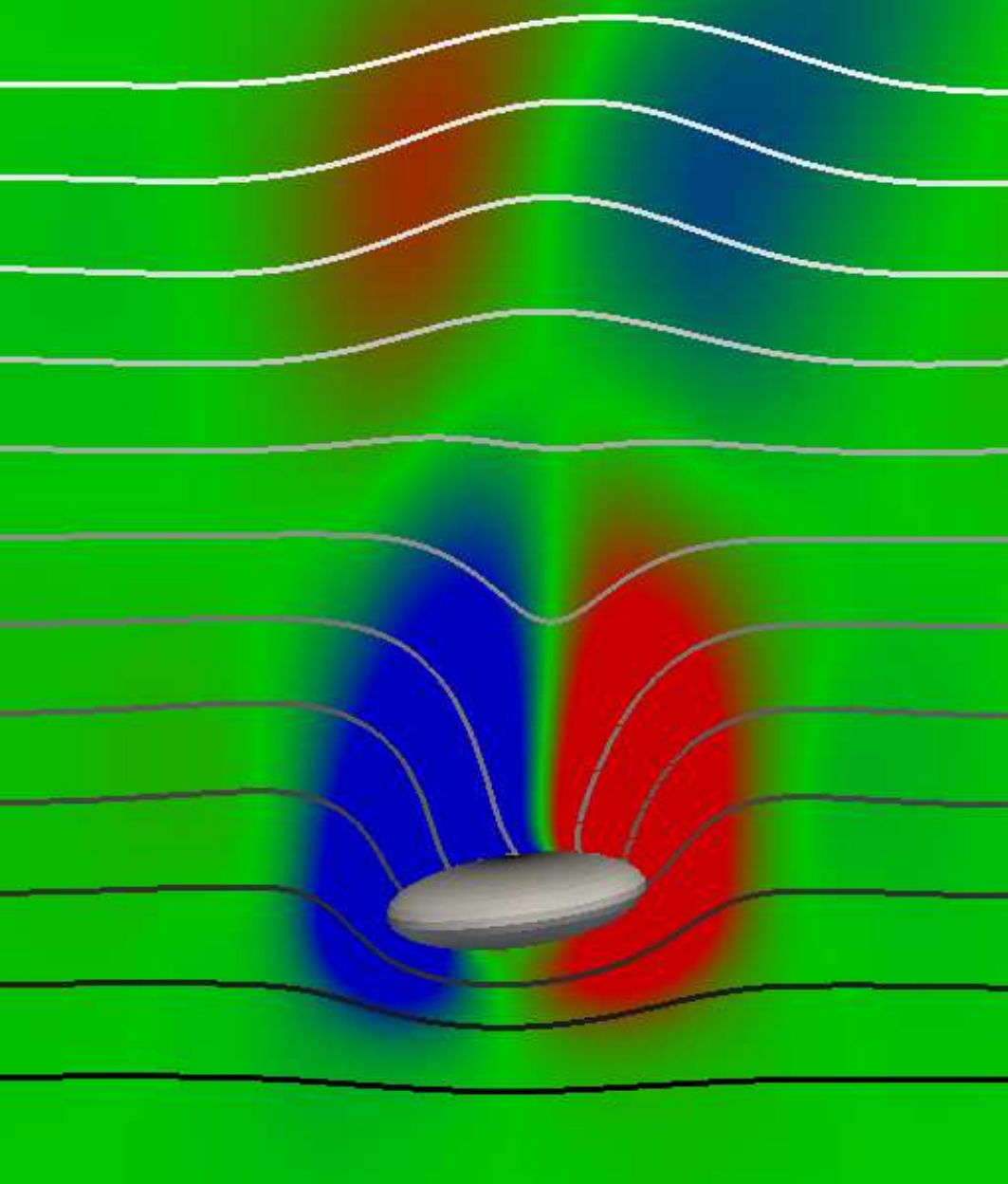}
   
    \end{subfigure} \\
     \begin{subfigure}[t]{0.19\textwidth}
        \centering
        \includegraphics[width=\textwidth]{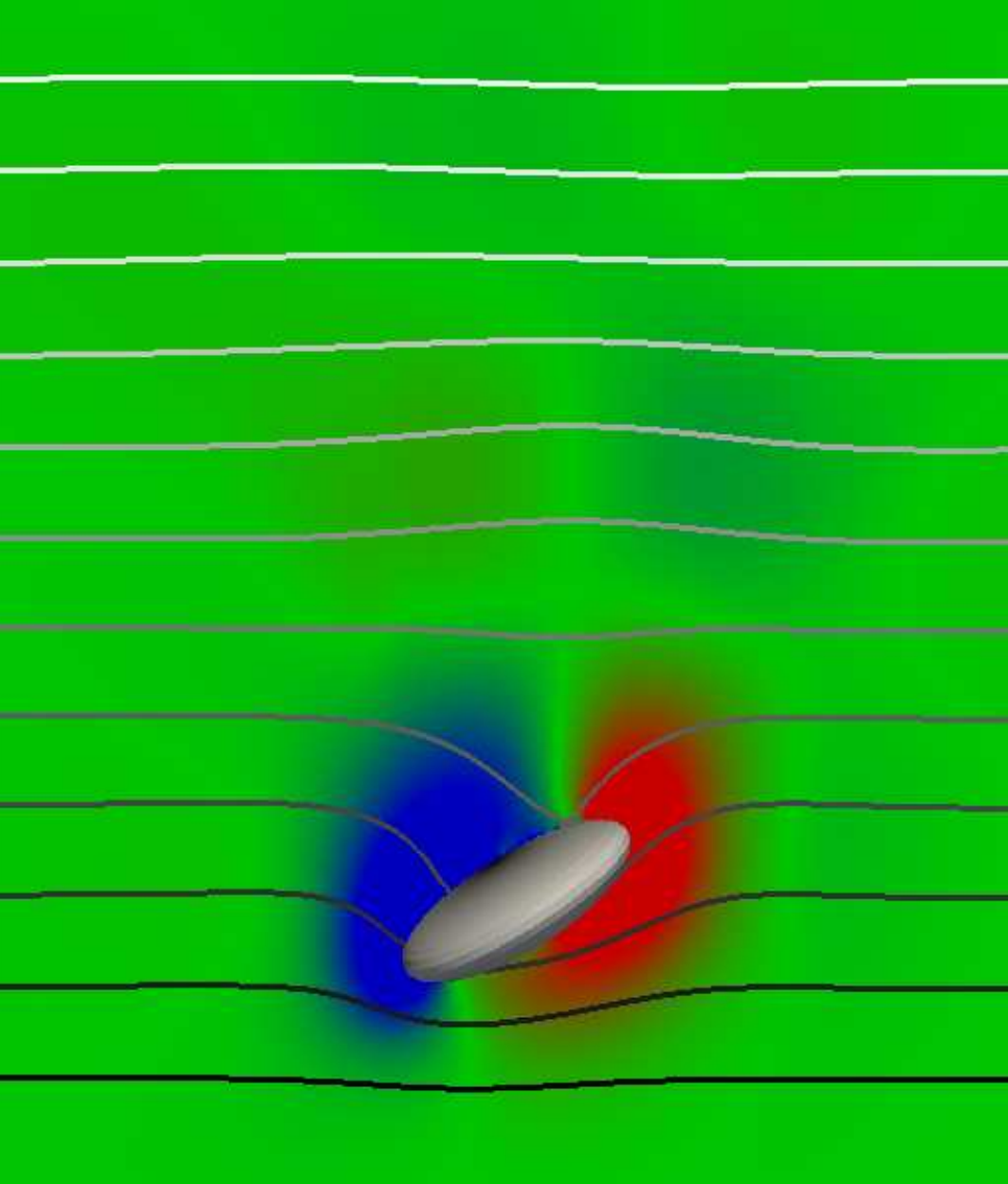}
    
    \end{subfigure} 
    \end{tabular}
    \hspace{-4mm}
    & 
    \begin{tabular}{c}
     \begin{subfigure}[t]{0.19\textwidth}
        \centering
        \includegraphics[width=\textwidth]{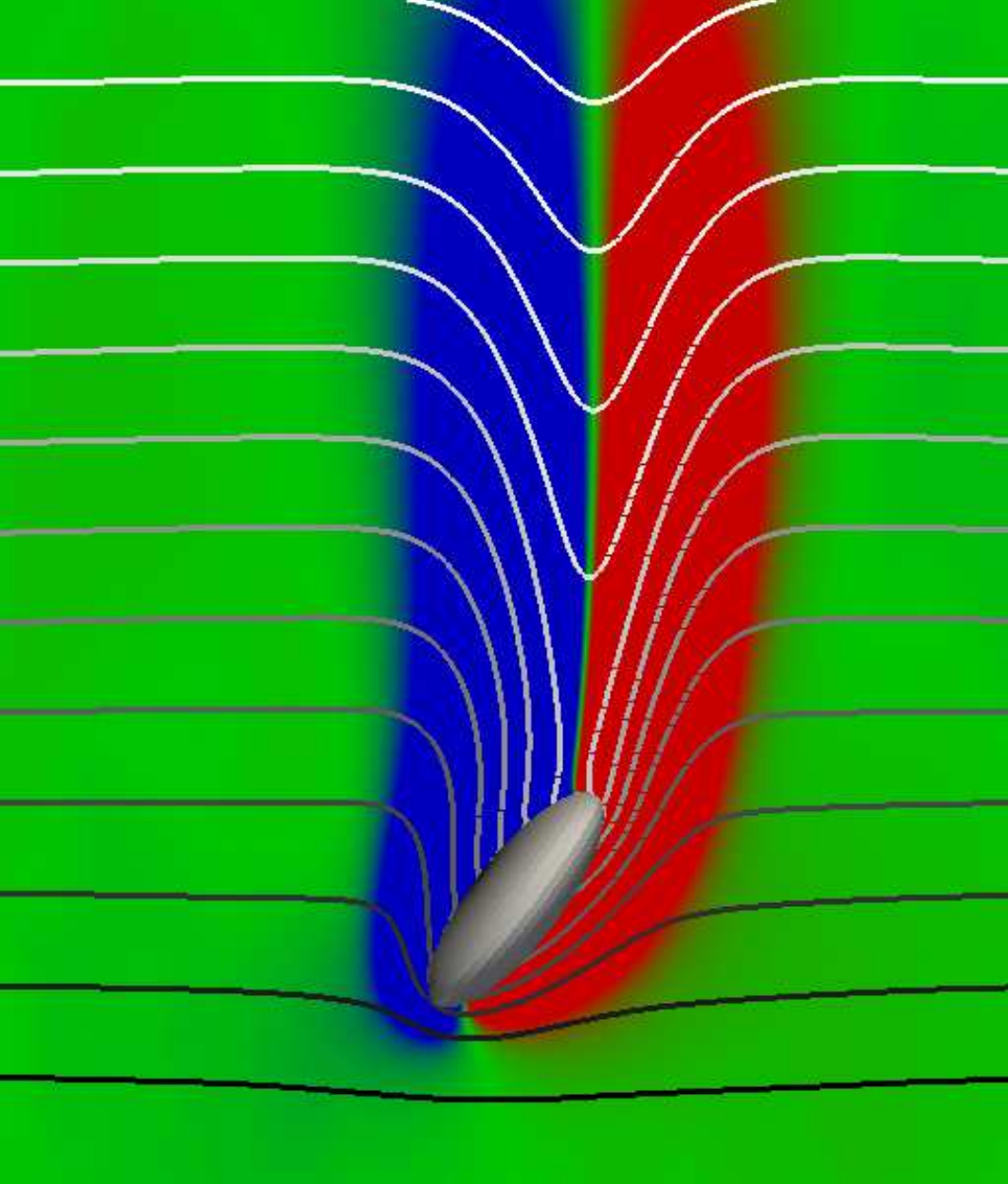}
   
    \end{subfigure} \\
    
     \begin{subfigure}[t]{0.19\textwidth}
        \centering
        \includegraphics[width=\textwidth]{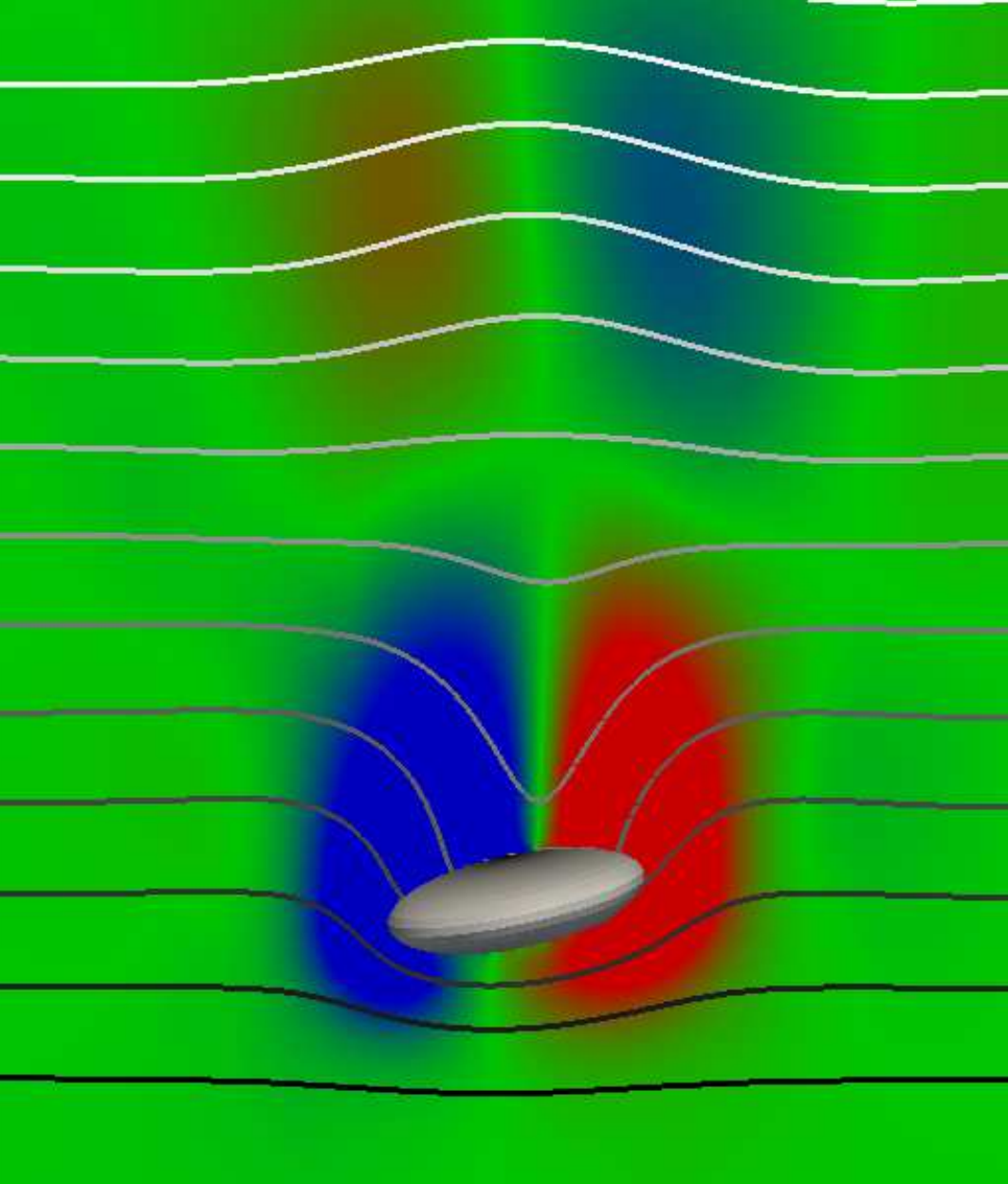}
   
    \end{subfigure} \\
     \begin{subfigure}[t]{0.19\textwidth}
        \centering
        \includegraphics[width=\textwidth]{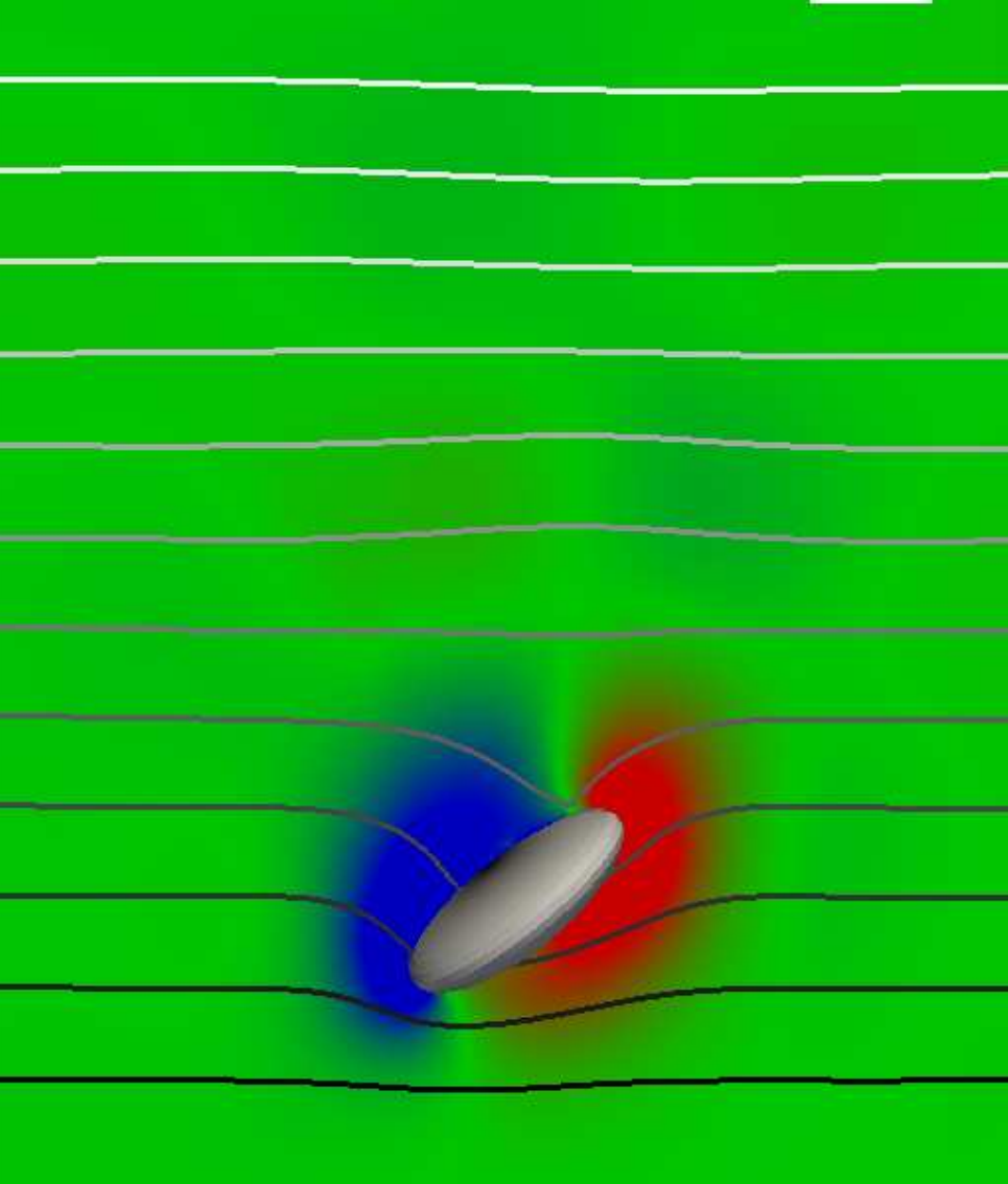}
  
    \end{subfigure} 
    \end{tabular}
    \hspace{-4mm}
    & 
    \begin{tabular}{c}
     \begin{subfigure}[t]{0.19\textwidth}
        \centering
        \includegraphics[width=\textwidth]{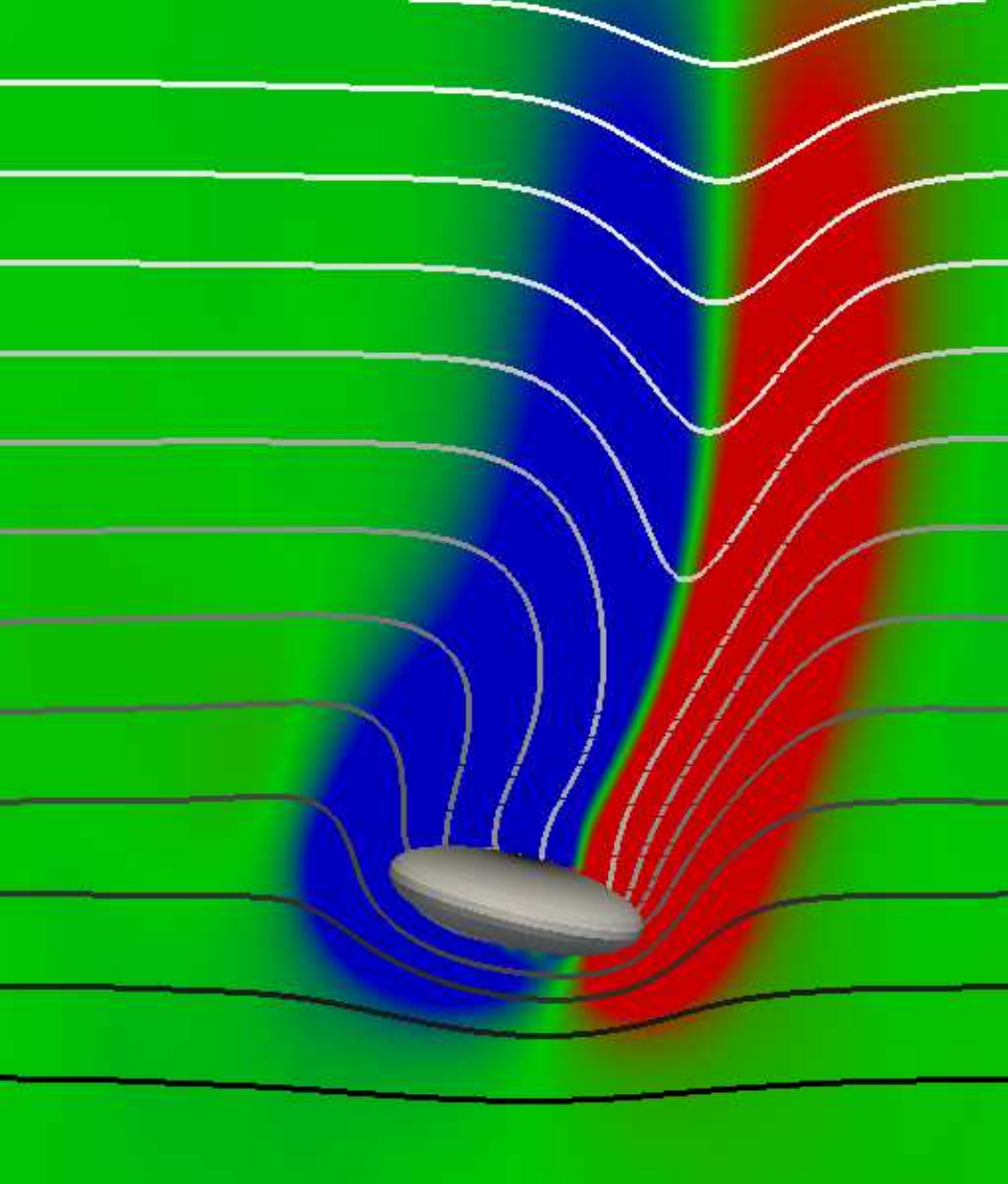}
    
    \end{subfigure} \\
    
     \begin{subfigure}[t]{0.19\textwidth}
        \centering
        \includegraphics[width=\textwidth]{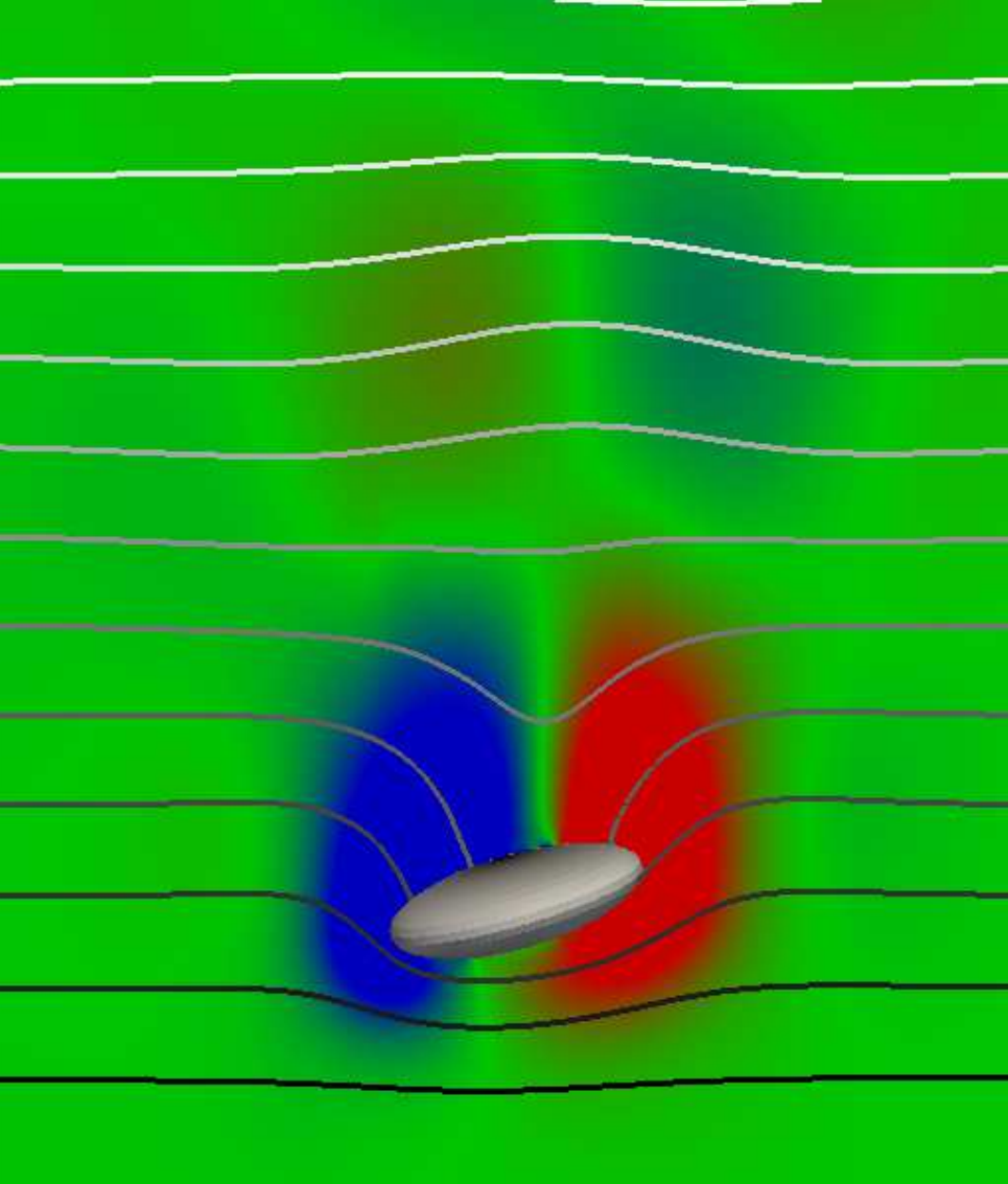}
  
    \end{subfigure} \\
     \begin{subfigure}[t]{0.19\textwidth}
        \centering
        \includegraphics[width=\textwidth]{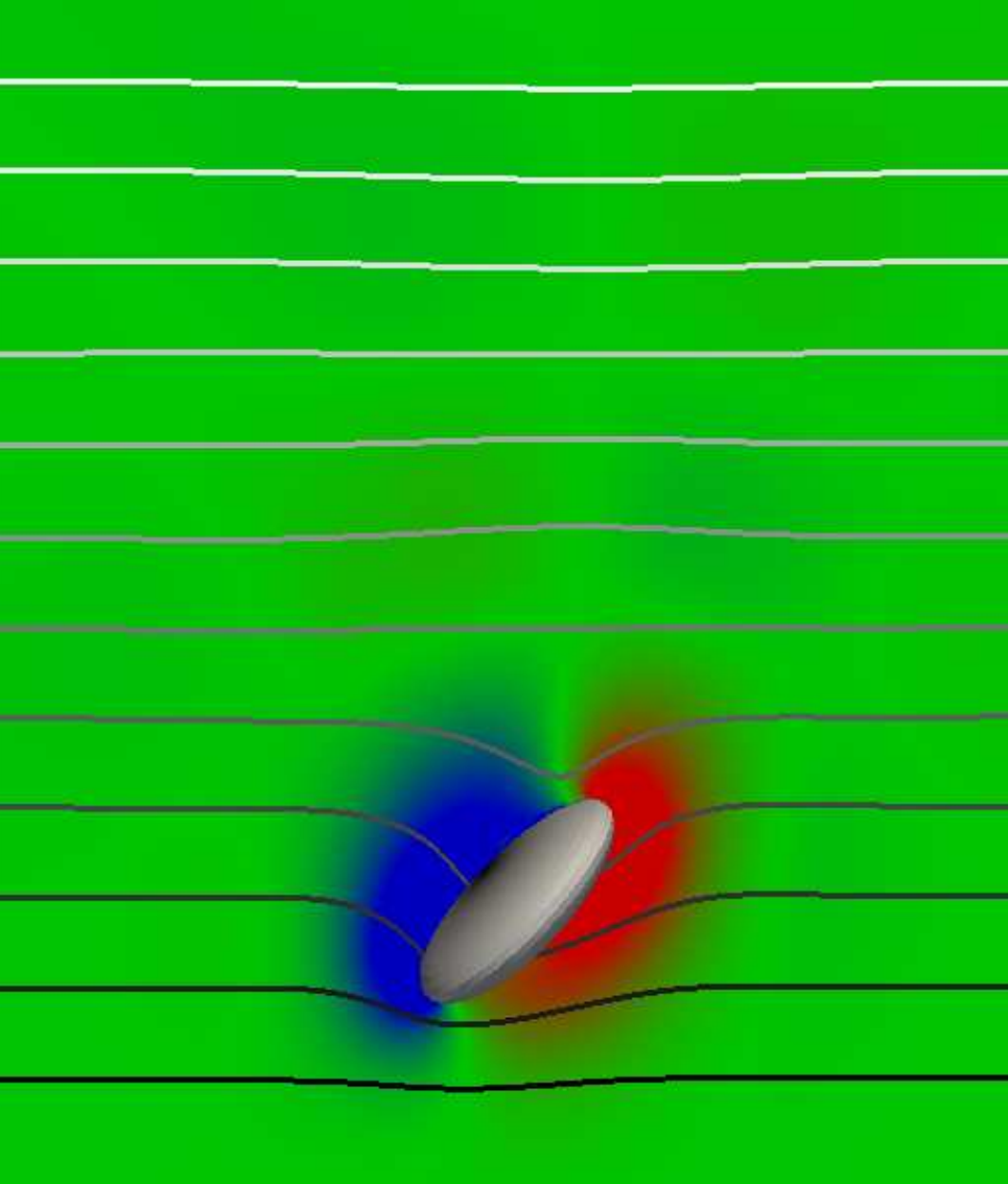}
   
    \end{subfigure} 
    \end{tabular}
    \hspace{-4mm}
    &
    \begin{tabular}{c}
    \begin{subfigure}[t]{0.19\textwidth}
        \centering
        \includegraphics[width=\textwidth]{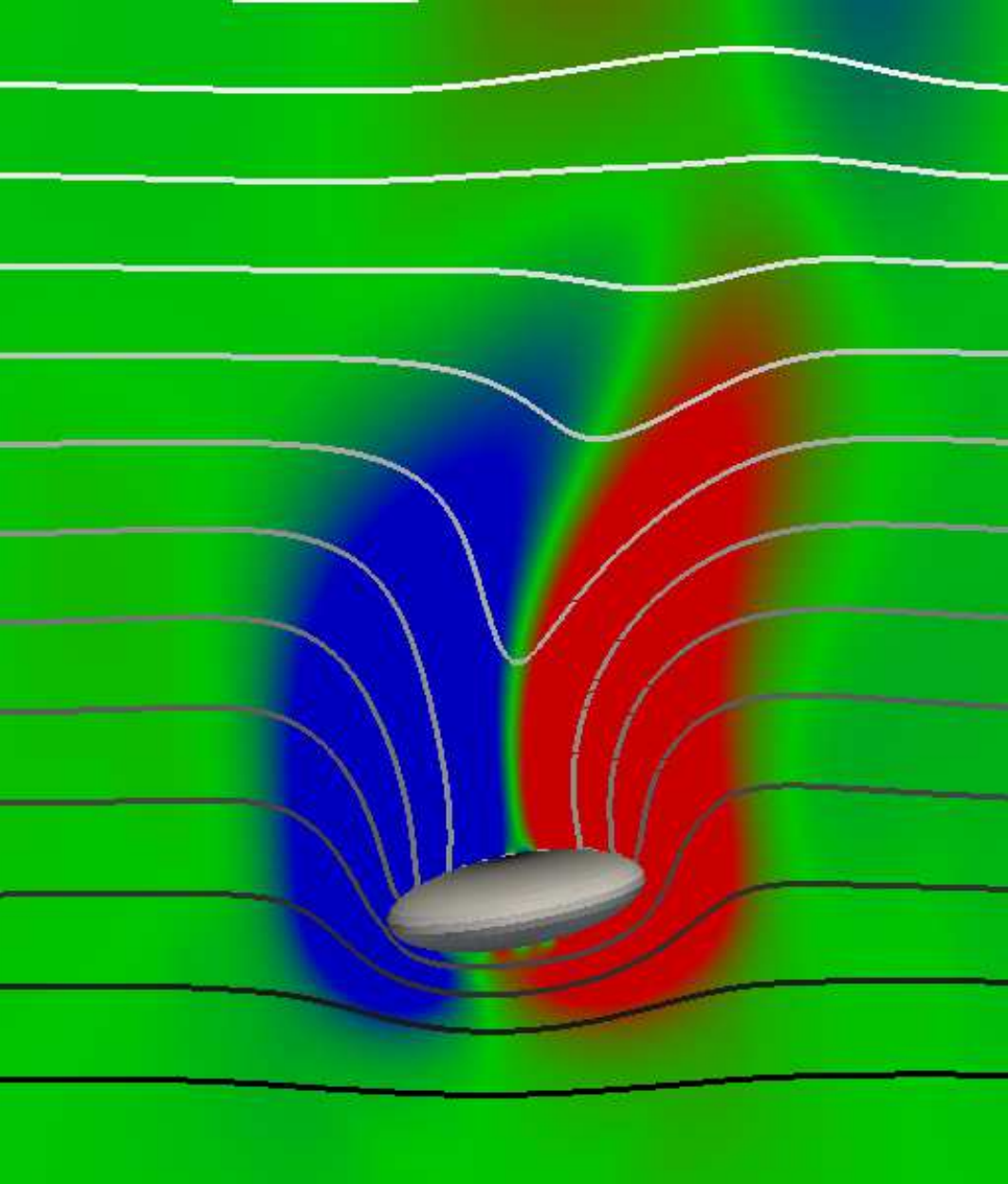}
  
    \end{subfigure} \\
     \begin{subfigure}[t]{0.19\textwidth}
        \centering
        \includegraphics[width=\textwidth]{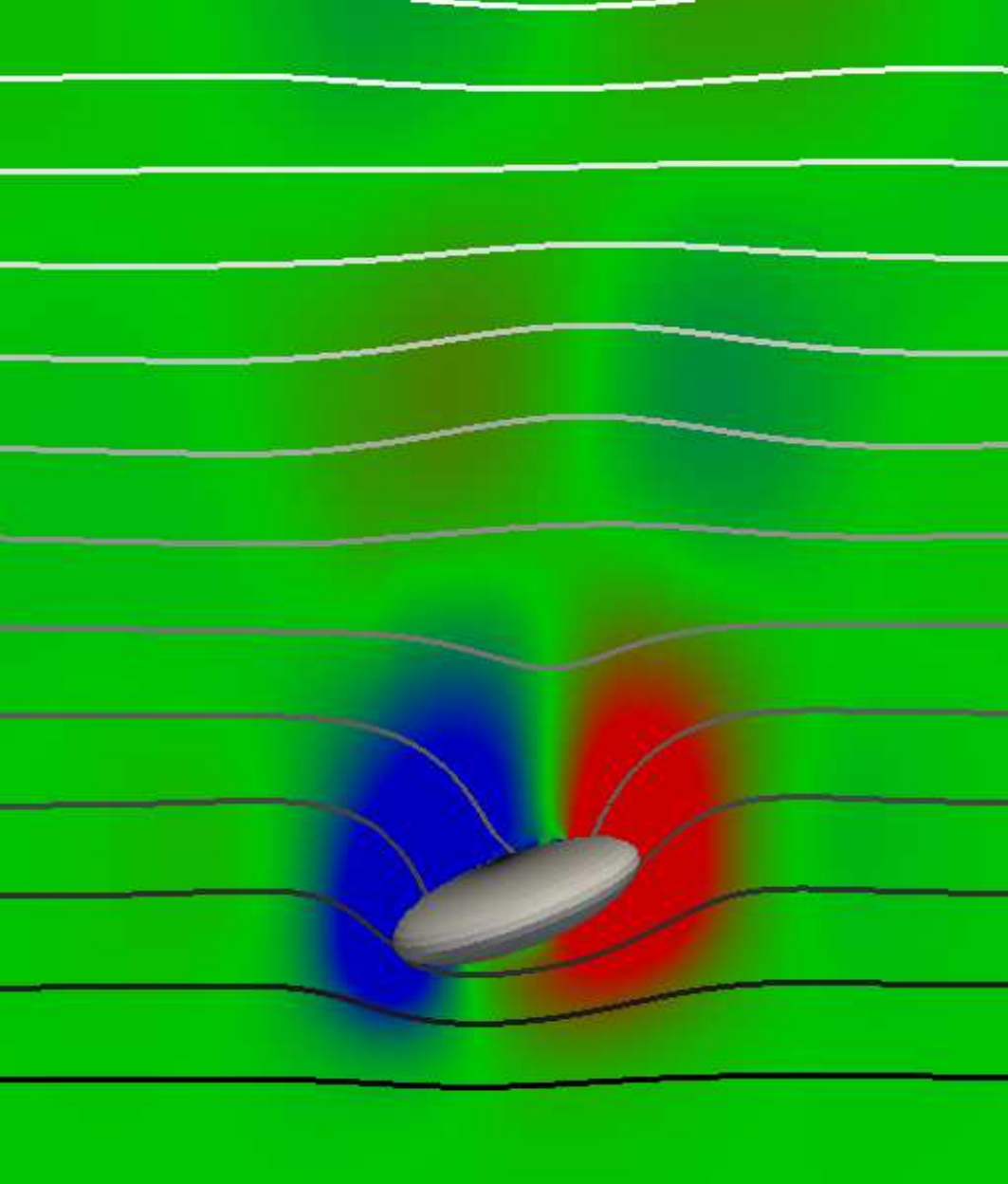}
  
    \end{subfigure} \\
     \begin{subfigure}[t]{0.19\textwidth}
        \centering
        \includegraphics[width=\textwidth]{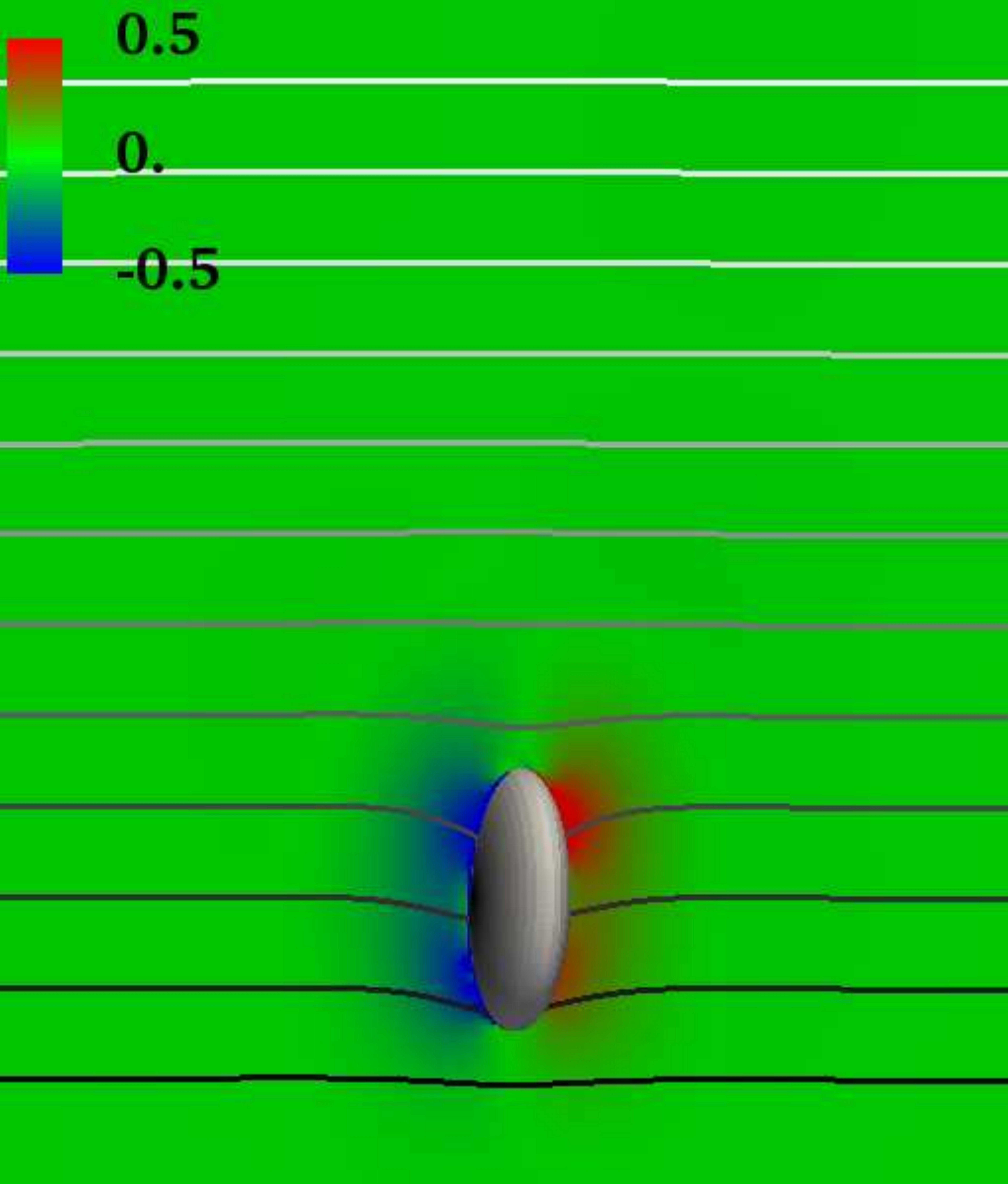}
   
    \end{subfigure}
    \end{tabular}
    \end{tabular}
    \caption{\label{fig:obl_wg}Evolution of the x-component of the dimensionless baroclinic vorticity generation term due to the mis-alignment of the density gradient vector with the direction of gravity, $\boldsymbol{\nabla}\rho_f \times \hat{\bf{k}}$, in the $x=0$ plane for an oblate spheroid with $\mathcal{AR}=1/3$, $Re=80$ and $Ri=5$. For a major clarity, colorbar for the baroclinic vorticity generation is shown only in the last panel. The solid lines indicate dimensionless isopycnals or equal density lines separated by a value of $0.5$. Darker shade of grey indicates a higher density. The panels are snapshots (row-wise) at specific time intervals with $t/\tau =$ 0, 2.69, 8.06, 13.43, 18.8, 24.17, 29.54, 34.91, 40.28, 45.65, 51.02, 56.39, 61.76, 67.13, and 107.4. The first panel shows the initial configuration and the last shows the settling configuration after the oblate reorients in the edge-wise orientation.}
\end{figure*}

The situation is completely different in the case of an oblate spheroid settling in a stratified fluid. This is due to the fact that stratification suppresses the vertical motion of the fluid (\cite{ardekani2010stratlets, Doostmohammadi2014b, more2018mixing,hilali2022sheared} as shown by the isopycnals in fig.~\ref{fig:obl_wg}) and prevents the particle from attaining any steady state speed. As a result, there is no mechanism which can lead to periodic vortex shedding as described above. Conversely, we observe two toroidal vortices, one attached to the particle and one detached from the particle, as shown in fig.~\ref{fig:obl_q}. Once the particle velocity falls below the threshold velocity for reorientation, the detached vortex is asymmetric and does not oscillate from one side to the other unlike the case of an oblate spheroid sedimenting in a homogeneous fluid. As a result, there is a consistent low pressure region behind the oblate spheroid which predominantly remains on one side. This results in a torque on the particle which reorients it until it reaches its neutrally buoyant position. Eventually, as the oblate stops, the torque acting on it also vanishes and it stops in the edge-wise orientation. 

{To make this point clear, we measure the forces and torques acting on the spheroid. As shown in the methodology section, the force (torque) acting on the spheroid can be split into two components (eq.~\ref{eq:NewtonEulerHeat} and~\ref{eq:NewtonEulerHeat2}): 1) $\mathbf{F}_h$ ($\mathbf{T}_h$) , arising from the hydrodynamic stresses acting on the particle surface, denoted as the hydrodynamic force (hydrodynamic torque), and 2) $\mathbf{F}_{b}$ ($\mathbf{T}_b$), arising from the buoyancy or the density disturbance at the particle surface, denoted as the buoyancy force (buoyancy torque). The reason behind the deceleration of the spheroid and its reorientation becomes clear by looking at the $z$-component of the forces and the $x$-component of the torques acting on the spheroid shown for an oblate spheroid with $\mathcal{AR}=1/3$, $Re=80$ and $Ri=5$ in fig.~\ref{fig:frc}. }

{Initially, the density difference between the particle and the local surrounding fluid results in a high buoyancy force (high $F_{b,z}$) on the spheroid resulting in its acceleration (negative $F_z$ at the initial $t/\tau$ in fig.~\ref{fig:force}). As the spheroid accelerates, the magnitude of the hydrodynamic drag increases ($F_{h,z}$ increases) and the buoyancy force decreases in a region with increasing fluid density. Hence, the spheroid accelerates till the magnitude of the hydrodynamic drag becomes larger than the buoyancy force ($F_{h,z} > F_{b,z}$) at which point it attains the maximum velocity. The buoyancy force is unable to overcome this increasing hydrodynamic drag which leads to the deceleration of the spheroid ($F_z > 0$ meaning the net force acting on the spheroid is in the opposite direction to its motion). Eventually, as the particle reaches its neutrally buoyant position, it stops as there is no net force acting on it. In a homogeneous fluid, i.e., $Ri=0$, the buoyancy force acting on the particle is constant, $F_{b,z} = (\rho_p-\rho_f)V_pg$ and the hydrodynamic drag balances the buoyancy force at steady state, resulting in a constant terminal velocity.  }

{To understand the reason behind the reorientation, we plot the x-component of the torques acting on the spheroid in fig.~\ref{fig:torq}. Initially, as the particle accelerates, it topples from edge-wise orientation to a broad-side orientation because of the increasing magnitude of the hydrodynamic torque ($T_{h,x}$) compared to the buoyancy torque ($T_{b,x}$). Because of inertia, $T_{h,x}$ changes sign and and the oblate oscillates about its broad-side on configuration. This is shown by the oscillating $T_{h,x}$ and $\omega_{p,x}$ in fig.~\ref{fig:torq} at initial times. Meanwhile, the buoyancy torque ($T_{b,x}$) increases gradually and is always $> 0$, which leads to dampening of the oscillations of the oblate spheroid about the broad-side on orientation as can be seen from the diminishing magnitude of the rotational velocity in fig.~\ref{fig:torq}. }

{The spheroid keeps oscillating about the broad-side on orientation as long as the inertial effects (or $T_{h,x}$) are stronger compared than the buoyancy effects (or $T_{b,x}$). However, as the spheroid decelerates, inertial effects start to weaken. In addition, the isopycnals resist further deformation as will be explained below. As the spheroid velocity falls below the threshold for reorientation ($U_p(t)/U < 0.15$), the destabilizing buoyancy torque dominates over the stabilizing hydrodynamic torque, i.e., $T_{b,x} > |T_{h,x}|$. This transition in the dominating torque is demarcated by a dotted vertical line in fig.~\ref{fig:torq} which also corresponds to the time when $U_p(t)/U < 0.15$. As a result, the spheroid stops oscillating about the broad-side on orientation and starts to reorient to the edge-wise orientation since $T_{b,x} > |T_{h,x}|$ implies a net positive torque on the spheroid which results in a net positive rotational velocity ($\omega_{p,x} > 0$) as shown in fig.~\ref{fig:torq}. In a homogeneous fluid, the buoyancy/baroclinic torque is absent. Hence, the inertia and $T_h$ acting on the spheroid results in a broad-side on orientation at steady state. The competition between the stabilizing hydrodynamic torque and the destabilizing buoyancy torque can be understood by looking at the flow field and the isopycnals around the spheroid as it sediments, as discussed below.  }

The equation for the vorticity, $\boldsymbol{\omega}$, can be obtained by taking the curl of the momentum equation~\ref{eq:NS}. 

\begin{equation}
 \rho_f \frac{D\boldsymbol{\omega}}{D{t}}  = \left( \boldsymbol{\omega} \cdot \boldsymbol{\nabla} \right) \mathbf{u}+ \mu \nabla^2 \boldsymbol{\omega} - g \boldsymbol{\nabla} \rho_f \times \hat{\mathbf{k}}.
  \label{eq:vort}
\end{equation}
The last term on the right hand side of equation~\ref{eq:vort}, i.e., $\boldsymbol{\omega}_g = - g \boldsymbol{\nabla} \rho_f \times \hat{\mathbf{k}}$, is the vorticity generation due to the displacement of isopycnals caused by the settling motion of the particle. This term is also known as the baroclinic vorticity generation. This contribution arises due to the mis-alignment of the density gradient with the direction of gravity. This term will be exactly $\bf{0}$ in a homogeneous fluid. This contribution is thus specific to particles sedimenting in a stratified fluid as the vorticity around the particle is very different in a homogeneous and a stratified fluid \citep{Doostmohammadi2014b}. 

We plot the x-component of the baroclinic vorticity generation, $\boldsymbol{\omega}_g$, in the $yz$ plane around a settling oblate spheroid in fig.~\ref{fig:obl_wg}. This term reveals the reason behind the onset of instability and the reorientation of an oblate spheroid in a stratified fluid. Initially this vorticity generation term is symmetric with a thin region of zero $\boldsymbol{\omega}_g$ separating regions of positive and negative baroclinic vorticity (blue and red regions in fig.~\ref{fig:obl_wg}) exactly along the center-line of the spheroid. We call this the plume of zero baroclinic vorticity or ``the plume" for simplicity. The plume also acts as the axis of symmetry for $\boldsymbol{\omega}_g$. We call the point at which the plume intersects the particle surface as the origin of the plume. A vertically straight plume with its origin on one of the center-lines of the oblate signifies a symmetric $\boldsymbol{\omega}_g$ around the particle. 

As the particle settles and slows down, the oblate spheroid topples from an edge-wise to broad-side on orientation due to inertial effects. Since the particle is accelerating, the vorticity generation region expands as the isopycnals deform in the long wake behind the particle till it reaches the peak velocity. After reaching the peak velocity, the particle decelerates due to increasing buoyancy effects because of the tendency of the displaced isopycnals to return to their original levels as shown by evolution of isopycnals in fig.~\ref{fig:obl_wg}. As a result, the region of vorticity generation shrinks. The origin of the plume shifts along the longer face of the oblate towards the other end as it oscillates about the broad-side on orientation. 

As the inertial effects decrease with the particle deceleration, the oscillations of the oblate spheroid about the broad-side on orientation are dampened. The isopycnals that were deformed earlier (in the wake of the particle) do not completely return to their original form, thus opposing further deformation as the oblate particles tries to oscillate. Hence, the oscillations die out.  This prevents the origin of the thin plume from shifting completely to the middle of the spheroid, thus preventing $\boldsymbol{\omega}_g$ to become symmetric. Since the origin of the plume is not at the center of the oblate, the generated vorticity field is asymmetric. The origin of the plume does not cross the center of the spheroid and remains on one side. In addition, because of the reduced inertia, there is no mechanism to keep the spheroid oscillating about the horizontal. Thus, the $\boldsymbol{\omega}_g$ distribution around the oblate remains asymmetric. This results in the onset of instability in the oblate orientation as the origin of the plume tries to return to its earlier position on the spheroid, i.e., on the edge. The net torque on the oblate spheroid slowly reorients it to the edge-wise orientation (fig.~\ref{fig:torq}). The same process will occur irrespective of the initial orientation of the oblate spheroid which will eventually reorient in the edge-wise orientation.

\subsection{Settling dynamics of a prolate spheroid in a stratified fluid}
Similar to the case of an oblate spheroid, we report the simulation results on the settling dynamics of a prolate spheroid with $\mathcal{AR}=2$ in a stratified fluid. We present the results for the settling dynamics in a homogeneous fluid as well for comparison.

\subsubsection{Fluid stratification slows down and partially reorients a settling prolate spheroid}\label{sec:prolmain}
\begin{figure*}
    \centering
     \begin{subfigure}[t]{0.45\textwidth}
        \centering
        \includegraphics[width=\textwidth]{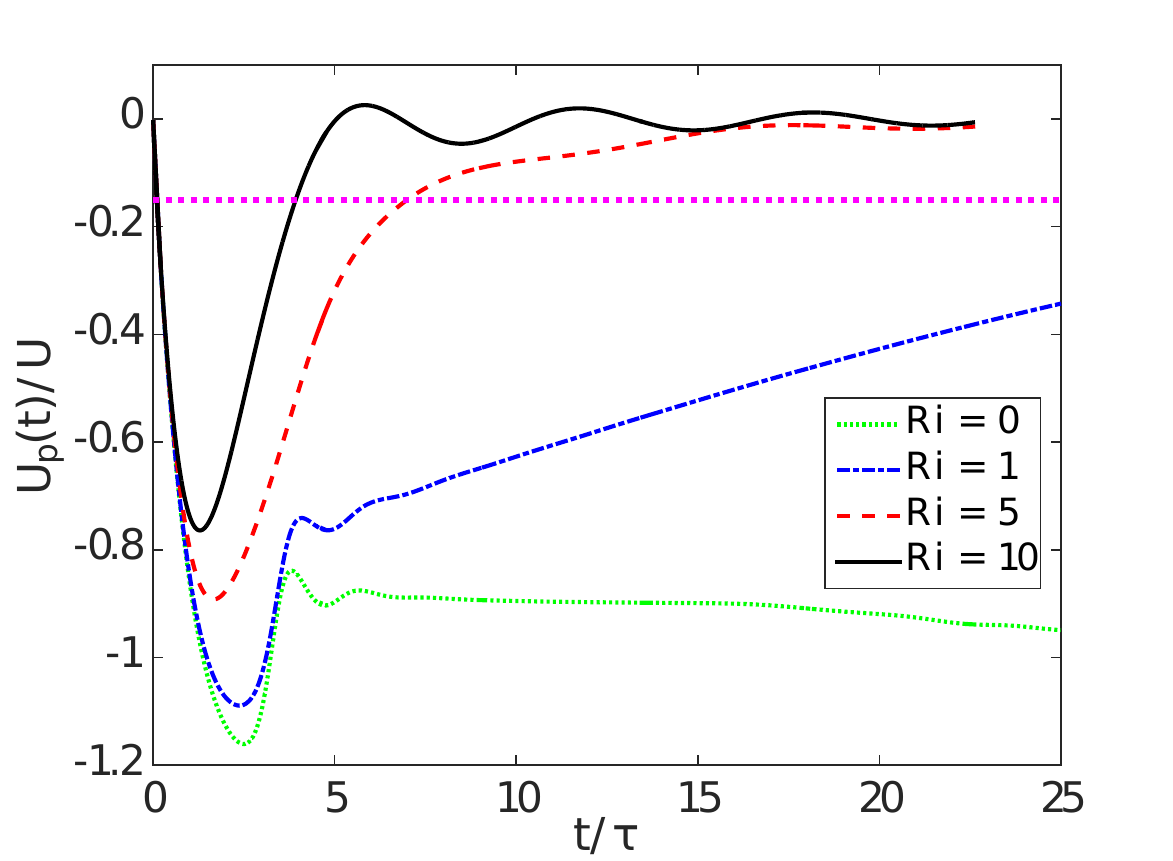}
    	\caption{Settling velocity, $Re=80$\label{fig:80_vel}}
     
    \end{subfigure}
    ~ 
    \begin{subfigure}[t]{0.45\textwidth}
        \centering
        \includegraphics[width=\textwidth]{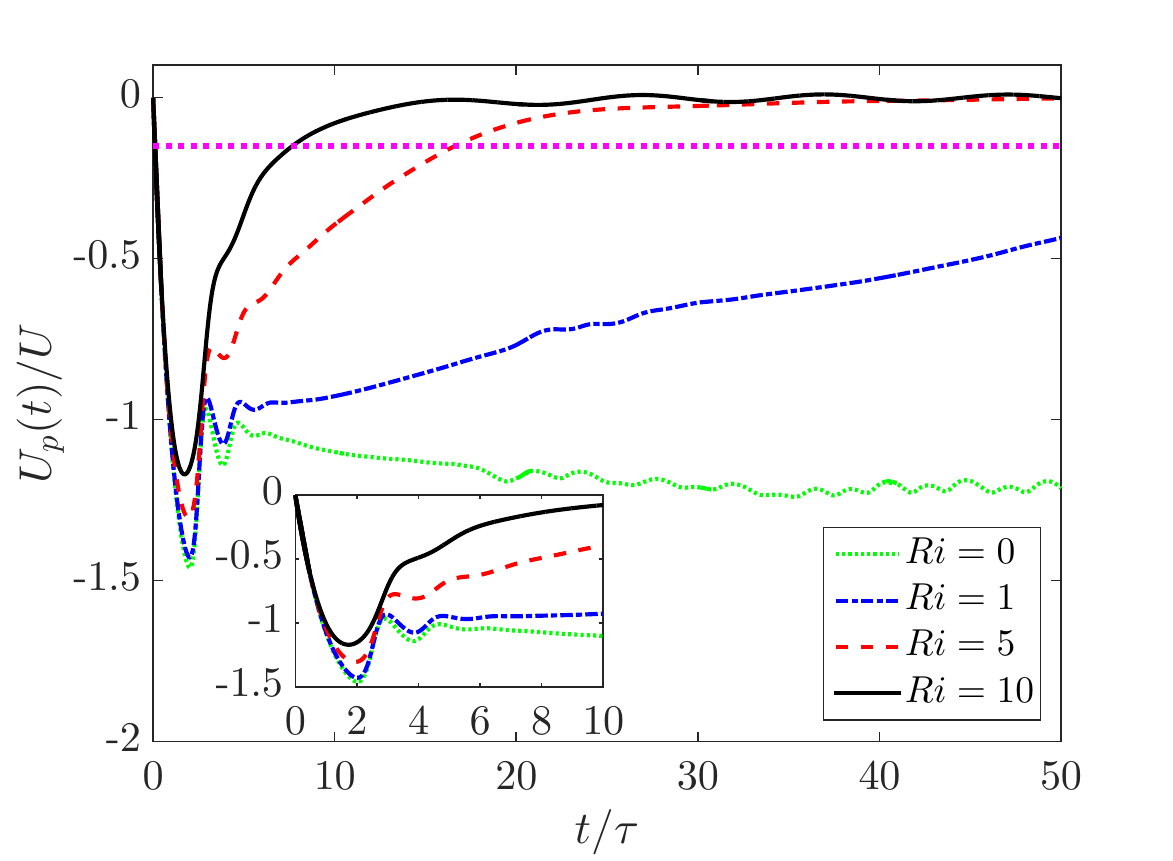}
    	\caption{Settling velocity, $Re=180$\label{fig:180_vel}}
      
    \end{subfigure}
    
    \begin{subfigure}[t]{0.45\textwidth}
        \centering
        \includegraphics[width=\textwidth]{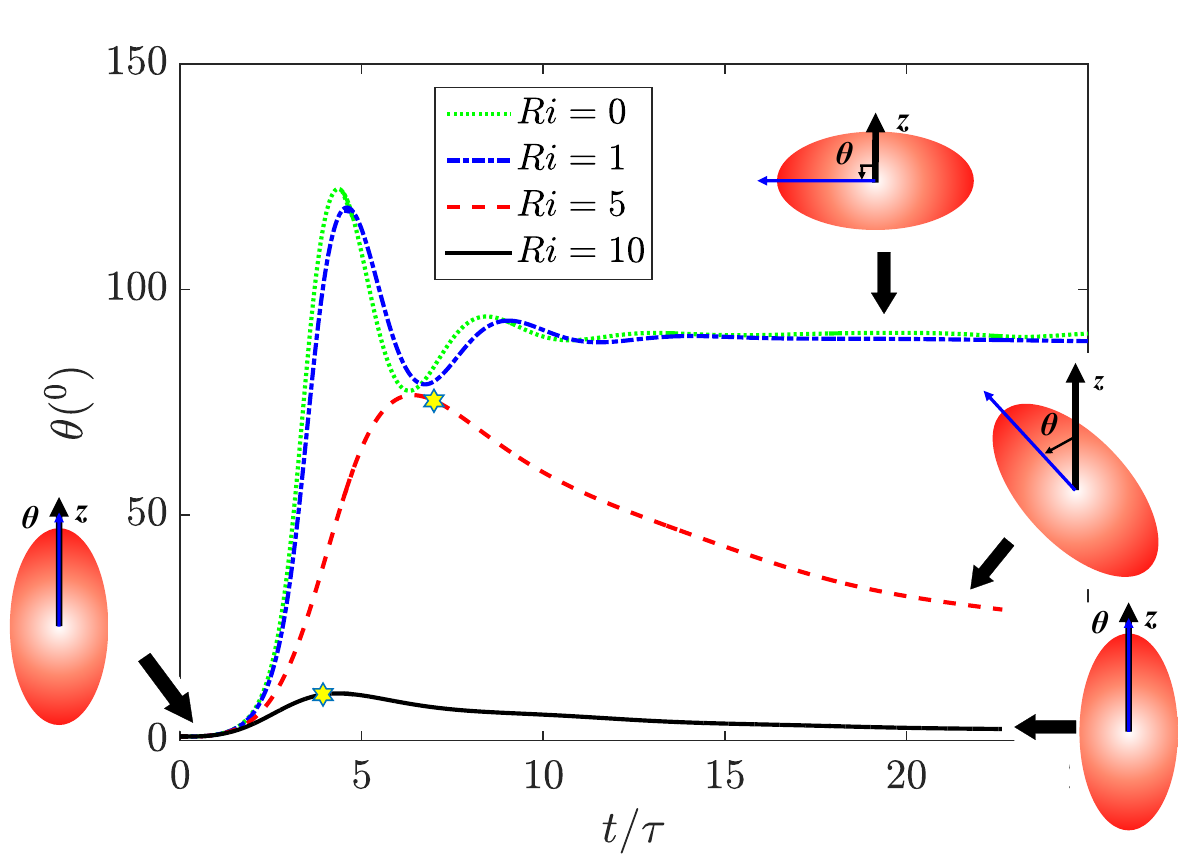}
    	\caption{Orientation, $Re=80$\label{fig:80_or}}
     
    \end{subfigure}
    ~ 
    \begin{subfigure}[t]{0.45\textwidth}
        \centering
        \includegraphics[width=\textwidth]{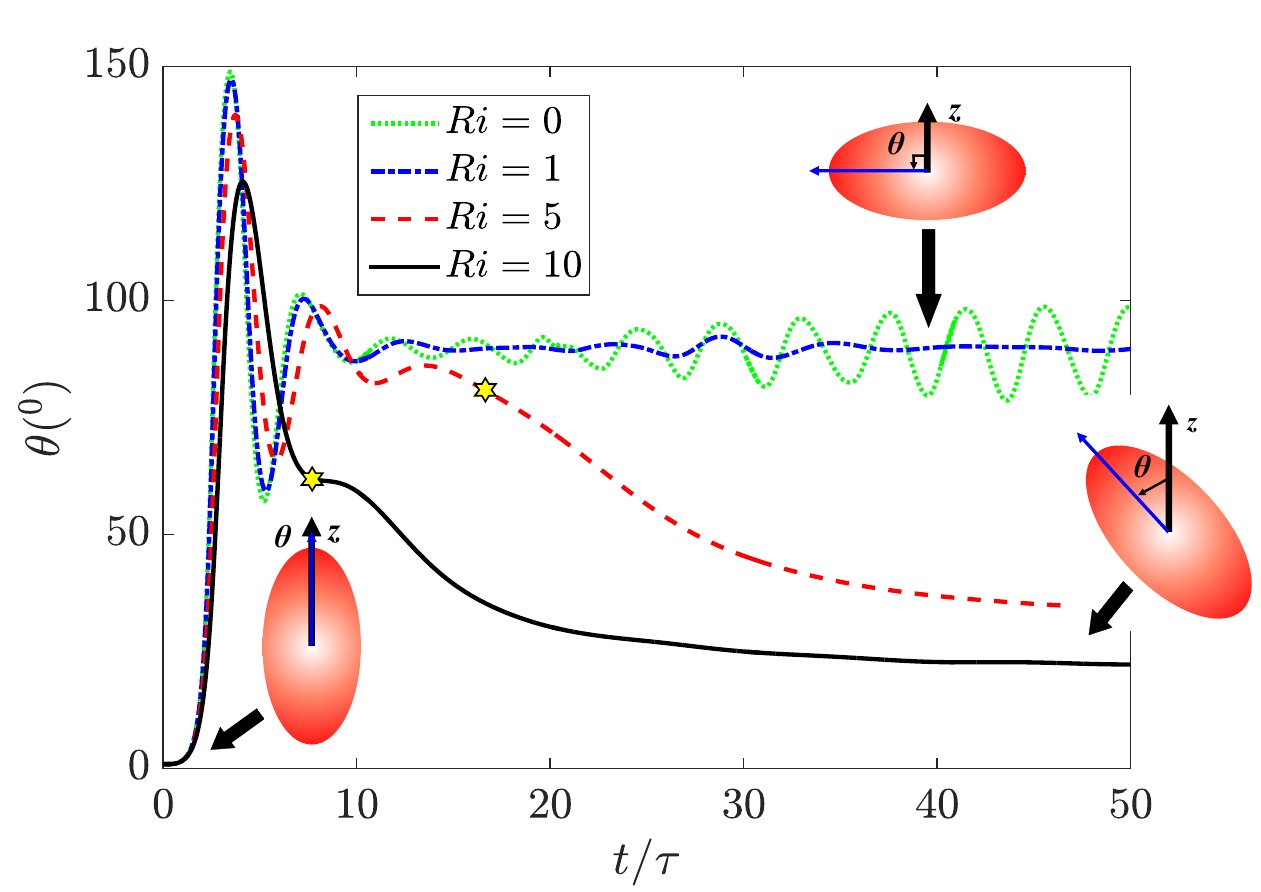}
    	\caption{Orientation, $Re=180$\label{fig:180_or}}
   
    \end{subfigure}
    
    \caption{\label{fig:prol_vel}Time evolution of the settling velocity of a prolate spheroid with $\mathcal{AR}=2$ in a homogeneous fluid ($Ri=0$) and a stratified fluid with different $Ri$: a) $Re=80$, b) $Re=180$. Evolution of the prolate orientation for $\mathcal{AR}=2$ in a homogeneous fluid ($Ri=0$) and a stratified fluid with different $Ri$: c) $Re=80$, d) $Re=180$. The inset in (b) shows the initial oscillations with decreasing amplitudes in the velocity and orientation of the spheroid. The prolate spheroid attains a steady state terminal velocity and orientation (broad-side on) in a homogeneous fluid. Stratification leads to a reduction in the spheroid velocity and a continuous deceleration of the spheroid velocity until it stops. The magnitude of the deceleration increases with stratification. The onset of reorientation given by $|U_p/U| < 0.15$ and is denoted by a dotted horizontal line in (a,b) and correspondingly by yellow stars in (c,d).}
\end{figure*}

Fig.~\ref{fig:prol_vel} shows the settling velocity of a prolate spheroid with $\mathcal{AR}=2$ in a homogeneous and stratified fluid with different stratification strengths for $Re$ = 80 and 180. The prolate spheroid starts from rest in an initially quiescent fluid. It then accelerates to reach a maximum velocity depending on its $Re$ and $Ri$. In a homogeneous fluid, i.e., $Ri=0$, the prolate spheroid reaches a terminal settling velocity as shown in fig.~\ref{fig:80_vel} and~\ref{fig:180_vel}. 

\begin{figure*}
    \centering
    \begin{tabular}[t]{ccccc}
    \hspace{-4mm}
     \begin{tabular}{c}
    \begin{subfigure}[t]{0.19\textwidth}
        \centering
        \includegraphics[width=\textwidth]{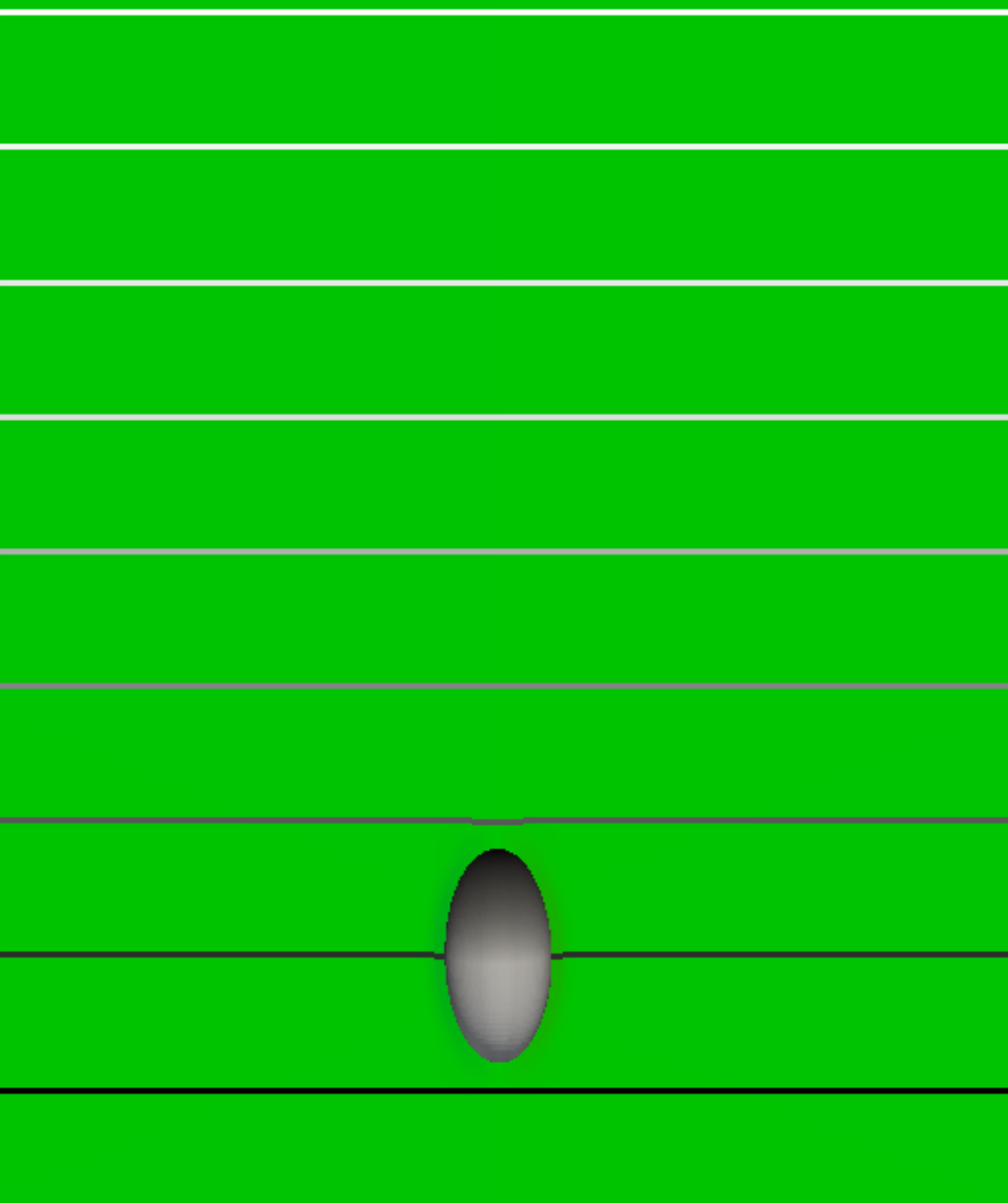}
      
    \end{subfigure} \\
     \begin{subfigure}[t]{0.19\textwidth}
        \centering
        \includegraphics[width=\textwidth]{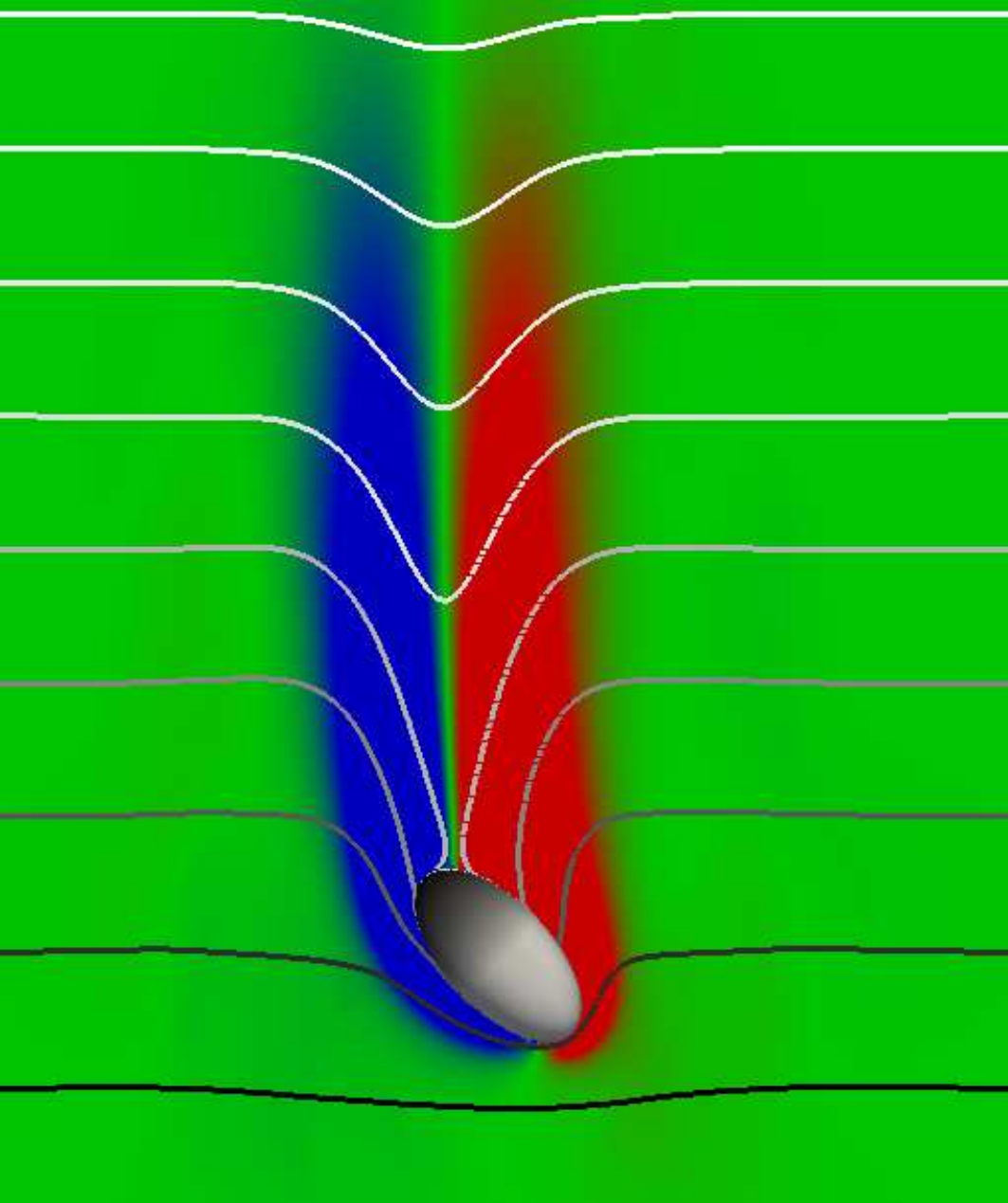}
  
    \end{subfigure} \\
     \begin{subfigure}[t]{0.19\textwidth}
        \centering
        \includegraphics[width=\textwidth]{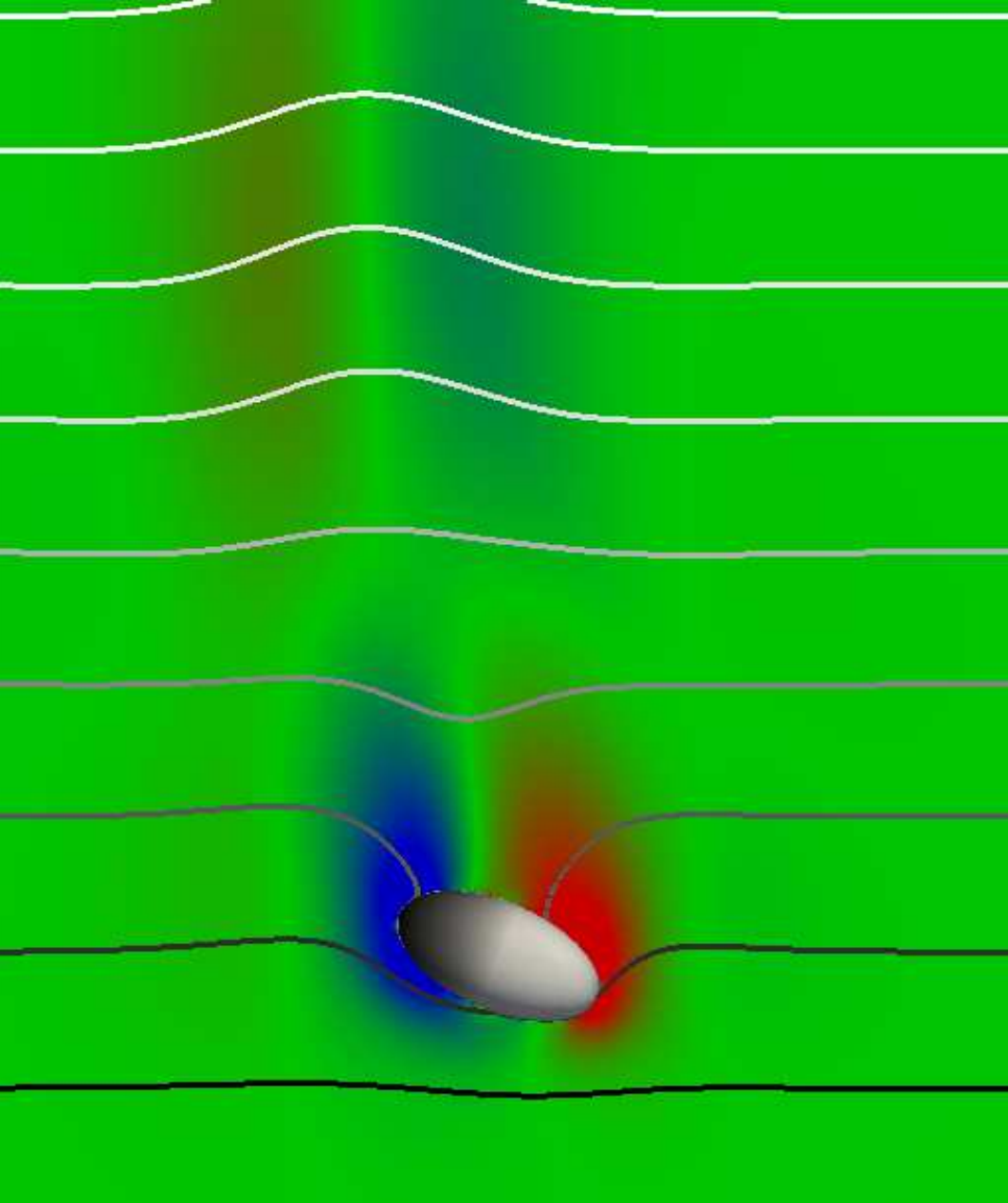}
  
    \end{subfigure} 
    \end{tabular}
    \hspace{-4mm}
   &
   \begin{tabular}{c}
     \begin{subfigure}[t]{0.19\textwidth}
        \centering
        \includegraphics[width=\textwidth]{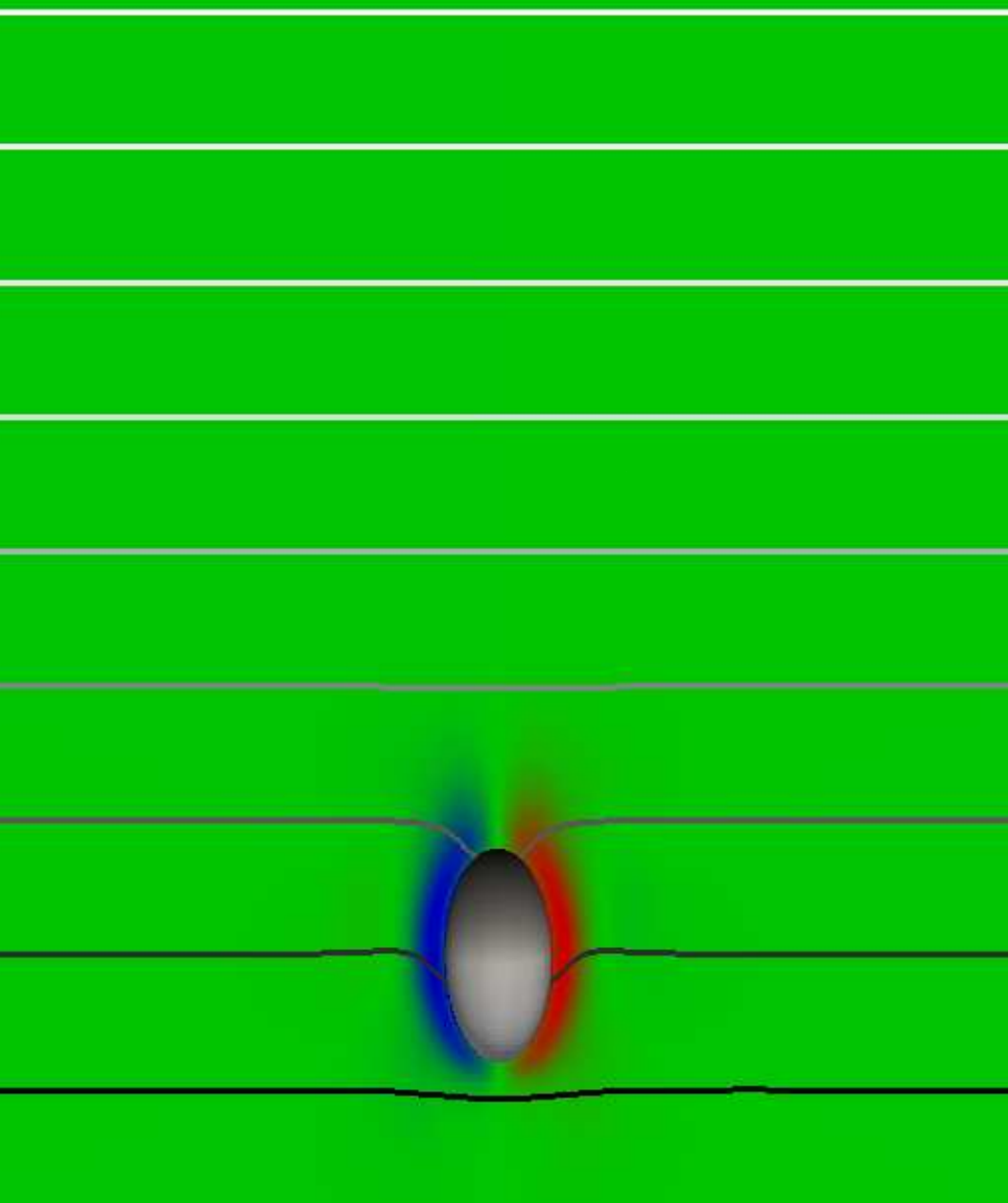}
   
    \end{subfigure} \\
     \begin{subfigure}[t]{0.19\textwidth}
        \centering
        \includegraphics[width=\textwidth]{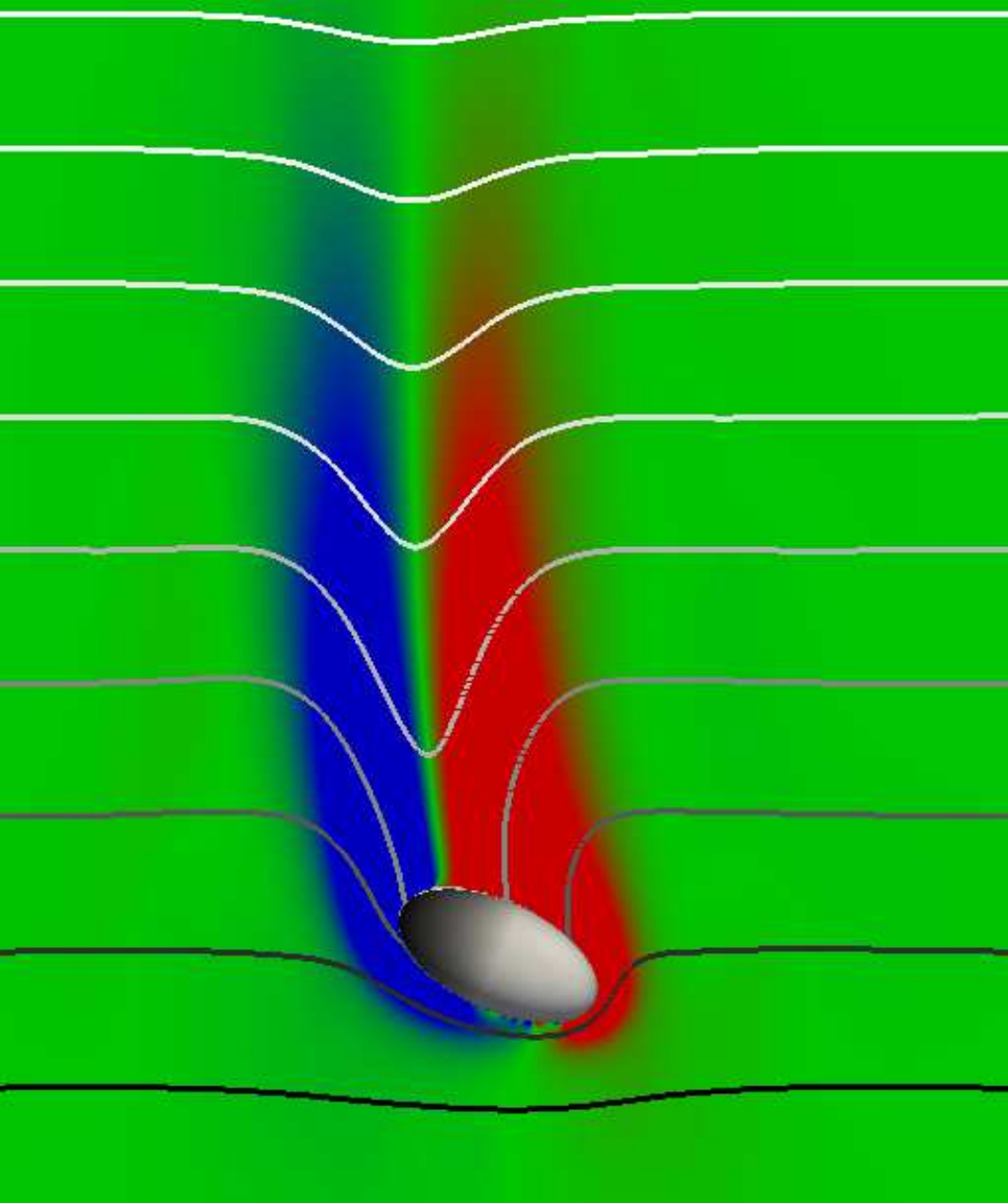}
    
    \end{subfigure} \\
     \begin{subfigure}[t]{0.19\textwidth}
        \centering
        \includegraphics[width=\textwidth]{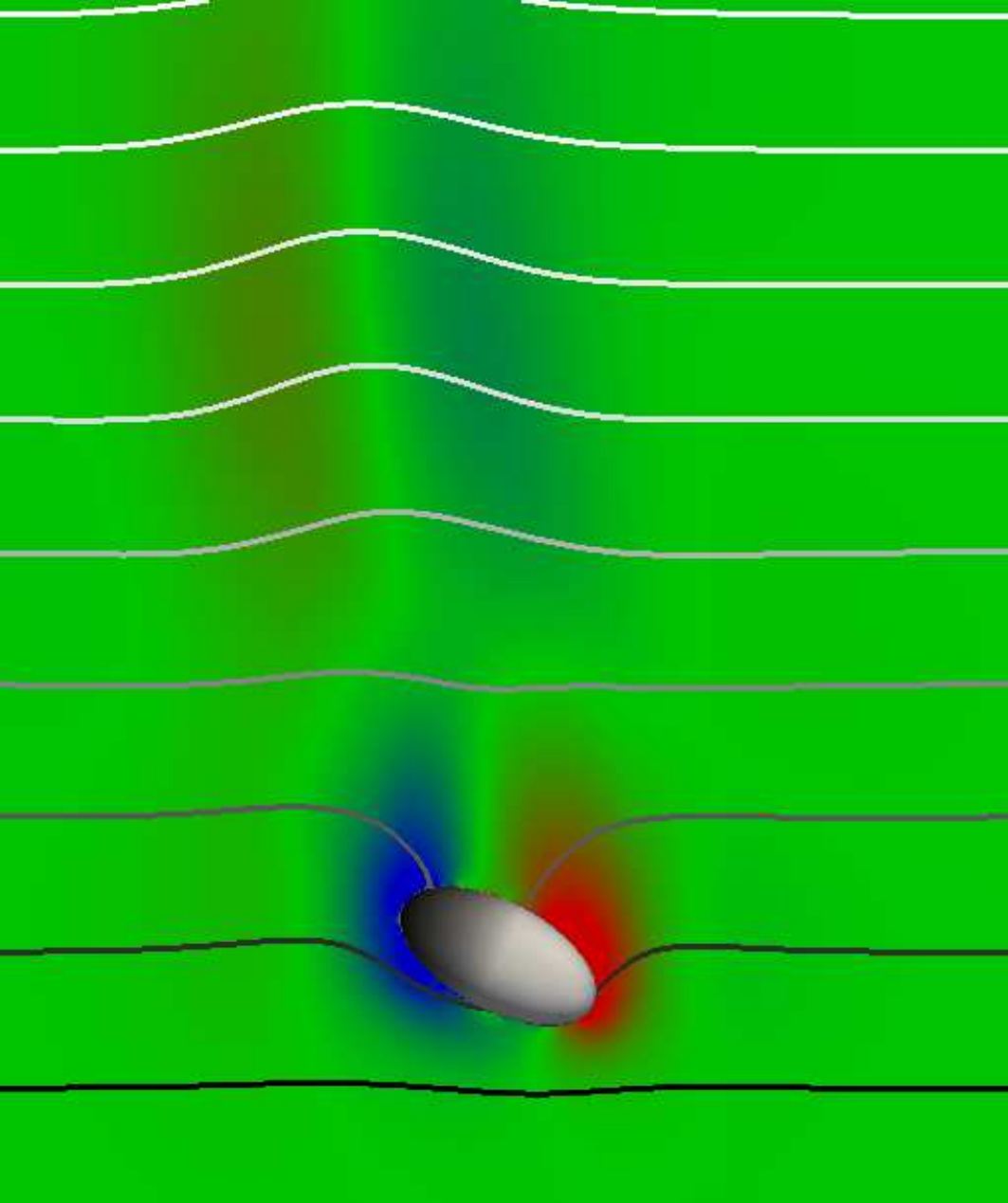}
    
    \end{subfigure} 
    \end{tabular}
    \hspace{-4mm}
    & 
    \begin{tabular}{c}
     \begin{subfigure}[t]{0.19\textwidth}
        \centering
        \includegraphics[width=\textwidth]{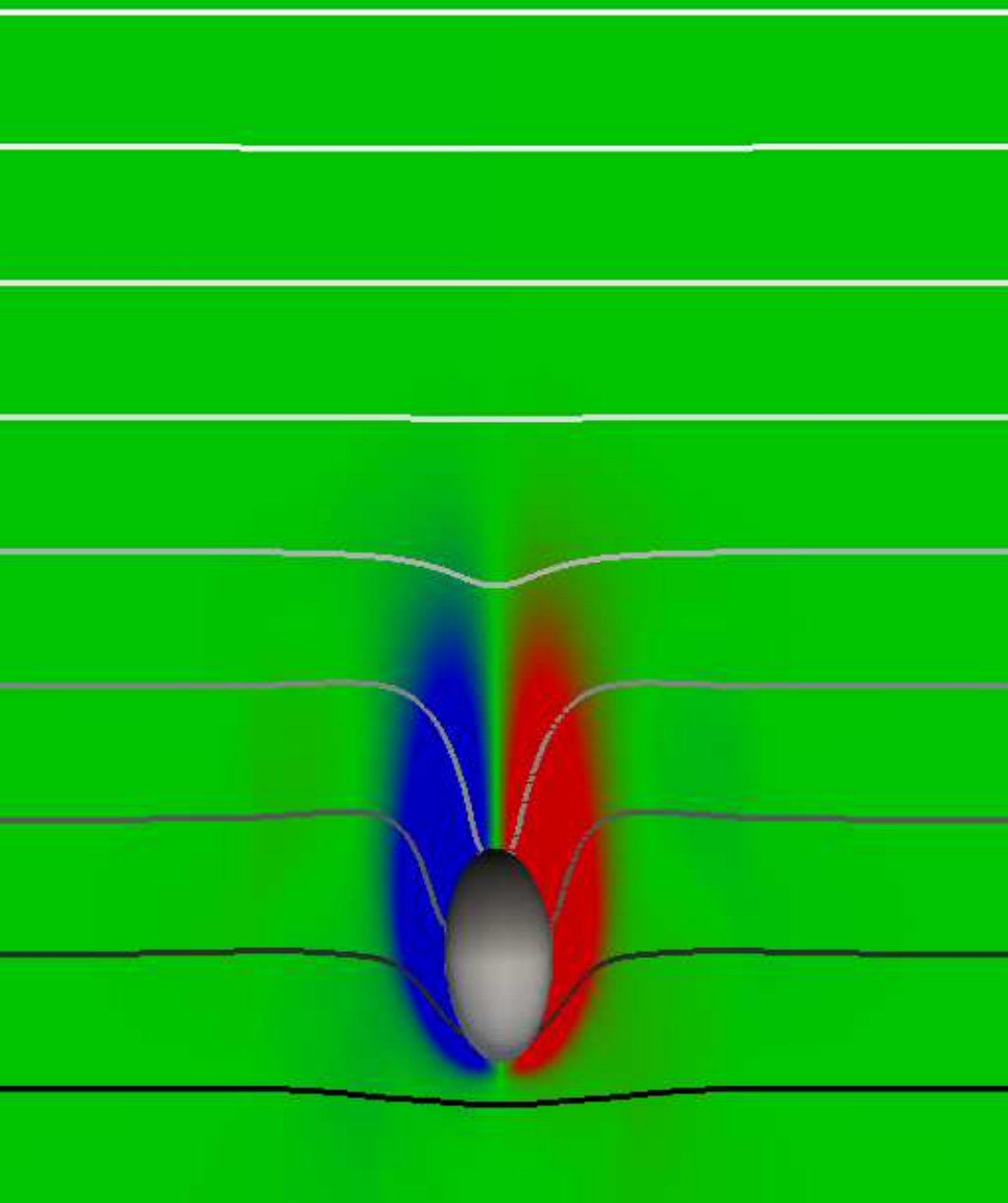}
 
    \end{subfigure} \\
     \begin{subfigure}[t]{0.19\textwidth}
        \centering
        \includegraphics[width=\textwidth]{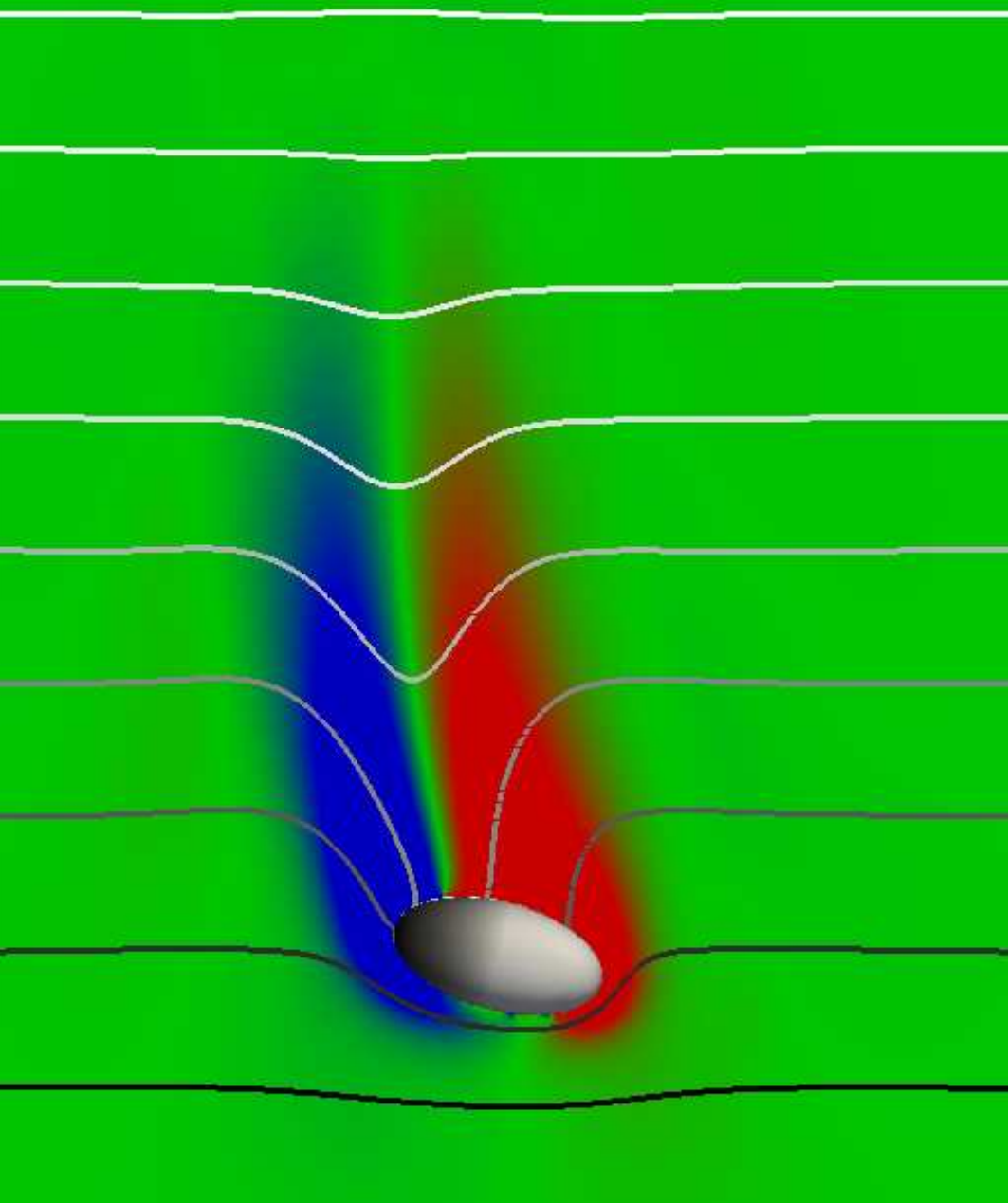}
  
    \end{subfigure} \\
     \begin{subfigure}[t]{0.19\textwidth}
        \centering
        \includegraphics[width=\textwidth]{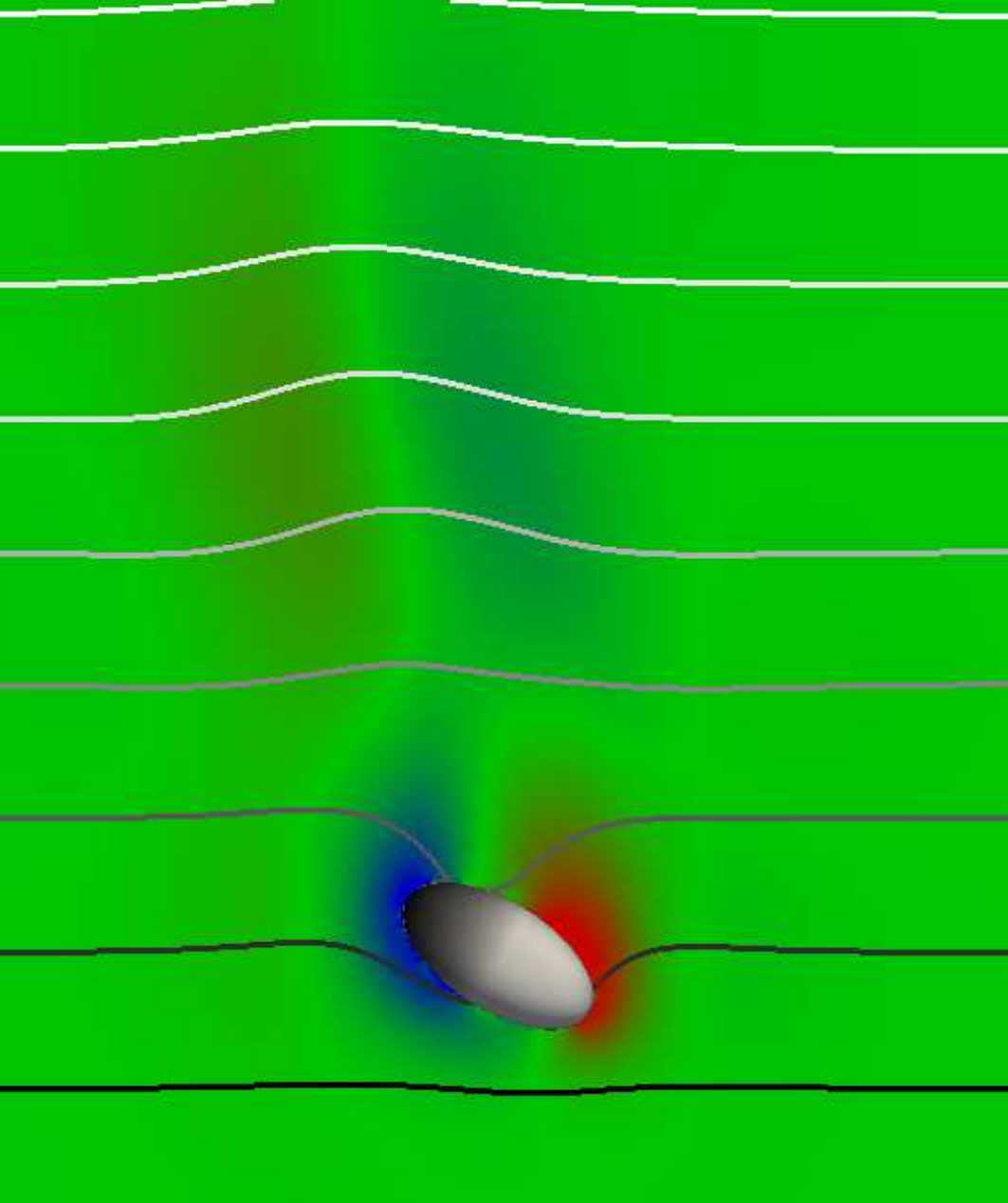}
  
    \end{subfigure} 
    \end{tabular}
    \hspace{-4mm}
    & 
    \begin{tabular}{c}
     \begin{subfigure}[t]{0.19\textwidth}
        \centering
        \includegraphics[width=\textwidth]{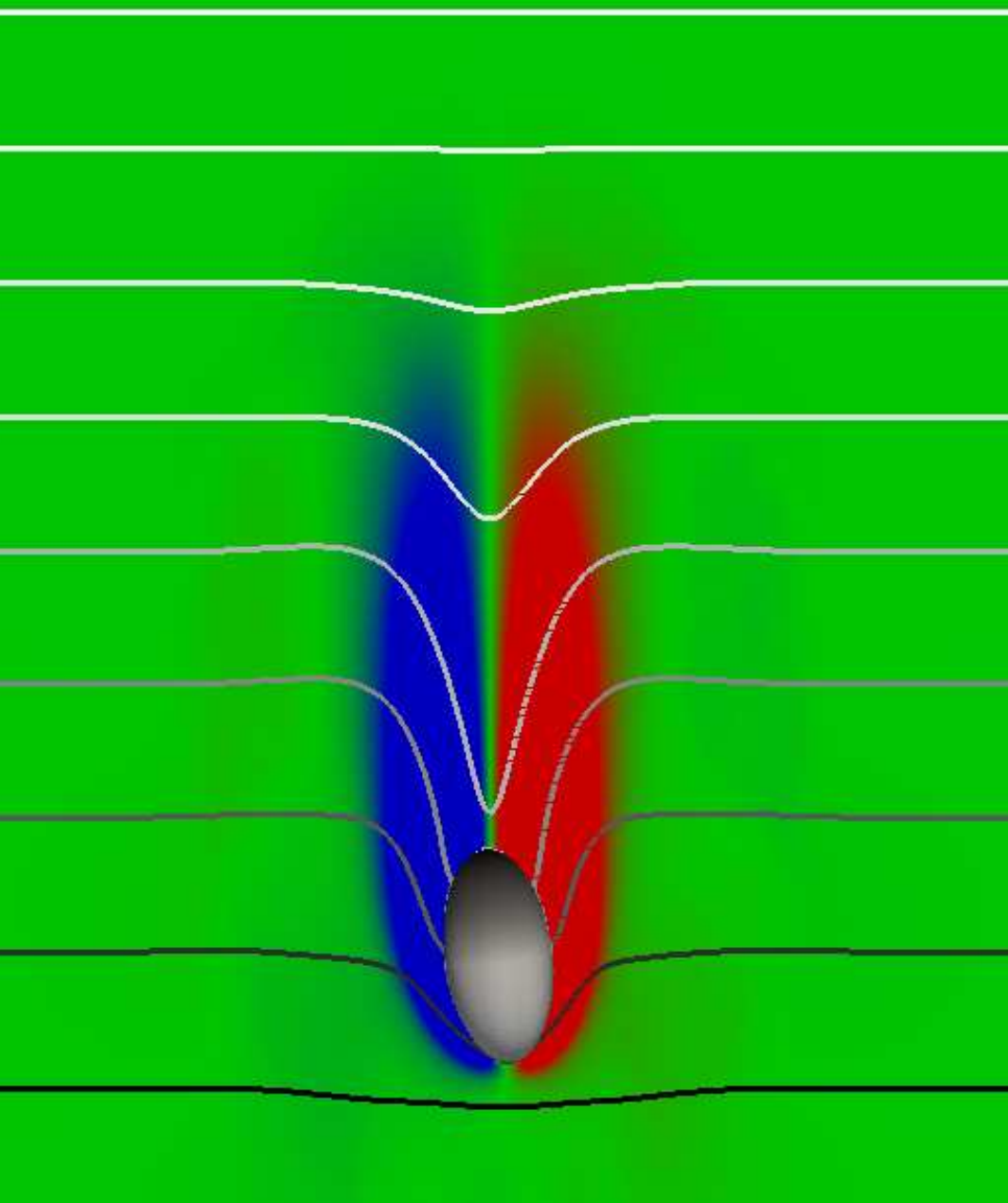}
   
    \end{subfigure} \\
     \begin{subfigure}[t]{0.19\textwidth}
        \centering
        \includegraphics[width=\textwidth]{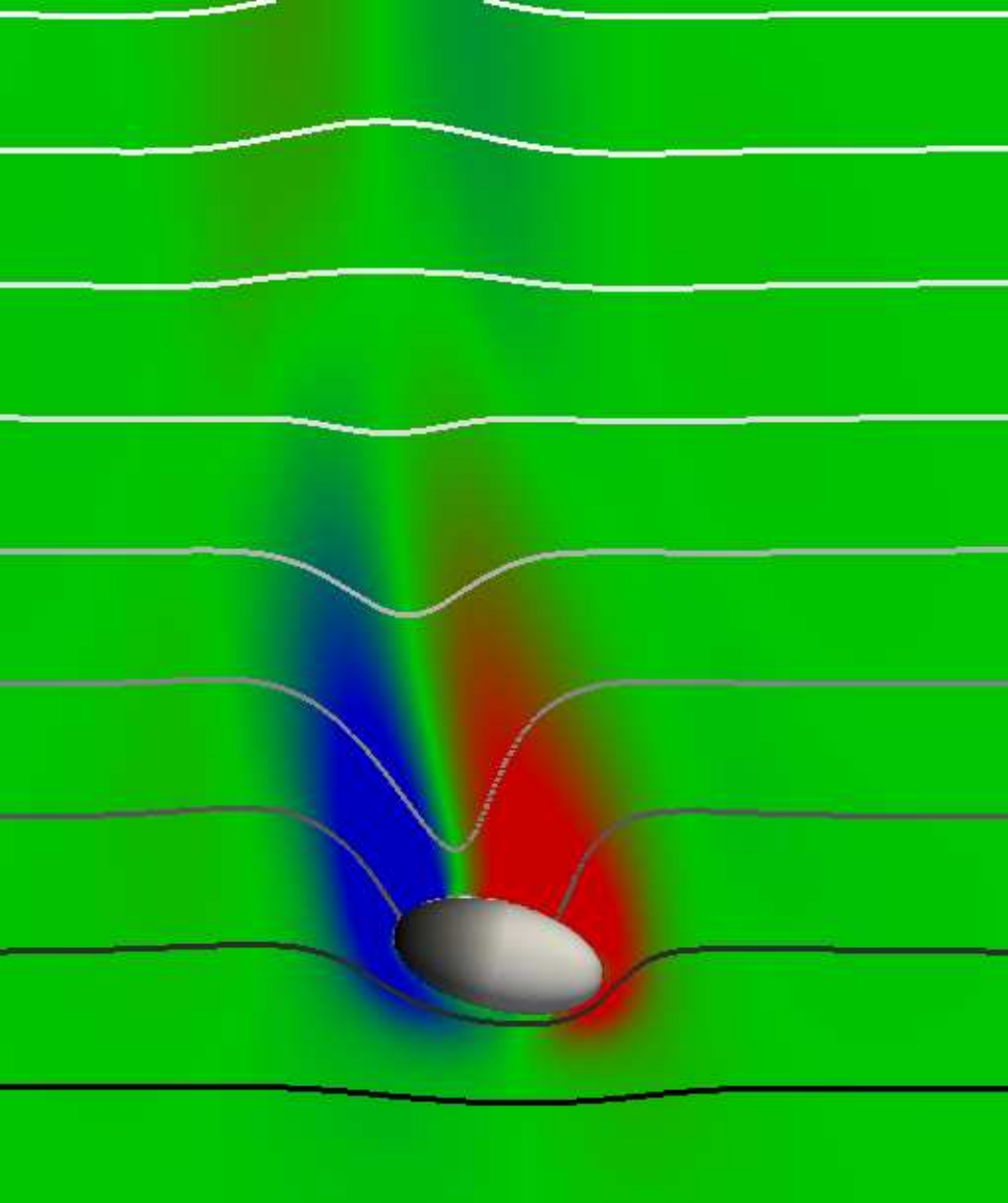}
  
    \end{subfigure} \\
     \begin{subfigure}[t]{0.19\textwidth}
        \centering
        \includegraphics[width=\textwidth]{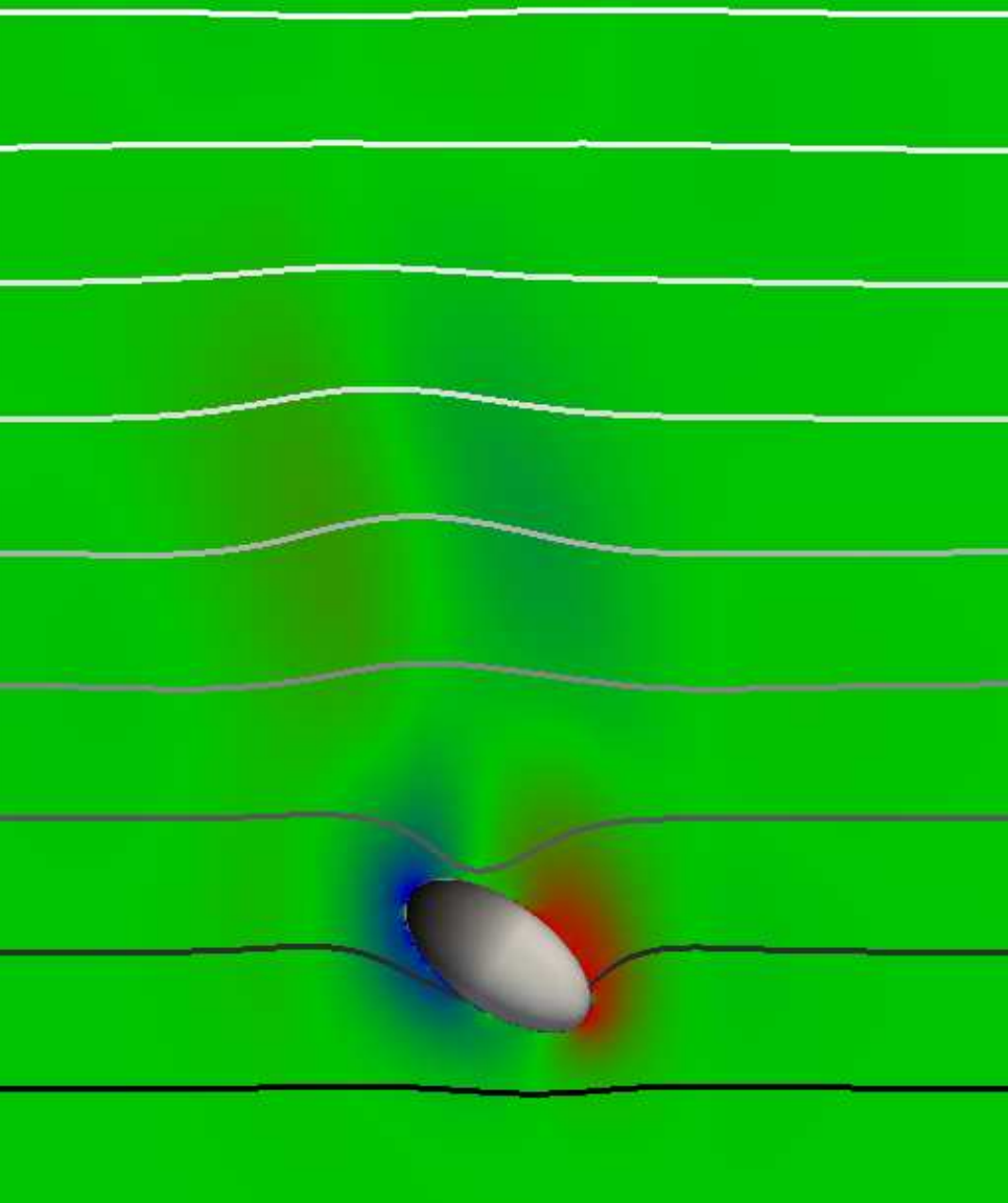}
  
    \end{subfigure} 
    \end{tabular}
    \hspace{-4mm}
    &
    \begin{tabular}{c}
    \begin{subfigure}[t]{0.19\textwidth}
        \centering
        \includegraphics[width=\textwidth]{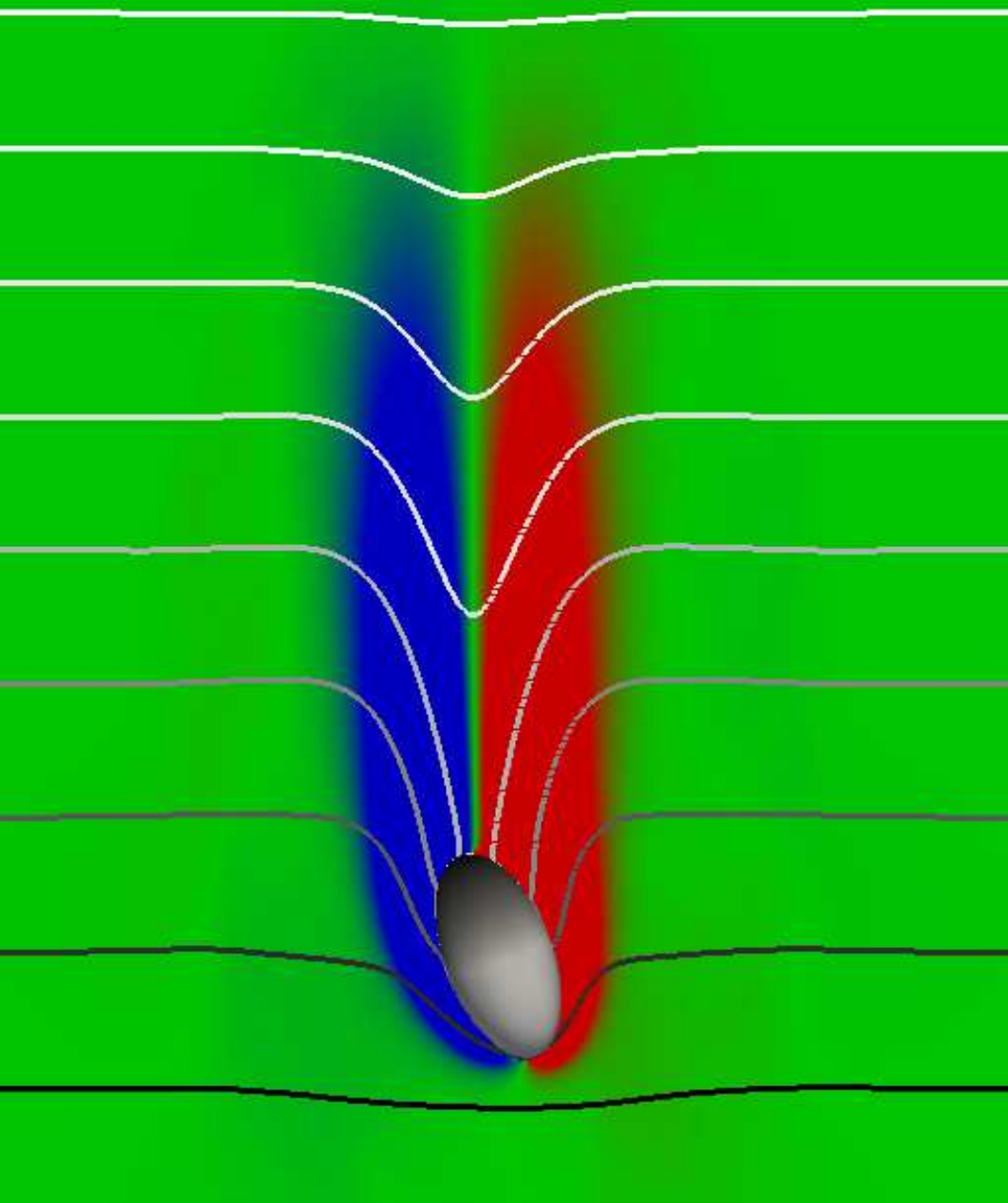}
    
    \end{subfigure} \\
     \begin{subfigure}[t]{0.19\textwidth}
        \centering
        \includegraphics[width=\textwidth]{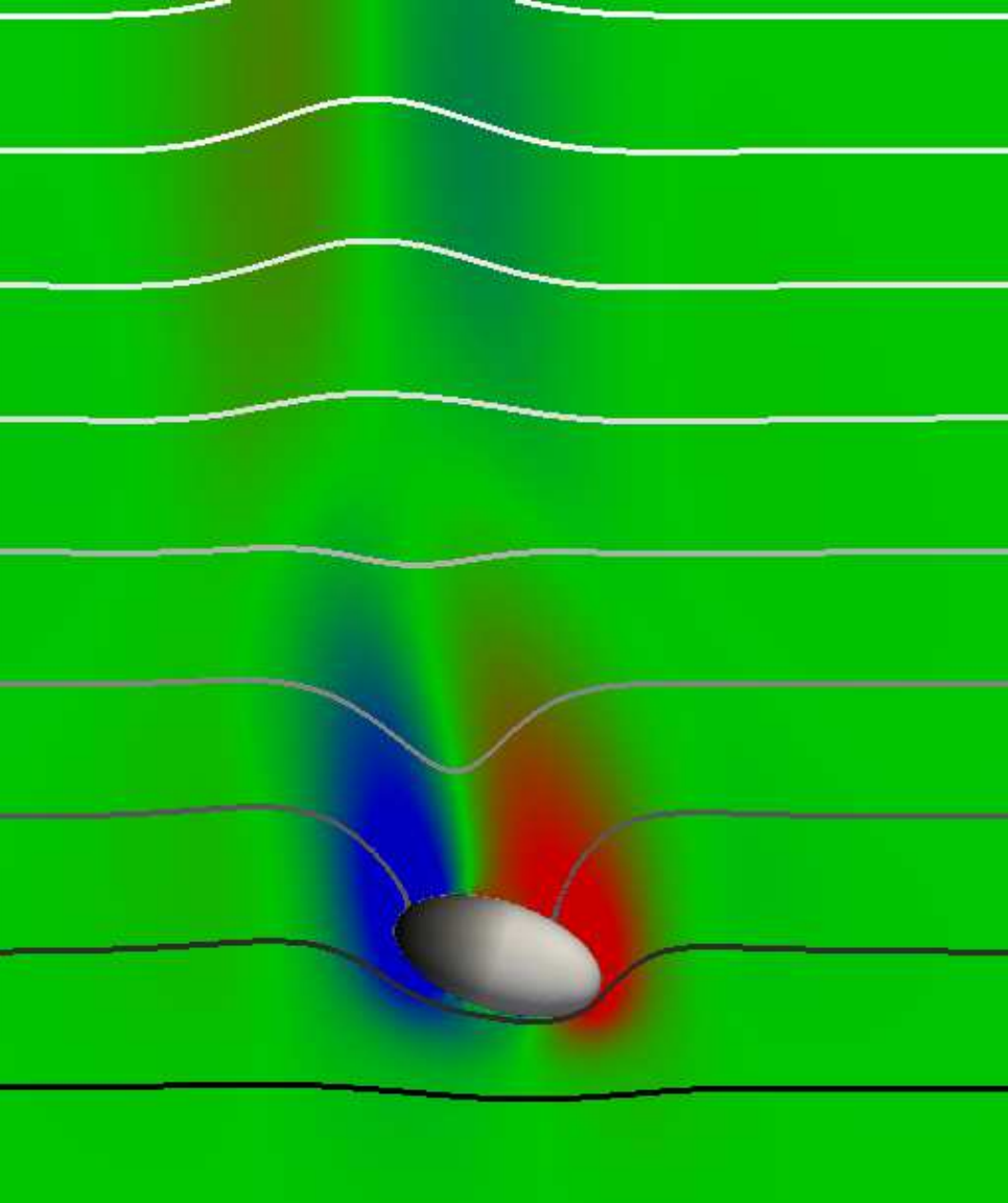}
 
    \end{subfigure} \\
     \begin{subfigure}[t]{0.19\textwidth}
        \centering
        \includegraphics[width=\textwidth]{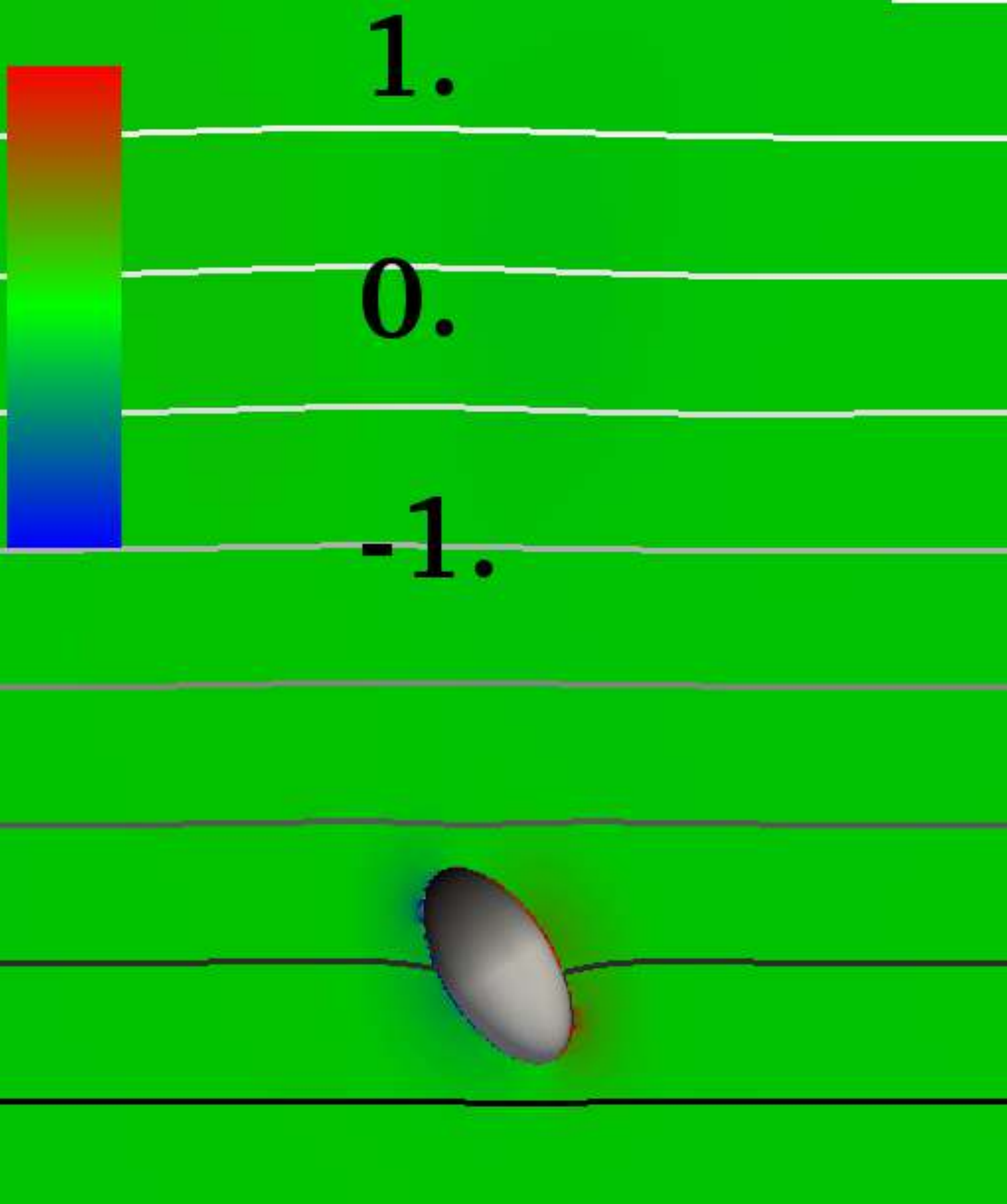}
    
    \end{subfigure}
    \end{tabular}
    \end{tabular}
    \caption{\label{fig:prl_wg}Evolution of the x-component of the dimensionless vorticity generation term due to the mis-alignment of the density gradient vector with the direction of gravity, $\boldsymbol{\nabla}\rho_f \times \hat{\bf{k}}$, in the $x=0$ plane for a prolate spheroid with $\mathcal{AR}=2$, $Re=80$ and $Ri=5$. The solid lines indicate dimensionless isopycnals or equal density lines separated by a value of $0.5$. Darker shade of grey indicates a higher density. The panels are snapshots (row-wise) at specific time intervals with $t/\tau =$ 0, 4.77, 14.32, 23.87, 33.42, 42.97, 52.52, 62.07, 71.62, 81.17, 90.72, 100.27, 109.82, 119.37, and 219.65. The first panel shows the initial configuration and the last shows the settling configuration after the prolate stops.}
\end{figure*}

The stratification has the same effect on the settling velocity of a prolate spheroid as it has on an oblate spheroid. In particular, the stratification causes a continuous deceleration of the settling velocity after the initial transients. In addition, the settling velocity magnitude reduces with the stratification strength for prolate spheroids with same $Re$. This is expected as the prolate  moves from a region with a lighter fluid into a region with a heavier fluid while settling. As a result, it experiences an enhanced drag force. Finally, when it reaches its neutrally buoyant position it stops moving altogether. Similar to the case of an oblate spheroid, the suppression of the fluid flow due to the tendency of the displaced isopycnals to return to their original positions is also one of the reasons for this behavior (discussed in detail in Sec.~\ref{sec:prlres} and fig.~\ref{fig:prl_wg}).

To study the effect of stratification on the particle orientation, we initialize the prolate spheroid in an edge-wise orientation, i.e., $\theta = 0^{\circ}$. In a homogeneous fluid, we find that, the prolate eventually settles down in a broad-side on, i.e., $\theta = 90^{\circ}$ orientation, once it attains its terminal velocity. However, similar to the case of an oblate spheroid, fluid stratification significantly changes the settling orientation as shown in fig.~\ref{fig:80_or} and~\ref{fig:180_or}. 

As the prolate accelerates from rest, it topples from an edge-wise orientation to a broad-side on orientation. However, this orientation is stable only in a homogeneous fluid. In a stratified fluid, once the velocity magnitude falls below a particular threshold (we find that to be 0.15), the prolate spheroid starts to reorient. But unlike an oblate spheroid, it can only reorient partially, i.e., it does not exactly go back to $\theta= 0^{\circ}$. The final settling orientation depends on the stratification strength and the Reynolds number. This becomes clear when examining the final orientations at $Re=80$ and $Re=180$ for increasing stratification strengths in fig.~\ref{fig:80_or} and~\ref{fig:180_or}. At low Re, i.e., $Re=80$, the prolate spheroid reorients almost completely at high stratification ($Ri=10$) such that $\theta \approx 0^{\circ}$ at the final times. However, for a lower stratification strength, i.e., $Ri=5$, the prolate reaches a final orientation of $\theta \approx 30^{\circ}$. At a higher $Re$, i.e., $Re=180$, the final orientation is $\theta \approx 22^{\circ}$ and $\theta \approx 35^{\circ}$ for $Ri=10$ and $Ri=5$, respectively. Thus, the final orientation angle increases if we increase $Re$ at fixed $Ri$, i.e., final orientation progressively leaves the edge-wise orientation ($\theta = 0^{\circ}$).

\begin{figure*}
    \centering
     \begin{subfigure}[t]{0.45\textwidth}
        \centering
        \includegraphics[width=\textwidth]{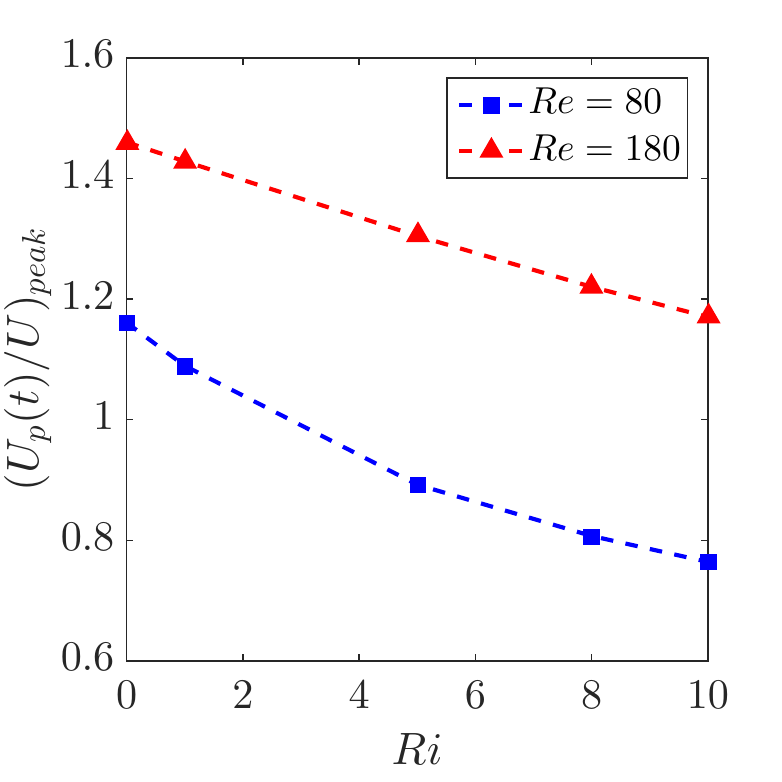}
    	\caption{\label{fig:AR2_peak}}
    
    \end{subfigure}
    ~ 
    \begin{subfigure}[t]{0.45\textwidth}
        \centering
        \includegraphics[width=\textwidth]{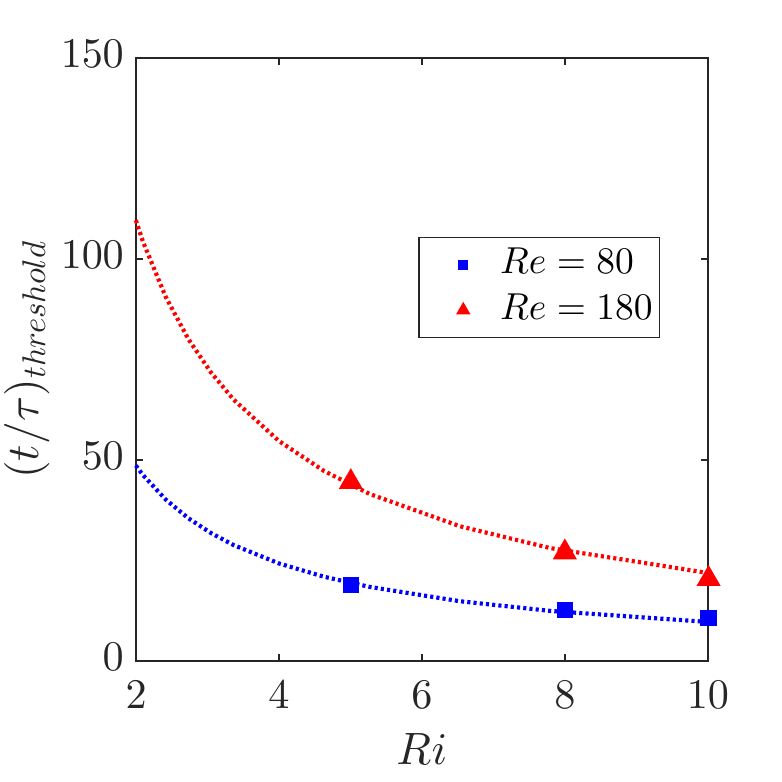}
    	\caption{\label{fig:AR2_th}}
      
    \end{subfigure}
    \caption{\label{fig:AR2}Effect of inertia and stratification strength on a) the peak velocity, $(U_p{t}/U)_{peak}$, of a settling prolate spheroid with $\mathcal{AR}=2$. The peak velocity attained by the particle decreases with increasing stratification and increases with particle inertia, and b) the time ($(t/\tau)_{threshold}$) at which $|U_p(t)/U| < 0.15$. The dashed line in (a) is a guide to the eye. The dotted line in (b) is the $(t/\tau)_{threshold} = A*Ri^{-1}$ fit with A = 97.0 and 218.8 for $Re=80$ and $Re=180$, respectively. The $O(Ri^{-1})$ fit in (b) is consistent with the case of an oblate spheroid in Sec.~\ref{sec:oblmain}.}
\end{figure*}

Next, we quantify the effects of fluid density stratification on the peak velocity of the prolate (see fig.~\ref{fig:AR2_peak}). The results are similar to the case of an oblate spheroid. We observe that the peak velocity decreases monotonically with the fluid stratification strength. The peak velocity increases with increasing $Re$. Also, the relative decrease in the peak velocity for the lowest to highest stratification strength explored reduces with $Re$. For $Re=80$, it decreases by $\approx 33 \%$ while for $Re=180$ it decreases by $\approx 18 \%$. This is due to the increase of the inertial effects as compared to the stratification effects with increasing $Re$ for the same $Ri$. As shown in fig.~\ref{fig:80_vel} and~\ref{fig:180_vel}, increasing the stratification strength or reducing the inertia of the particle results in the earlier onset of the reorientation instability. To conclude, we observe that, the time ($(t/\tau)_{threshold}$) at which the particle velocity falls below the threshold velocity for the onset of reorientation decreases as $O(Ri^{-1})$, see fig.~\ref{fig:AR2_th} where we display the time for the onset of the instability for different particle Reynolds number and stratifications.

\subsubsection{Stratification drag on a prolate spheroid}\label{sec:prl_drag}
\begin{figure*}
    \centering
     \begin{subfigure}[t]{0.45\textwidth}
        \centering
        \includegraphics[width=\textwidth]{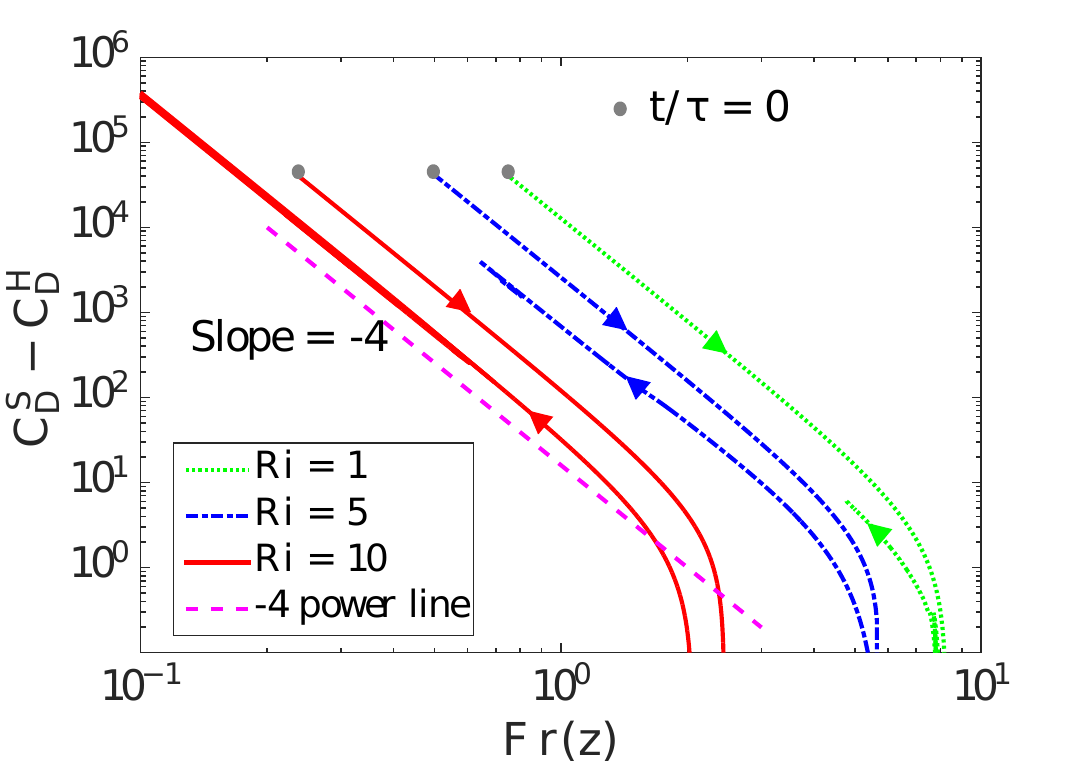}
    	\caption{\label{fig:drag_80}}
       
    \end{subfigure}
    ~ 
    \begin{subfigure}[t]{0.45\textwidth}
        \centering
        \includegraphics[width=\textwidth]{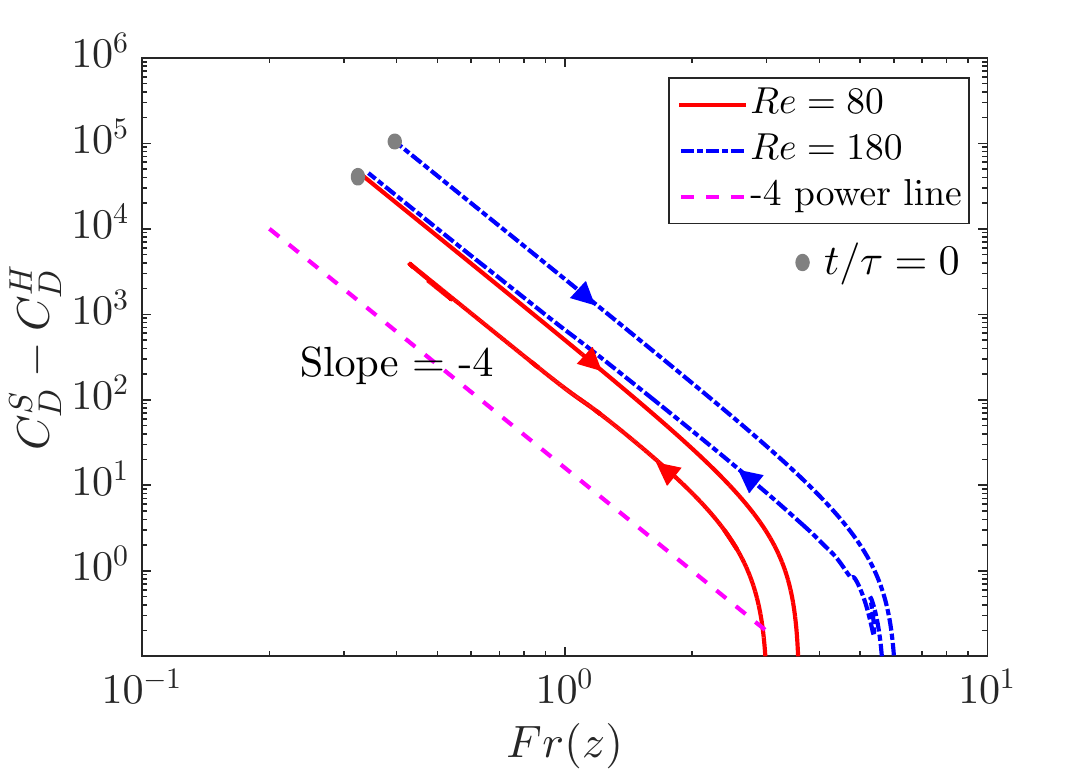}
    	\caption{\label{fig:drag_5}}
      
    \end{subfigure}

    \caption{\label{fig:drag_prl}{Added drag due to stratification, $C_D^S-C_D^H$, for a prolate spheroid with $\mathcal{AR}=2$ as a function of the instantaneous particle Froude number, $Fr(z)$, for a) $Re=80$ and different stratification strengths. b) Added drag at $Ri=3$ for different $Re$. The arrows show the direction of increasing time and the filled dots show the simulation start time. The dashed pink line shows the $-4$ power line to indicate a $Fr(z)^{-4}$ scaling of $C_D^S-S_D^H$.}}
\end{figure*}

{Fig.~\ref{fig:drag_prl} shows the added drag due to stratification, $C_D^S-C_D^H$ on a prolate spheroid sedimenting in a stratified fluid. The drag due to stratification behaves similarly to the case of an oblate spheroid discussed in Sec.~\ref{sec:obl_drag}. The stratification drag on the prolate decreases as it accelerates. $C_D^S-C_D^H$ is minimum when the prolate attains a peak velocity and starts to increase again as the buoyancy/stratification effects take over inertial effects and slow down the prolate. As in the case of an oblate spheroid, $C_D^S-C_D^H$ scales as $\approx O(Fr(z)^{-4})$. }

\subsubsection{Settling trajectory of a prolate spheroid in a stratified fluid}



\begin{figure*}
    \centering
    \begin{tabular}[t]{cc}
     \begin{tabular}{c}
            \begin{subfigure}[t]{0.35\textwidth}
                \centering
                \includegraphics[width=\textwidth]{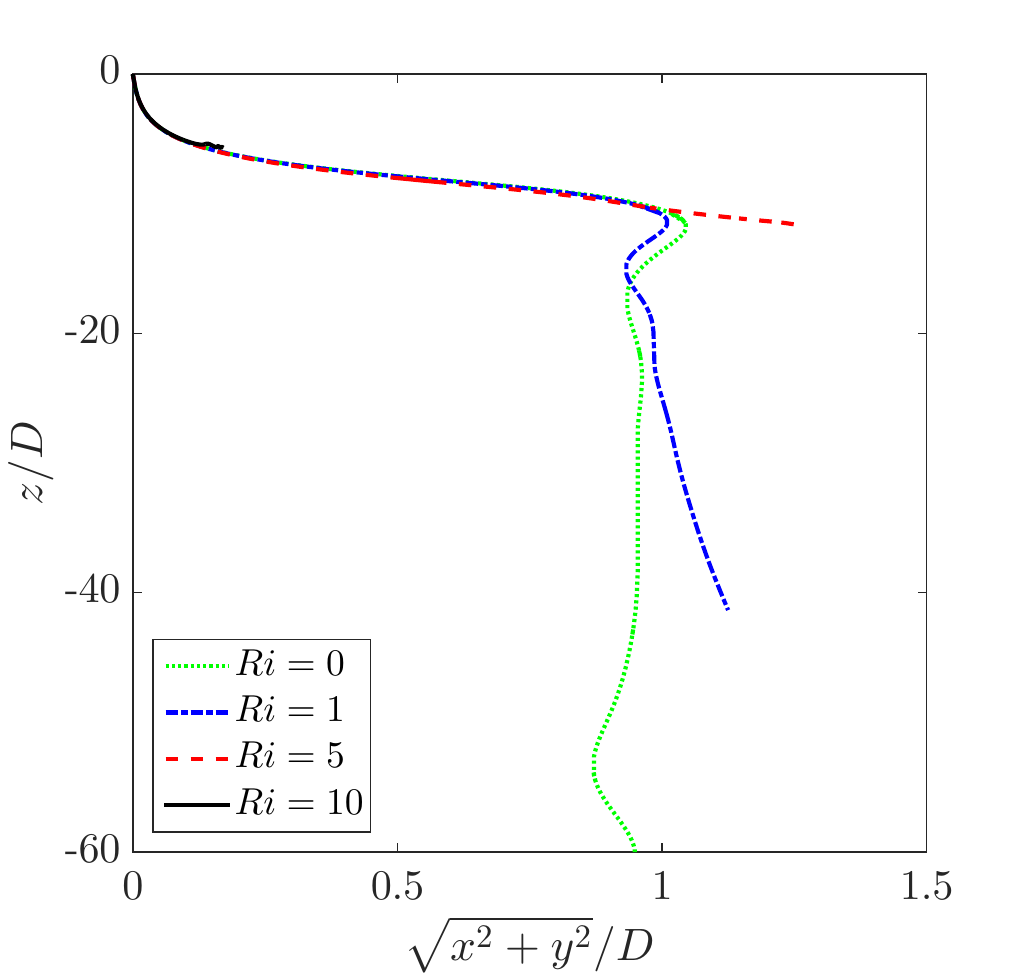}
                \caption{$Re=80$\label{fig:80_traj}}
            \end{subfigure}\\
            \begin{subfigure}[t]{0.35\textwidth}
                \centering
                \includegraphics[width=\textwidth]{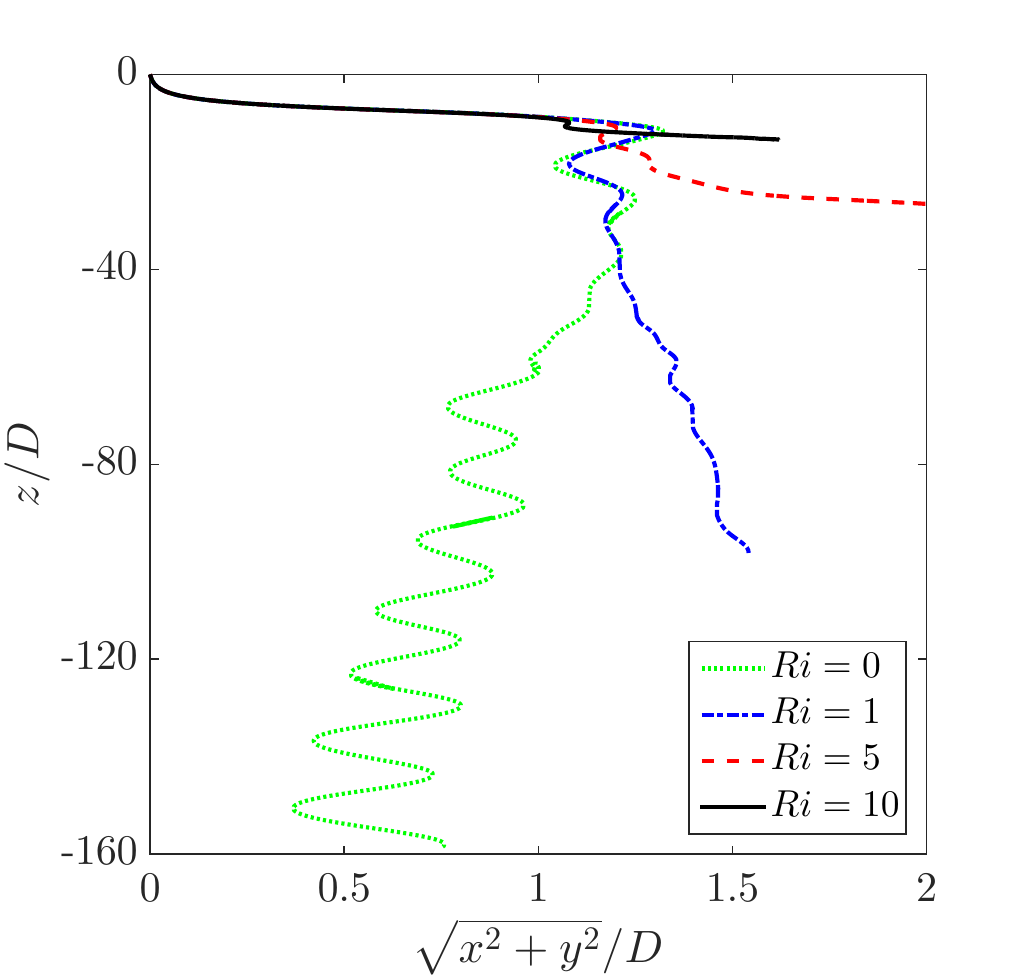}
                \caption{$Re=180$\label{fig:180_traj}}
            \end{subfigure}
        \end{tabular} 
    &
    \begin{subfigure}{0.45\textwidth}
        \centering
        \includegraphics[width=\textwidth]{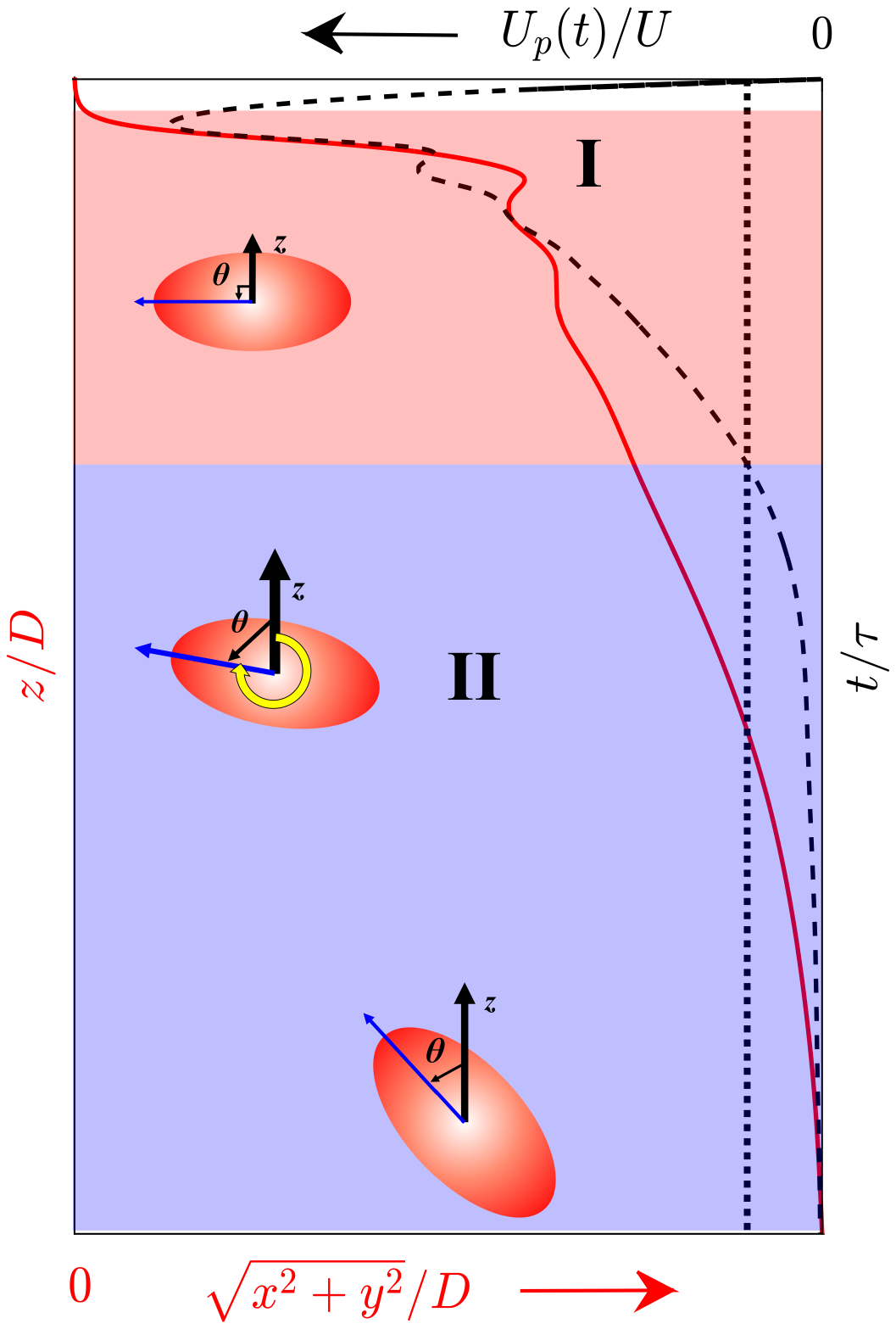}
        \caption{\label{fig:vel_traj_prol_cartoon}}
    \end{subfigure}
\end{tabular}
    \caption{\label{fig:prol_traj}Trajectories of a prolate spheroid with $\mathcal{AR}=2$ in a homogeneous and a stratified fluid for different $Re$ and $Ri$. a) $Re=80$, b) $Re=180$, and c) a schematic summarizing the settling velocity, particle trajectory and the orientation in the two regimes observed in the settling motion. Left vertical axis and bottom horizontal axis indicate the spheroid position (solid line is the settling trajectory). Right vertical axis and top horizontal axis display the particle settling velocity vs time (dashed line is the settling velocity).}
\end{figure*}

Similarly to the case of an oblate spheroid, stratification suppresses the oscillatory trajectories of a prolate spheroid in a homogeneous fluid at high $Re$. The settling trajectories of a prolate spheroid with $\mathcal{AR} = 2$ in a homogeneous and stratified fluid are displayed in fig.~\ref{fig:prol_traj}. In a homogeneous fluid, the particle settles in a straight line at $Re=80$ and in an oscillatory path at $Re=180$. Since stratification results in a reduction of the settling velocity, the prolate spheroid stops at an earlier position as we increase the stratification strength. In addition, the oscillatory path observed for a prolate spheroid with $Re=180$ in a homogeneous fluid disappears in a stratified fluid.

A prolate spheroid goes through two regimes, unlike the three regimes reported above for the settling of an oblate. In the first regime, denoted by $I$, the prolate spheroid oscillate about its broad-side on orientation as it settles. In this regime, the magnitude of the settling velocity is still higher than the threshold below which the spheroid starts to reorient. However, once the settling velocity drops below the threshold for the onset of reorientation, the prolate starts to rotate from broad-side on to edge-wise orientation. This is regime $II$. Unlike an oblate spheroid which rotates quickly in regime $II$ and settles at a final edge-wise orientation in regime $III$, a prolate spheroid reorients slowly in regime $II$. Furthermore, the prolate spheroid does not reorient completely, but attains a final oblique orientation with $\theta$ between $0^{\circ}$ and $35^{\circ}$. The exact value of the final $\theta$ depends on $Re$ and $Ri$ as explained before. The settling path along with the settling velocity are sketched in fig.~\ref{fig:vel_traj_prol_cartoon}.

\subsection{What causes deceleration and reorientation of a prolate spheroid in a stratified fluid?}\label{sec:prlres}

As for the case of an oblate spheroid, we analyse the wake vortices to gain insight into the mechanisms leading to the reorientation of a prolate spheroid. The reasons for the deceleration and the reorientation of a prolate spheroid in a stratified fluid are similar to that of an oblate spheroid as will be discussed in this subsection. For a prolate spheroid settling in a homogeneous fluid, a single vortex attached to the particle is initially observed as in the case of an oblate spheroid. As we increase $Re$, this vortex grows in size. At low $Re$ this vortical structure is still symmetric, however, it becomes helical resulting in an instability for a prolate spheroid with $\mathcal{AR}=3$ for $Re > 70$. As a result, a prolate spheroid with $\mathcal{AR}=3$ rotates about the vertical axis for $Re > 70$ \citep{Ardekani2016}. This is also clear in fig.~\ref{fig:prol_traj} as the prolate spheroid with $Re=80$ settles in a straight line while the prolate spheroid with $Re=180$ has an oscillatory path. As shown in \cite{Ardekani2016} the vortical structures for a prolate spheroid in a homogeneous fluid result in a broad-side on orientation.

\begin{figure*}
    \centering
    \begin{tabular}[t]{ccccc}
    
    \begin{subfigure}[t]{0.19\textwidth}
        \centering
        \includegraphics[width=\textwidth]{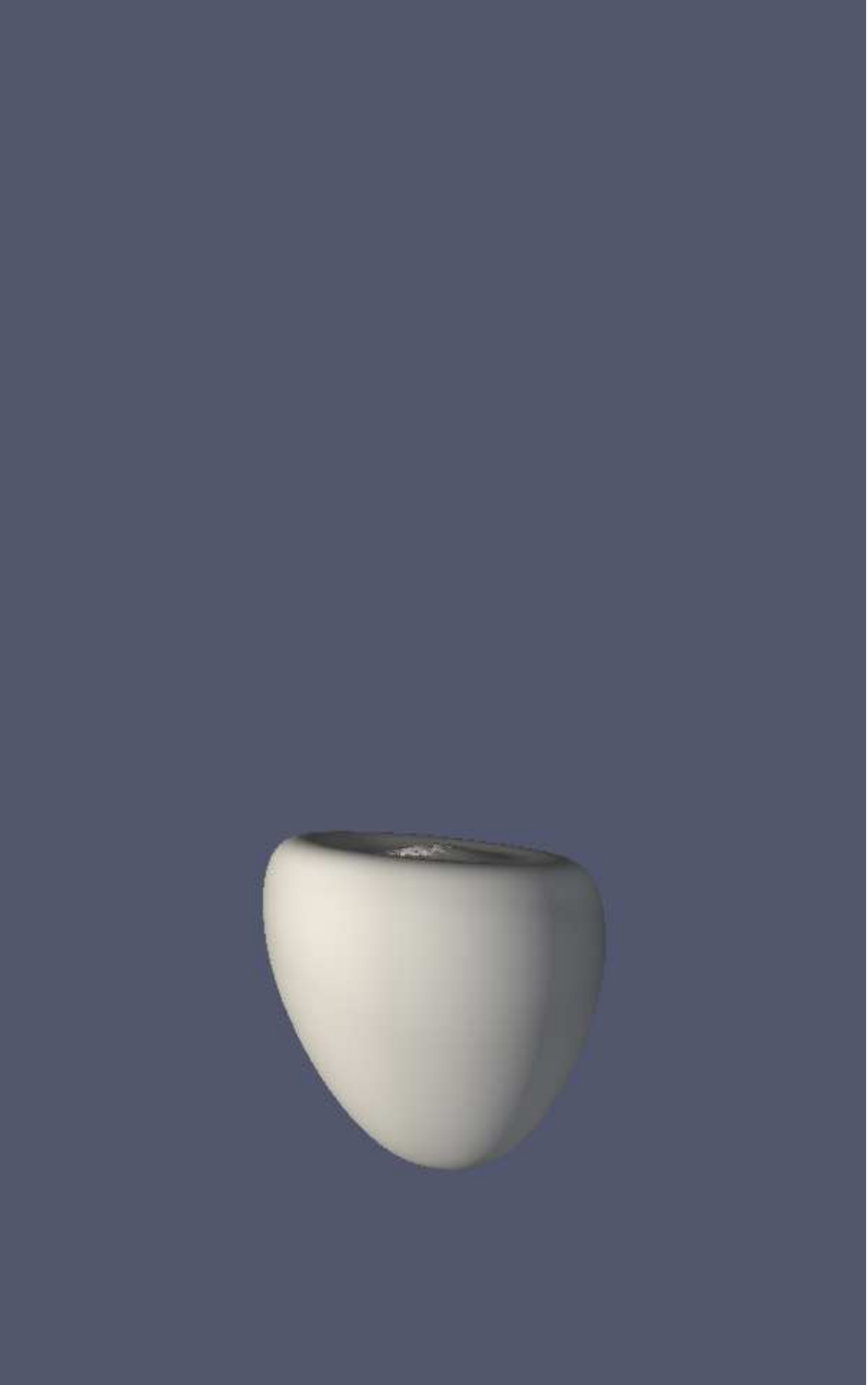}
    	\caption{\label{fig:2}}
 
    \end{subfigure}
   &
    \begin{subfigure}[t]{0.19\textwidth}
        \centering
        \includegraphics[width=\textwidth]{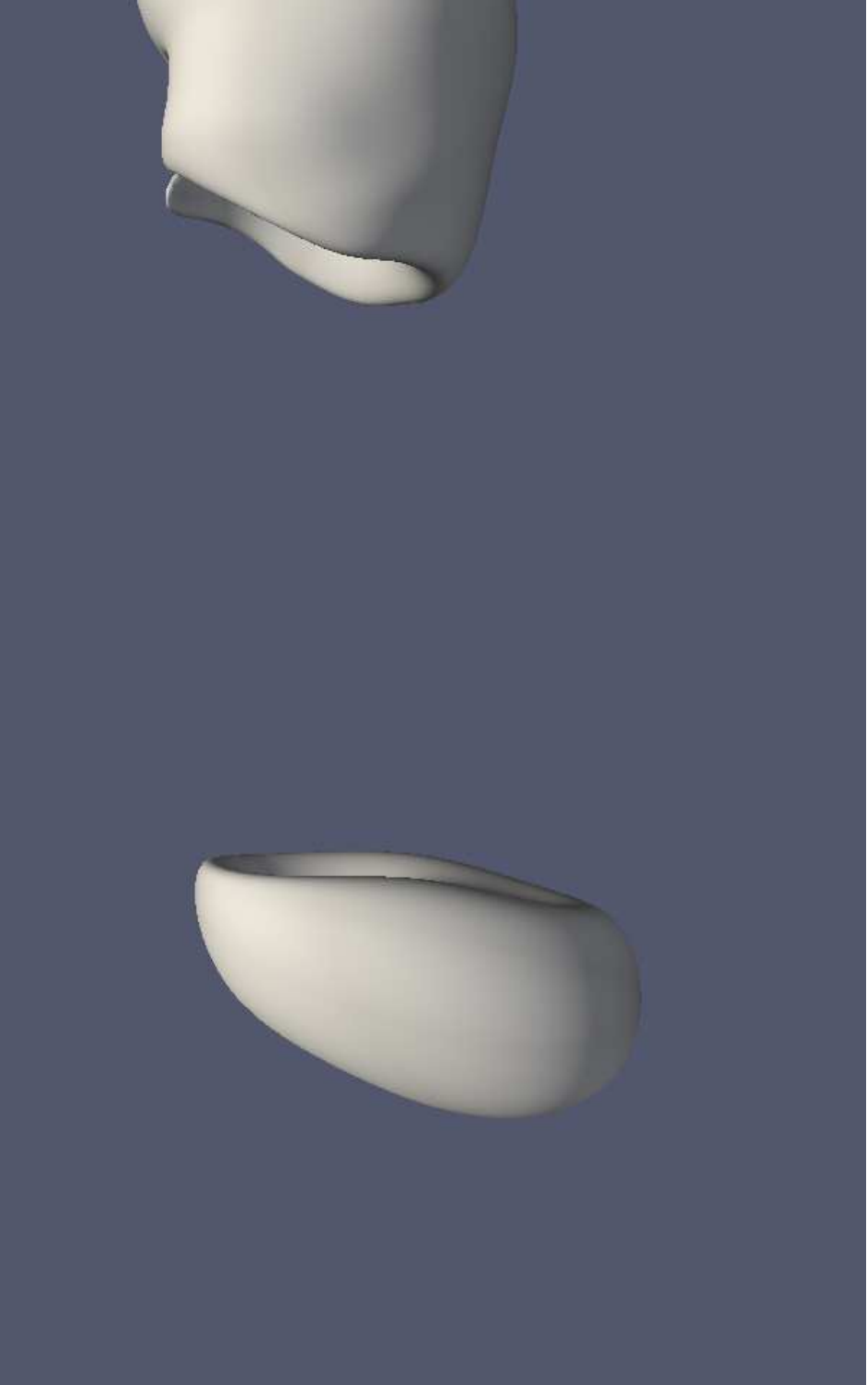}
    	\caption{\label{fig:3}}
       
    \end{subfigure}
    & 
    \begin{subfigure}[t]{0.19\textwidth}
        \centering
        \includegraphics[width=\textwidth]{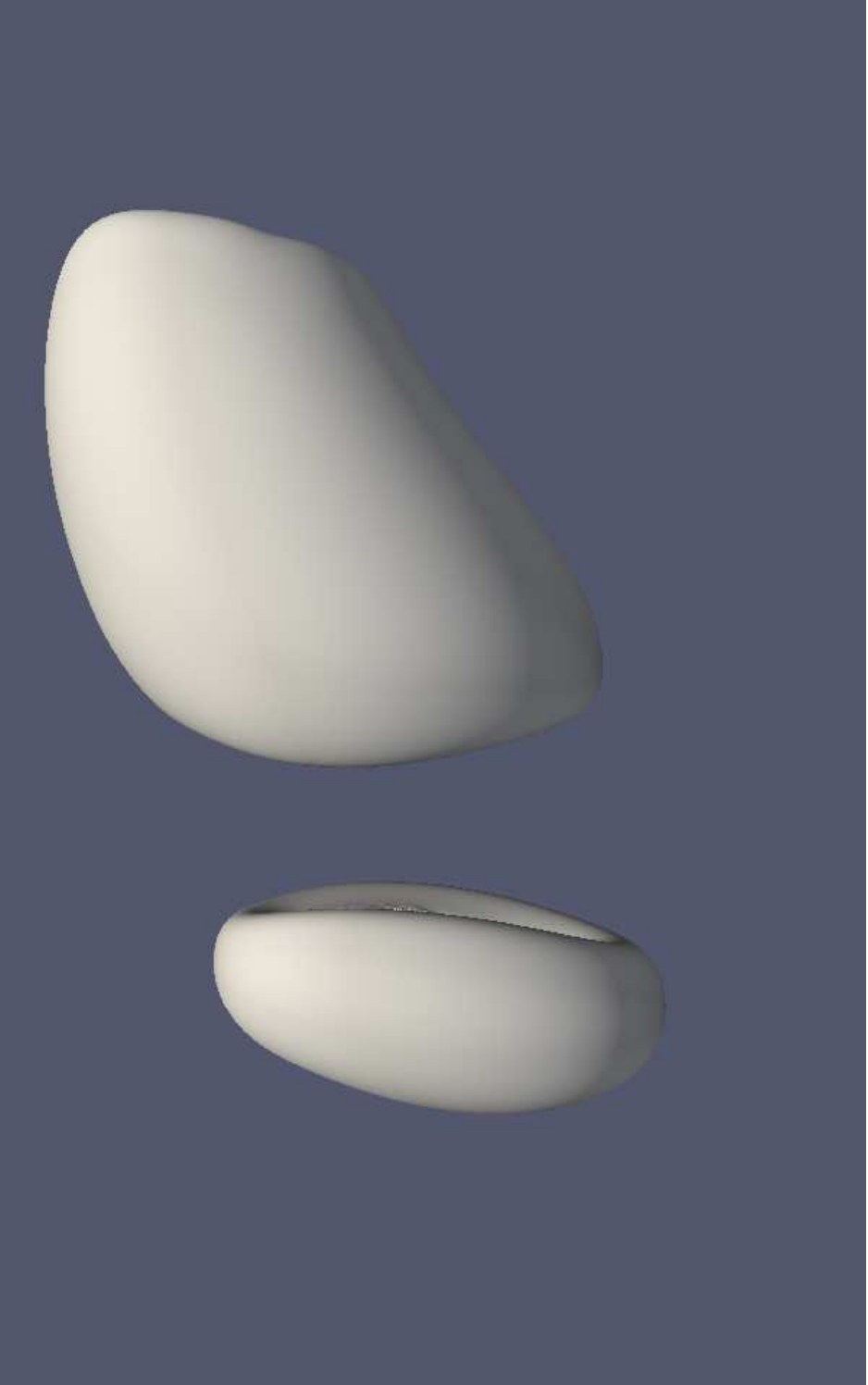}
    	\caption{\label{fig:4}}
      
    \end{subfigure}
    & 
    \begin{subfigure}[t]{0.19\textwidth}
        \centering
        \includegraphics[width=\textwidth]{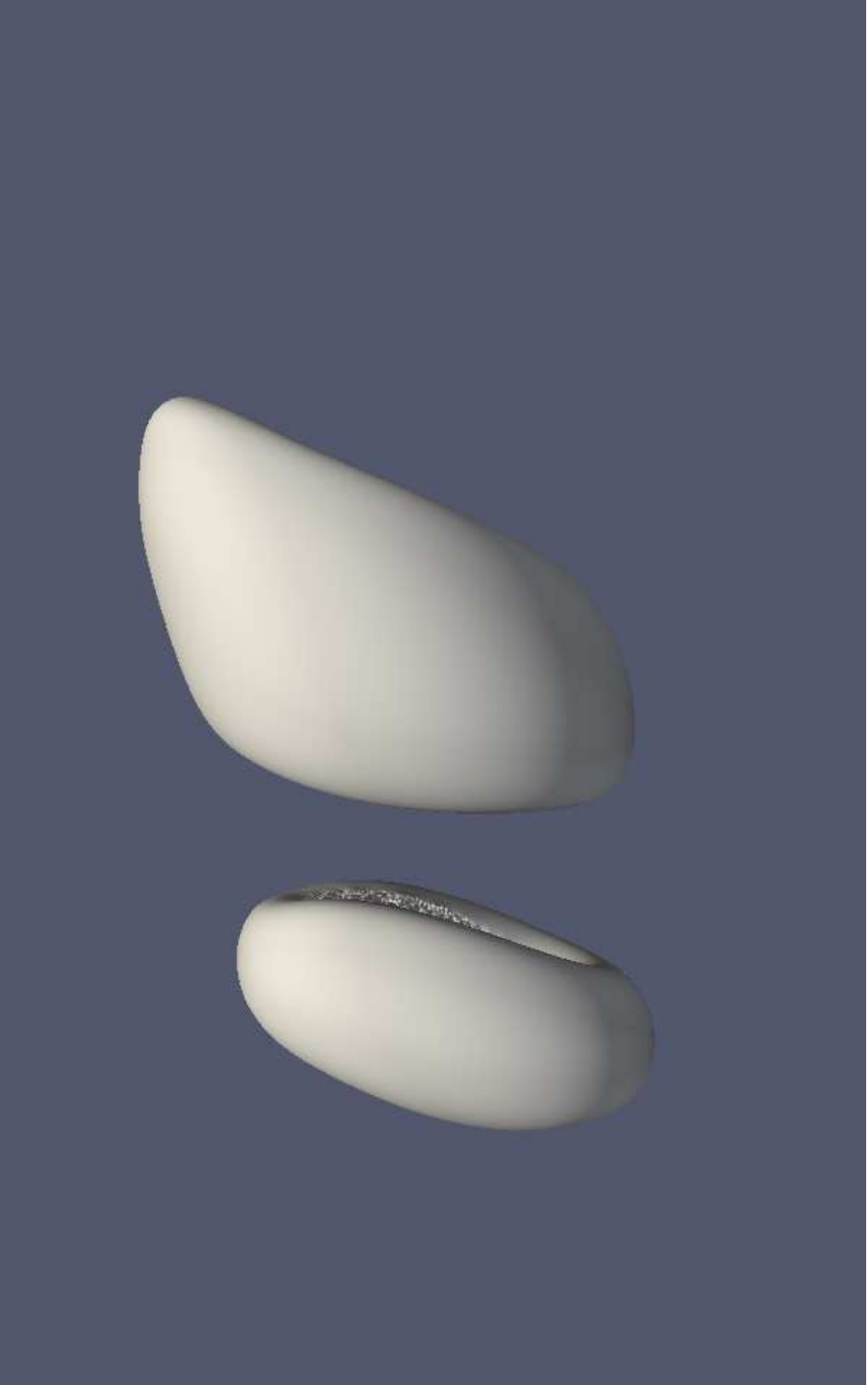}
    	\caption{\label{fig:5}}
  
    \end{subfigure}
    &
    \begin{subfigure}[t]{0.19\textwidth}
        \centering
        \includegraphics[width=\textwidth]{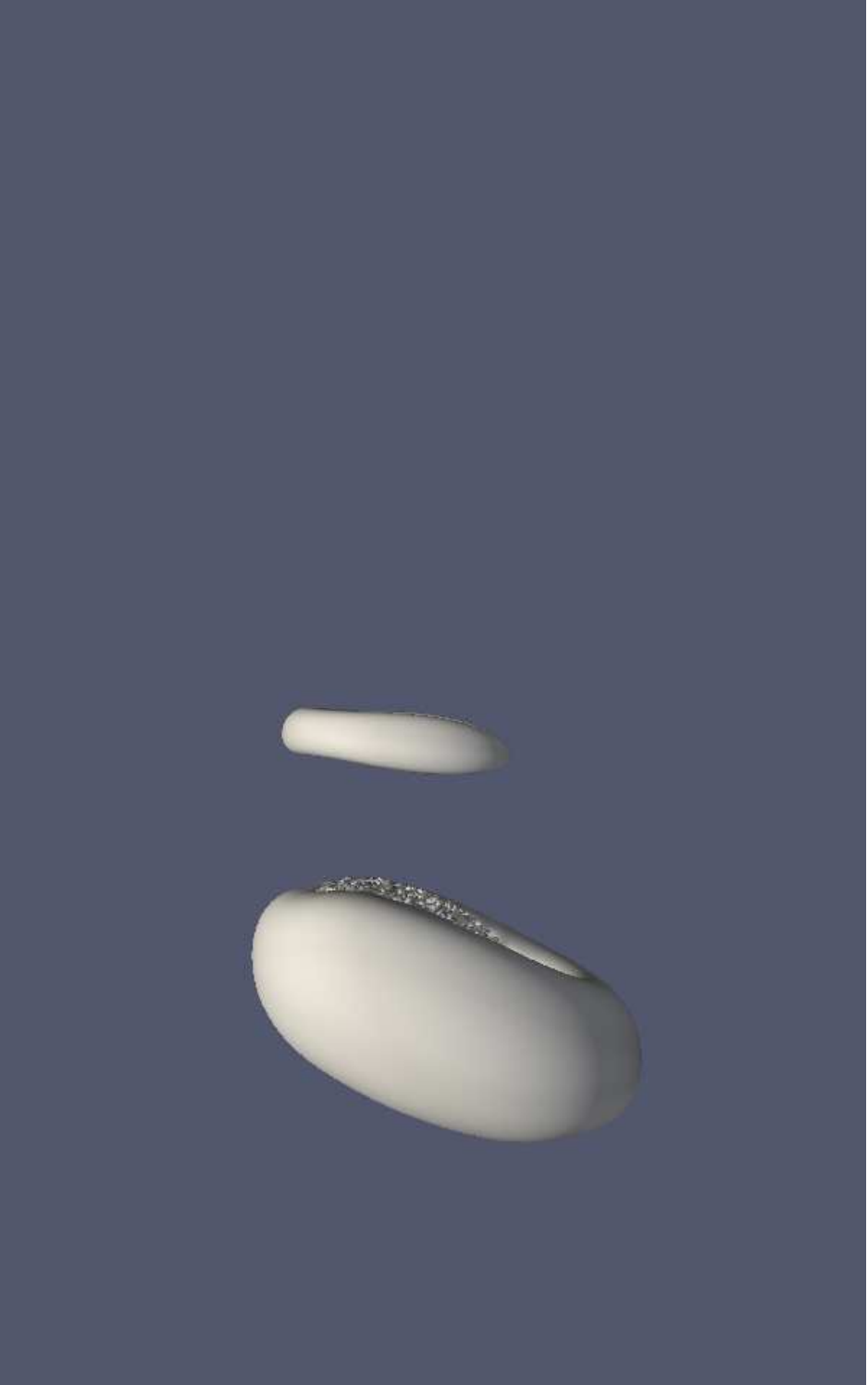}
    	\caption{\label{fig:1}}
       
    \end{subfigure}

    \end{tabular}
    \caption{\label{fig:prl_q}Dimensionless iso-surfaces of Q-criterion equal to $5 \times 10^{-4}$ for a prolate spheroid with $\mathcal{AR}=2$, $Re=80$ and $Ri=5$ at equal time intervals of $t/\tau=28.65$. $t/\tau = 23.87$ for the first panel. The vortical structures identified by the positive Q-criterion are associated with a lower pressure region behind the particle.}
\end{figure*}

The situation is completely different in the case of a prolate spheroid settling in a stratified fluid. In this configuration, the stratification suppresses the vertical motion of the fluid and prevents the particle from attaining any steady state speed. In a stratified fluid, initially there is one vortex attached to the particle as shown in fig.~\ref{fig:prl_q}. As time passes, a part of this vortex detaches and remains predominantly on one side of the prolate spheroid as also shown in fig.~\ref{fig:prl_q}. As a result of this, there is a significant asymmetric low pressure region behind the prolate spheroid. This results in a torque which reorients the particle with its major axis aligned with the density gradient until it settles at its neutrally buoyant position. As discussed above, a prolate settles at an angle between $0^{\circ}$ and $90^{\circ}$ depending on the Reynolds number.

\begin{figure*}
    \centering
     \begin{subfigure}[t]{0.45\textwidth}
        \centering
        \includegraphics[width=\textwidth]{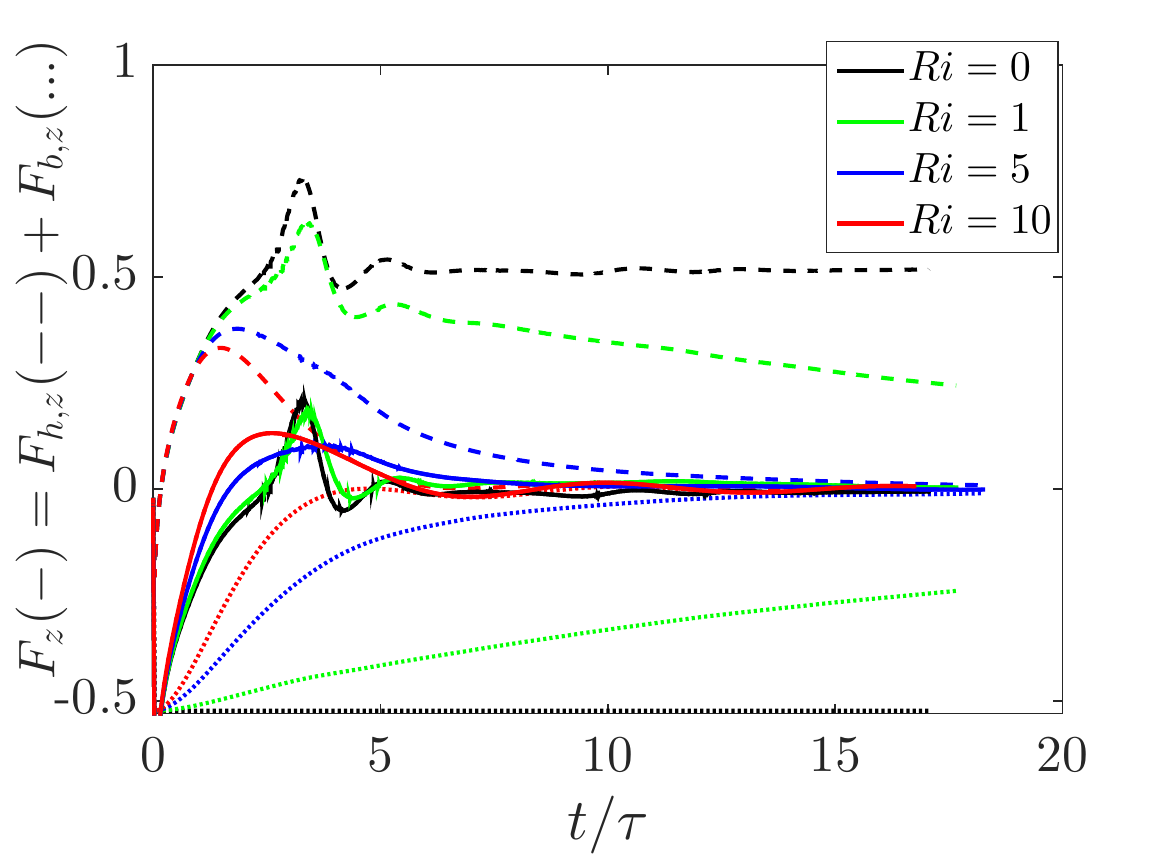}
    	\caption{\label{fig:forcep}}
    
    \end{subfigure}
    ~ 
    \begin{subfigure}[t]{0.45\textwidth}
        \centering
        \includegraphics[width=\textwidth]{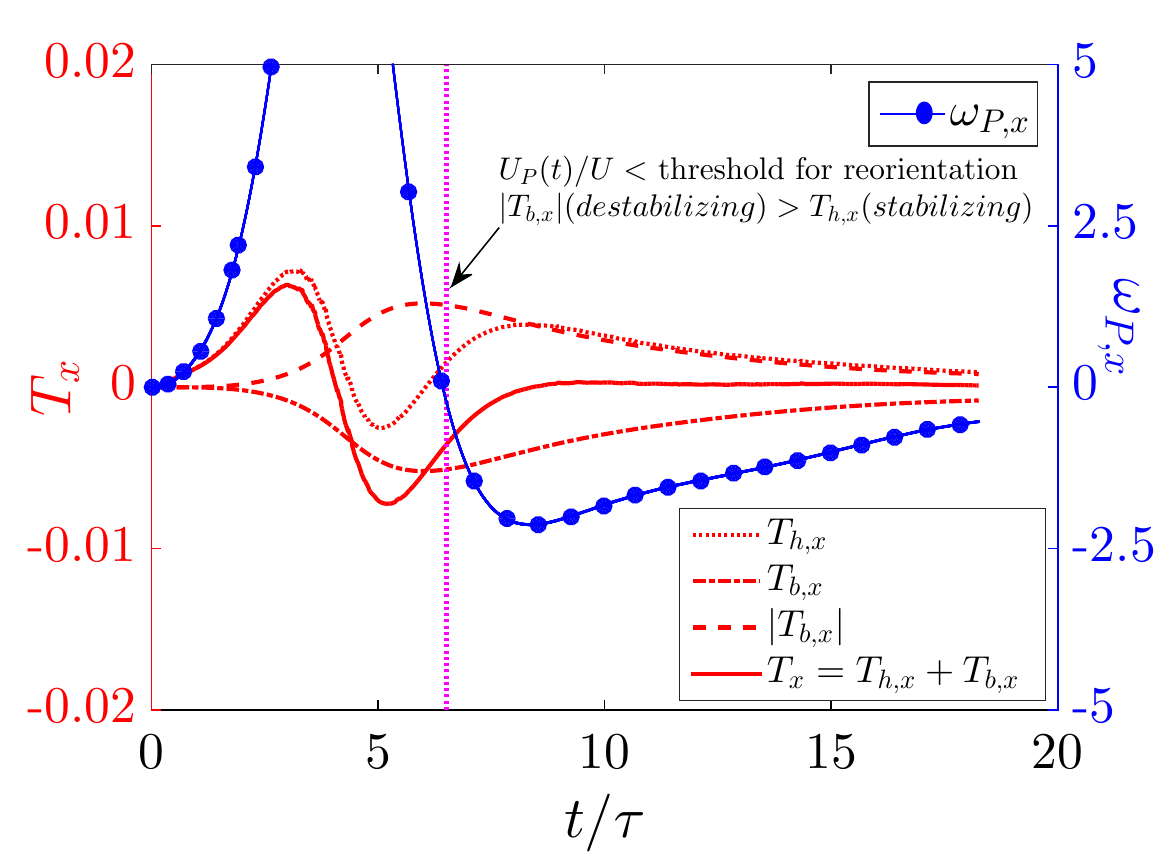}
    	\caption{\label{fig:torqp}}
      
    \end{subfigure}
    \caption{\label{fig:frcp}{ a) Forces acting on the prolate spheroid with $Re=80$ as it settles in a stratified fluid for different values of $Ri$ shown with different colors. The total force (solid line) can be split into two components, the hydrodynamic component (dashed line) and the buoyancy component (dotted line). b) x-component of the torque acting on a prolate spheroid with $Re=80$ as it sediments in a stratified fluid with $Ri=5$ along with the x-component of the angular velocity. The net torque (solid line) is split into two components, the hydrodynamic torque (dotted line) which tries to orient the prolate in a broadside on orientation (hence stabilizing) and the buoyancy component (dashed-dotted line) which is destabilizing and tries to reorient the prolate edgewise. The reorientation starts once the magnitude of the hydrodynamic torque falls below the buoyancy torque which happens when the prolate velocity falls below the threshold for reorientation discussed in section.~\ref{sec:prolmain}.}}
\end{figure*}

{We present for forces and torques acting on the prolate spheroid in fig.~\ref{fig:frcp}. The net force and the force components, $F_{h,z}$ and $F_{b,z}$ behave similarly to the case of an oblate spheroid discussed in Sec.~\ref{sec:oblres}. High magnitude of the buoyancy force compared to the hydrodynamic drag explains the initial acceleration of the prolate. However, the buoyancy force decreases as the prolate sediments in a region with higher fluid density causing it to slow down. The gradual increase in the magnitude of the destabilizing buoyancy torque compared to the stabilizing hydrodynamic torque as the prolate velocity decreases explains the onset of reorientation to the edgewise orientation below a threshold velocity as shown in fig.~\ref{fig:torqp}. This is similar to an oblate spheroid as shown in fig.~\ref{fig:torq}.}

To understand the reorientation mechanism, we again examine the x-component of the vorticity generation ($\boldsymbol{\omega}_g$) due to the deformation of the isopycnals. The dynamics are similar to what happens in the case of an oblate spheroid settling in a stratified fluid as discussed in Sec.~\ref{sec:oblres}. As the prolate spheroid accelerates initially in edge-wise orientation, the symmetric $\boldsymbol{\omega}_g$ region expands with the origin of the plume (as defined in Sec.~\ref{sec:oblres}) at the tip of the prolate spheroid. However, the prolate spheroid topples due to inertial effects which displaces the origin of the plume and makes $\boldsymbol{\omega}_g$ asymmetric. 

Once the prolate spheroid reaches the peak velocity, it starts to decelerate resulting in a decrease of the inertial effects. As a result, the region of non-zero $\boldsymbol{\omega}_g$ shrinks. In addition, the origin of the plume remains on one side of the broader face of the prolate. As the velocity falls below the threshold for reorientation, the buoyancy effects take over inertial effects and the origin of the plume starts to shift back to the closer tip of the prolate. This results in a net torque on the prolate spheroid eventually reorienting it in the edge-wise orientation. When the prolate spheroid reaches its neutrally buoyant location it stops moving. If this happens before the complete reorientation, it will remain partially reoriented as seen in fig.~\ref{fig:80_or} and~\ref{fig:180_or}.

\begin{figure*}
    \centering
     \begin{subfigure}[t]{0.45\textwidth}
        \centering
        \includegraphics[width=\textwidth]{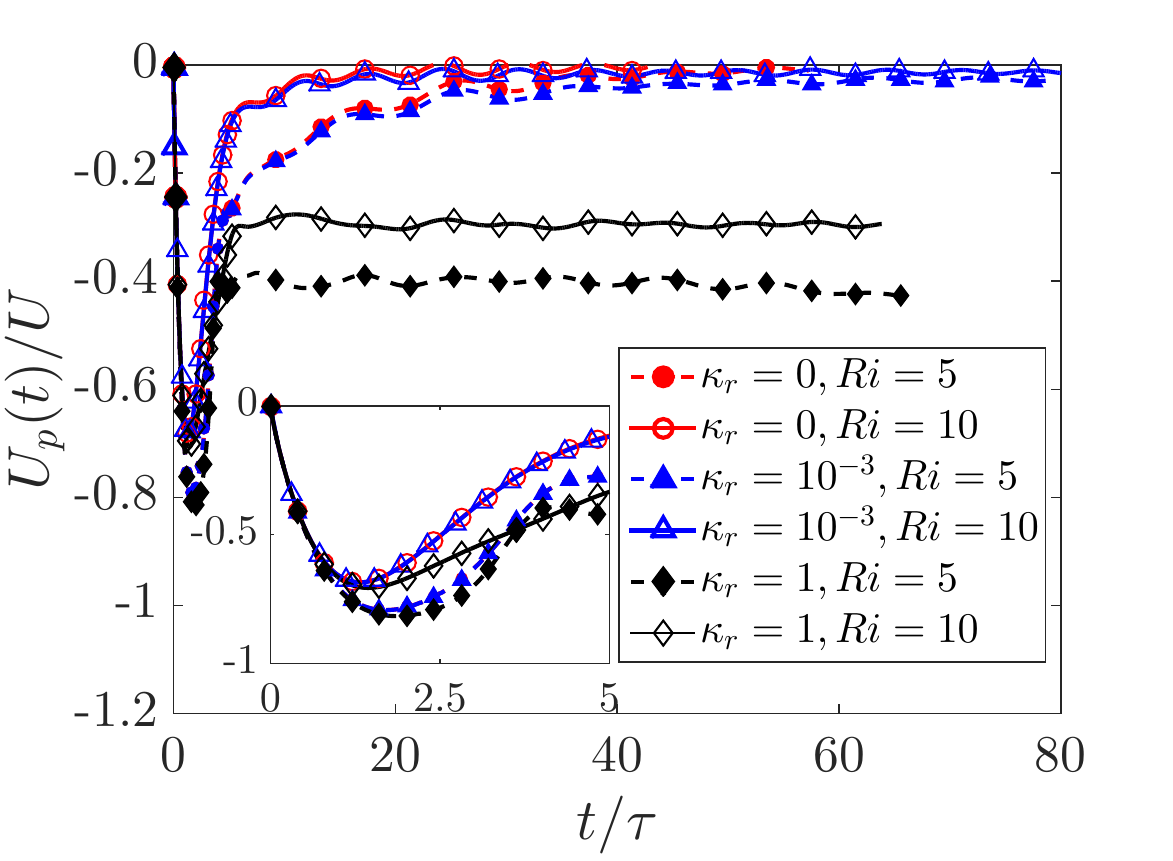}
    	\caption{\label{fig:vel_k}}
    
    \end{subfigure}
    ~ 
    \begin{subfigure}[t]{0.45\textwidth}
        \centering
        \includegraphics[width=\textwidth]{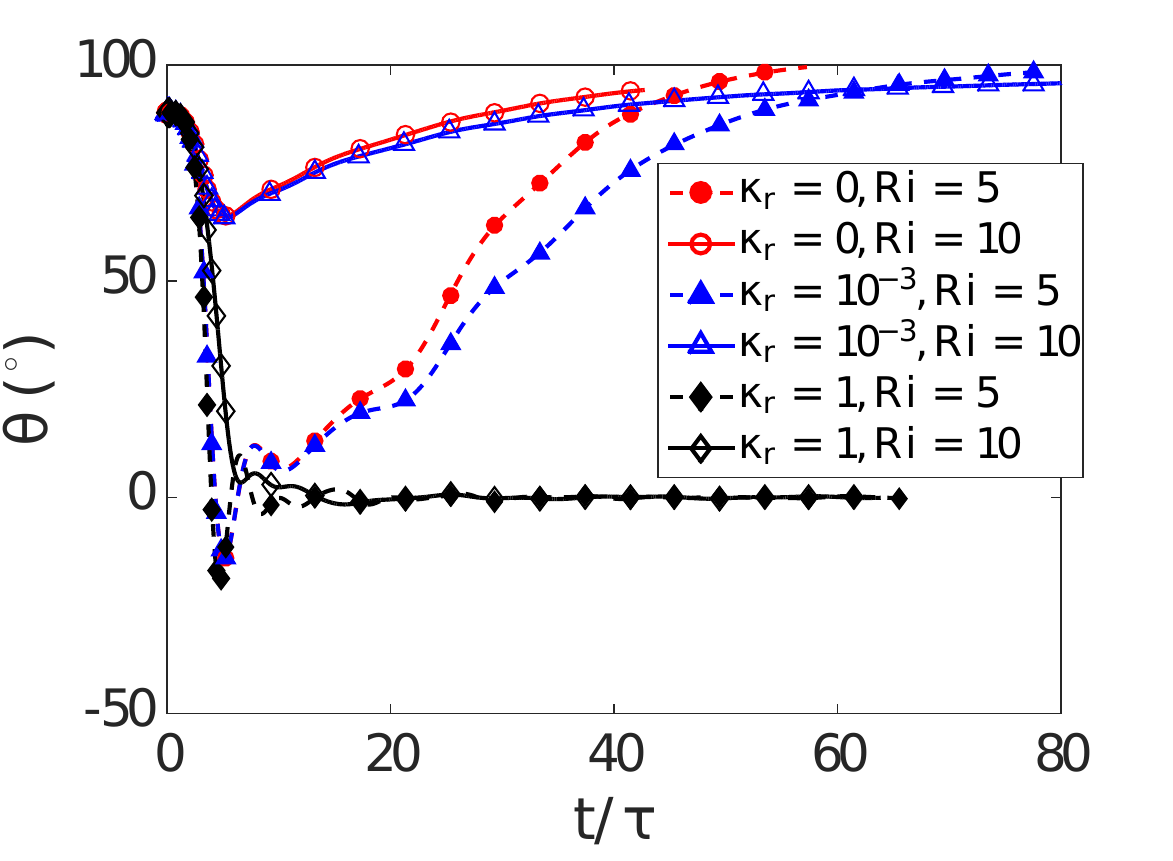}
    	\caption{\label{fig:or_k}}
      
    \end{subfigure}
    \caption{\label{fig:keff}{Effect of permeability of the particle of the stratifying agent on a) the settling velocity, $U_p(t)/U$, of a settling oblate spheroid with $\mathcal{AR}=1/3$, and b) the orientation, $\theta$, for $Ri = 5$ \& $10$. $k = 0$ inside the particle means the stratifying agent cannot diffuse into/ out of the spheroid. {A non-zero value for $k$ inside the particle results in increasing the temperature and decreasing the density of the boundary layer}. For a very small $k_r= 0.001$, the spheroid settling dynamics is similar to $k_r=0$ case. However, for a high $k_r=1$, the spheroid has a completely different settling dynamics. If the stratifying agent can diffuse inside the spheroid, then, the spheroid attains a terminal velocity and does not reorient. These results show that spheroids will reorient only in the case of salt stratified fluid or an adiabatic particle and not in a temperature stratified fluid with {conductive particles.}}}
\end{figure*}

{\section{The effect of heat conductivity ratio on the settling spheroid}}\label{sec:k}
{For this study, we have chosen a no flux boundary condition on the particle surface, i.e., the stratifying agent cannot diffuse inside the particle (adiabatic/impermeable or no flux) \citep{doostmohammadi2015suspension} and, as a consequence, pycnoclines must be normal to the particle surface. This is the case if the stratifying agent is salt or the particle is adiabatic. In this section, we investigate the settling dynamics when the fluid and particle temperature influence each other by changing the heat conductivity ratio $k_r$. Fig.~\ref{fig:keff} shows the settling dynamics of particle having a non-zero $k_r$. For a small $k_r=0.001$, the settling dynamics of an oblate spheroid is similar to the case $k_r=0$. The velocity is slightly higher for $k_r=0.001$. The particle accelerates initially, attaining a peak velocity after which it decelerates and stops when it reaches its neutrally buoyant position. Also, as its velocity falls below a threshold, it reorients to an edge-wise orientation. Since the flux of the stratifying agent into/out of the particle is much slower than the settling dynamics of such small value of $k_r$, {the surrounding fluid is not subjected to any significant heat exchange-induced density change.}
As for the cases studied above, the particle settles in a fluid region with increasing density, its velocity decreases as the net buoyancy force acting on it decreases and the isopycnals resist their deformation. No-flux boundary condition is typical for objects settling in a temperature or a salt stratified fluid, e.g., plastics, metals, organisms, etc. \citep{hanazaki2009schmidt, mehaddi2018inertial, Doostmohammadi2014b, mercier2020settling}. }

\begin{figure*}
    \centering
     \begin{subfigure}[t]{0.45\textwidth}
        \centering
        \includegraphics[width=\textwidth]{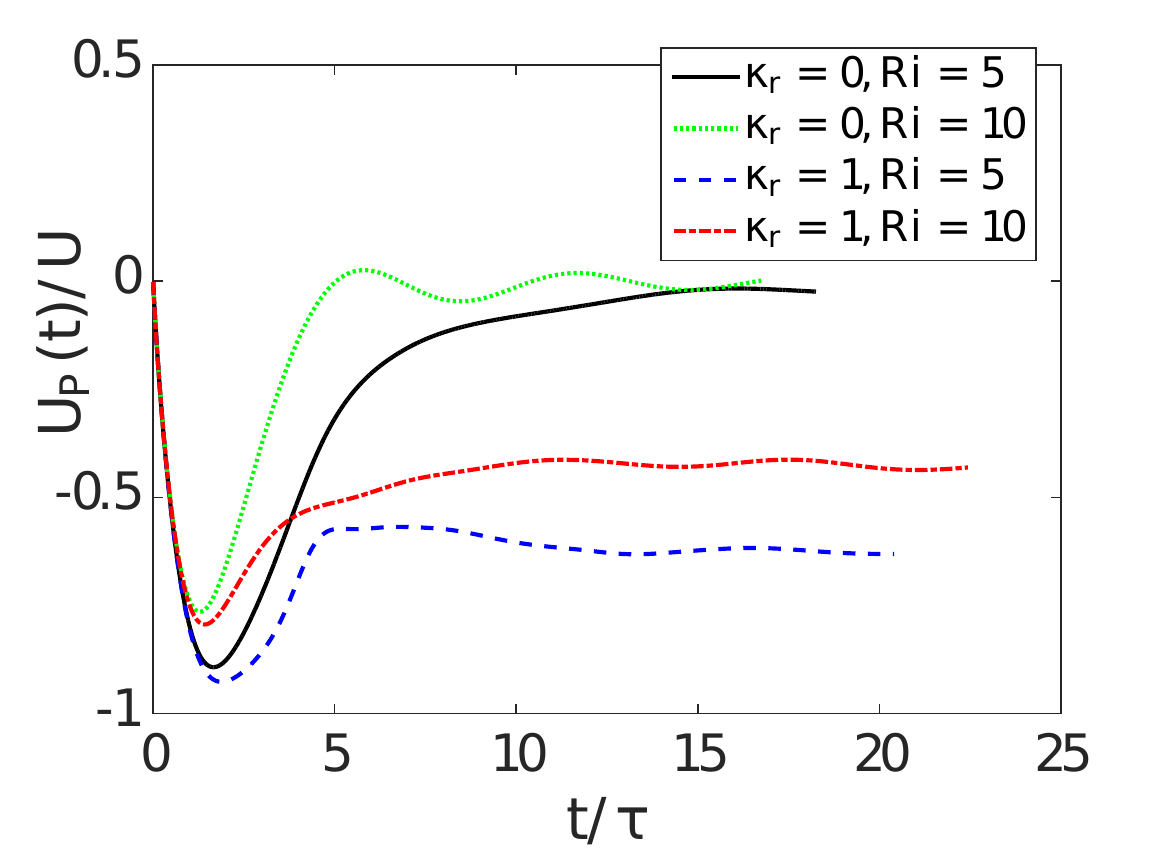}
    	\caption{\label{fig:vel_kp}}
    
    \end{subfigure}
    ~ 
    \begin{subfigure}[t]{0.45\textwidth}
        \centering
        \includegraphics[width=\textwidth]{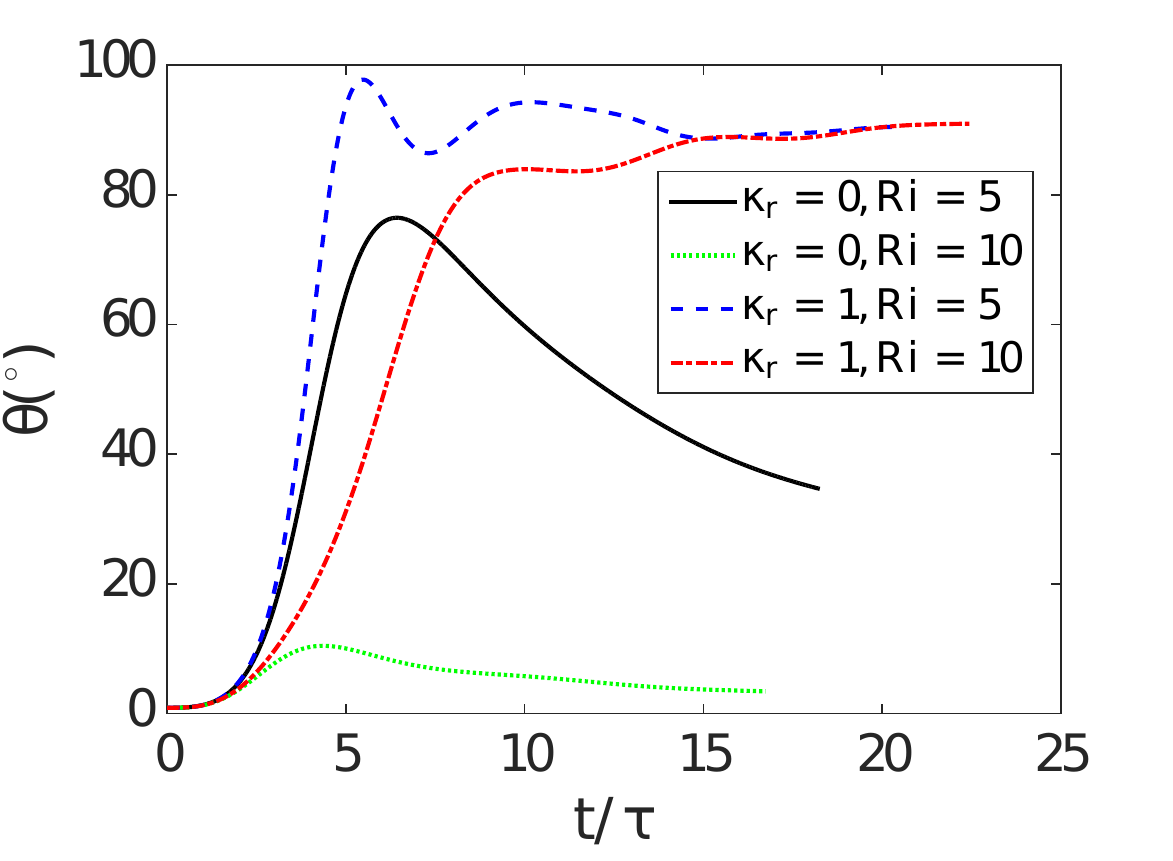}
    	\caption{\label{fig:or_kp}}
      
    \end{subfigure}
    \caption{\label{fig:prol_k}{Effect of permeability of the particle to the stratifying agent on a) the settling velocity, $U_p(t)/U$, of a settling prolate spheroid with $\mathcal{AR}=2$, and b) the orientation, $\theta$, for $Ri = 5$ \& $10$. $k = 0$ inside the particle means the stratifying agent cannot diffuse into the spheroid which results in no change in the density of the {surrounding boundary layer}. This is true in the case when the stratifying agent is salt. A non-zero value for $k$ inside the particle results {in diffusing heat to the surrounding fluid and thus decreasing the density of the boundary layer}. For a high $k_r=1$, the spheroid has a completely different settling dynamics, {with the spheroid attaining a terminal velocity and not reorienting}. These results show that spheroids will reorient only in the case of salt stratified fluid and not in a temperature stratified fluid {with conductive particles.}}}
\end{figure*}

{The settling dynamics changes for a high $k_r$ value. {A high $k_r$ value implies significant heat exchanges between the two phases and results in a warmer fluid close to the particle surface. The warmer boundary layer with decreased density accelerates upwards and thus creates a downforce that prevents particles from deceleration.} 
This scenario might occur in chemical processes, e.g. liquid fluidized beds, and marine snow settling in a temperature stratified water. For $k_r=1$, the settling dynamics during the initial time for an oblate spheroid is similar to $k_r=0$ case, however the particle does not keep decelerating as time passes in contrast to the case with $k_r=0$. For a high $k_r$, the particle attains a terminal velocity much like in the case of an oblate spheroid settling in a homogeneous fluid. The terminal velocity, however, decreases as we increase the stratification strength as shown in fig. \ref{fig:vel_k}. Furthermore, the oblate spheroid does not reorient to an edge-wise orientation as its velocity does not fall below the threshold for the onset of reorientation instability, but settles in a broad-side on orientation as shown in fig.~\ref{fig:or_k}. 
The same holds for a prolate spheroid as shown in fig.~\ref{fig:prol_k}. As discussed in Sec~\ref{sec:oblres} and~\ref{sec:prlres}, $k_r=0$ implies that the isopycnals are orthogonal to the particle surface, which creates a net torque on the spheroid. For a higher $k_r$ value, the pycnoclines {are not orthogonal to the particle surface} and hence do not result in a significant destabilizing buoyancy torque, $T_b$, on the spheroid. Thus, we conclude that no flux boundary condition is essential to observe the reorientation of spheroids settling in a stratified fluid. }

\section{Conclusions}
We investigated the settling dynamics of anisotropic shaped particles in a density stratified fluid using direct numerical simulations. The shapes considered are an oblate spheroid with $\mathcal{AR}=1/3$ and a prolate spheroid with $\mathcal{AR}=2$. We vary the Reynolds number $Re$ from $80-250$ and the Richardson number $Ri$ from $0-10$ while keeping the density ratio $\rho_r$ and Prandtl number $Pr$ constant. The results show that the settling dynamics of spheroids is significantly different in a stratified fluid than in a homogeneous fluid. 

Initially, the spheroids accelerate from rest and reach a maximum velocity. The peak velocity attained by the particles increases with their $Re$ while decreases monotonically when increasing the stratification. After the settling velocity attain its peak value, stratification dominates over inertia, because the inertial effects are not enough to sustain the deformation of the isopycnals once the particle reaches its peak velocity. Hence, due to the tendency of the isopycnals to return to their original positions, the fluid experiences a resistance to its motion. This results in an increased drag and hence a deceleration of the particle until it stops at its neutrally buoyant position. This evolution of the settling velocity is similar to that of a spherical particle settling in a stratified fluid. 

The fluid stratification alters the orientation of the spheroids compared to their orientations in a homogeneous fluid. The fluid stratification leads to reorientation instability as the particle settling velocity falls below a threshold. For the parameters considered here, the onset of the reorientation instability occurs at $|U_p(t)/U| < 0.15$. Interestingly, the dimensionless threshold velocity for the onset of reorientation instability is found to be the same for the oblate and prolate spheroids. This value might be different for different values of the density ratio and the Prandtl number. As a result of this instability, an oblate spheroid settles with its broader side aligned with the direction of the stratification. On the other hand, a prolate spheroid reorients partially or fully depending on its $Re$ and settles such that its longer edge is at an angle greater than $0^{\circ}$ and lower than $45^{\circ}$ with the horizontal direction. This is completely opposite to what happens in a homogeneous fluid as both an oblate and a prolate spheroid settle in a broad-side on orientation. Stratification also eliminates the oscillatory path instability observed for spheroids in a homogeneous fluid. This is due to the decreasing magnitude of the inertial effects as the particle decelerates while reaching regions of higher fluid density. 

The asymmetry in the low pressure region behind the spheroids due to an asymmetric wake results in the onset of the reorientation instability. This asymmetry results from the asymmetric distribution of the vorticity generation term due to the mis-alignment of the density gradient vector with the vertical direction (baroclinic vorticity generation). As a result, the destabilizing buoyancy torque becomes dominant over the stabilizing hydrodynamic torque as the spheroid velocity falls below a threshold value causing the onset of reorientation instability. We also report that the spheroids will only reorient in the case when they are impermeable to the stratifying agent ($\kappa_r<<1$) which is true in the case of a salt stratification or an adiabatic particle. If the stratifying agent can diffuse ($\kappa_r>>0$) inside the particle, then the spheroid won't reorient and the settling dynamics is similar to that in a homogeneous fluid with stratification causing a reduction in the terminal velocity. The results presented in this paper are a first contribution to the field of settling particles in a fluid, in particular for anisotropic particles and stratified fluids. As extensions of this work, it would be interesting to investigate the behavior of particle suspensions, the effect of the aspect ratios, the value of $Pr$ as well as the particle shape on the settling dynamics of anisotropic shaped particles in a stratified fluid. 

\section*{Acknowledgments}
A.M.A. would like to acknowledge financial support from National Science Foundation via grants CBET-1604423, and CBET-1705371. This work used the Extreme Science and Engineering Discovery Environment (XSEDE) \citep{towns2014xsede}, which is supported by the National Science Foundation grant number ACI-1548562 through allocations TG-CTS180066 and TG-CTS190041. L.B. acknowledges FORMAS (Swedish research council for sustainable development) for their financial support through the JPI-Oceans MicroplastiX project.

The authors report no conflict of interest.

\bibliographystyle{jfm}
\bibliography{jfm-instructions.bib}

\begin{thebibliography}{60}
\expandafter\ifx\csname natexlab\endcsname\relax\def\natexlab#1{#1}\fi
\def\au#1{#1} \def\ed#1{#1} \def\yr#1{#1}\def\at#1{#1}\def\jt#1{\textit{#1}}
  \def\bt#1{#1}\def\bvol#1{\textbf{#1}} \def\vol#1{#1} \def\pg#1{#1}
  \def\publ#1{#1}\def\arxiv#1{#1}\def\org#1{#1}\def\st#1{\textit{#1}}

\bibitem[Alldredge \& Gotschalk(1989)]{alldredge1989direct}
{\sc \au{Alldredge, A. } \& \au{Gotschalk, C. }} \yr{1989}  \at{Direct
  observations of the mass flocculation of diatom blooms: characteristics,
  settling velocities and formation of diatom aggregates}.  \jt{Deep Sea
  Research Part A. Oceanographic Research Papers}  \bvol{36}~(2),
  \pg{159--171}.

\bibitem[Ardekani \& Stocker(2010)]{ardekani2010stratlets}
{\sc \au{Ardekani, A. } \& \au{Stocker, R. }} \yr{2010}  \at{Stratlets: Low
  reynolds number point-force solutions in a stratified fluid}.  \jt{Physical
  review letters}  \bvol{105}~(8),  \pg{084502}.

\bibitem[Ardekani {\em et~al.\/}(2017)Ardekani, Doostmohammadi \&
  Desai]{ardekani2017transport}
{\sc \au{Ardekani, A.~M. }, \au{Doostmohammadi, A. } \& \au{Desai, N. }}
  \yr{2017}  \at{Transport of particles, drops, and small organisms in density
  stratified fluids}.  \jt{Physical Review Fluids}  \bvol{2}~(10),
  \pg{100503}.

\bibitem[Ardekani(2019)]{niazi2019numerical}
{\sc \au{Ardekani, M.~N. }} \yr{2019}  \at{Numerical study of transport
  phenomena in particle suspensions}. PhD thesis, KTH Royal Institute of
  Technology.

\bibitem[Ardekani {\em et~al.\/}(2018{\natexlab{{\em a\/}}})Ardekani, Abouali,
  Picano \& Brandt]{niazi_ardekani2018}
{\sc \au{Ardekani, M.~N. }, \au{Abouali, O. }, \au{Picano, F. } \& \au{Brandt,
  L. }} \yr{2018{\natexlab{{\em a\/}}}}  \at{Heat transfer in laminar couette
  flow laden with rigid spherical particles}.  \jt{Journal of Fluid Mechanics}
  \bvol{834},  \pg{308–334}.

\bibitem[Ardekani {\em et~al.\/}(2018{\natexlab{{\em b\/}}})Ardekani, Asmar,
  Picano \& Brandt]{Ardekani2018}
{\sc \au{Ardekani, M.~N. }, \au{Asmar, L.~A. }, \au{Picano, F. } \& \au{Brandt,
  L. }} \yr{2018{\natexlab{{\em b\/}}}}  \at{{Numerical study of heat transfer
  in laminar and turbulent pipe flow with finite-size spherical particles}}.
  \jt{International Journal of Heat and Fluid Flow}  \bvol{71},  \pg{189--199}.

\bibitem[Ardekani {\em et~al.\/}(2016)Ardekani, Costa, Breugem \&
  Brandt]{Ardekani2016}
{\sc \au{Ardekani, M.~N. }, \au{Costa, P. }, \au{Breugem, W.~P. } \&
  \au{Brandt, L. }} \yr{2016}  \at{{Numerical study of the sedimentation of
  spheroidal particles}}.  \jt{International Journal of Multiphase Flow}
  \bvol{87},  \pg{16--34}.

\bibitem[Auguste {\em et~al.\/}(2010)Auguste, Fabre \&
  Magnaudet]{auguste2010bifurcations}
{\sc \au{Auguste, F. }, \au{Fabre, D. } \& \au{Magnaudet, J. }} \yr{2010}
  \at{Bifurcations in the wake of a thick circular disk}.  \jt{Theoretical and
  Computational Fluid Dynamics}  \bvol{24}~(1-4),  \pg{305--313}.

\bibitem[Bainbridge(1957)]{bainbridge1957size}
{\sc \au{Bainbridge, R. }} \yr{1957}  \at{The size, shape and density of marine
  phytoplankton concentrations}.  \jt{Biological Reviews}  \bvol{32}~(1),
  \pg{91--115}.

\bibitem[Basset(1888)]{basset1888treatise}
{\sc \au{Basset, A.~B. }} \yr{1888} {\em A treatise on hydrodynamics: with
  numerous examples\/}, ,  \vol{vol.~2}.  \publ{Deighton, Bell and Company}.

\bibitem[Boussinesq(1985)]{boussinesq1985resistance}
{\sc \au{Boussinesq, J. }} \yr{1985}  \at{Sur la r{\'e}sistance qu’oppose...
  soient n{\'e}gligeables}.  \jt{CR Acad. Sci} .

\bibitem[Chrust(2012)]{chrust2012etude}
{\sc \au{Chrust, M. }} \yr{2012}  \at{Etude num{\'e}rique de la chute libre
  d'objets axisym{\'e}triques dans un fluide newtonien}. PhD thesis,
  Strasbourg.

\bibitem[Chrust {\em et~al.\/}(2013)Chrust, Bouchet \&
  Du{\v{s}}ek]{chrust2013numerical}
{\sc \au{Chrust, M. }, \au{Bouchet, G. } \& \au{Du{\v{s}}ek, J. }} \yr{2013}
  \at{Numerical simulation of the dynamics of freely falling discs}.
  \jt{Physics of Fluids}  \bvol{25}~(4),  \pg{044102}.

\bibitem[Cloern(1984)]{cloern1984temporal}
{\sc \au{Cloern, J. }} \yr{1984}  \at{Temporal dynamics and ecological
  significance of salinity stratification in an estuary (south san-francisco
  bay, usa)}.  \jt{Oceanologica Acta}  \bvol{7}~(1),  \pg{137--141}.

\bibitem[Doostmohammadi \& Ardekani(2014)]{doostmohammadi2014reorientation}
{\sc \au{Doostmohammadi, A. } \& \au{Ardekani, A. }} \yr{2014}
  \at{Reorientation of elongated particles at density interfaces}.
  \jt{Physical Review E}  \bvol{90}~(3),  \pg{033013}.

\bibitem[Doostmohammadi \& Ardekani(2015)]{doostmohammadi2015suspension}
{\sc \au{Doostmohammadi, A. } \& \au{Ardekani, A. }} \yr{2015}  \at{Suspension
  of solid particles in a density stratified fluid}.  \jt{Physics of Fluids}
  \bvol{27}~(2),  \pg{023302}.

\bibitem[Doostmohammadi {\em et~al.\/}(2014)Doostmohammadi, Dabiri \&
  Ardekani]{Doostmohammadi2014b}
{\sc \au{Doostmohammadi, A. }, \au{Dabiri, S. } \& \au{Ardekani, A.~M. }}
  \yr{2014}  \at{{A numerical study of the dynamics of a particle settling at
  moderate Reynolds numbers in a linearly stratified fluid}}.  \jt{Journal of
  Fluid Mechanics}  \bvol{750},  \pg{5--32}.

\bibitem[Doostmohammadi {\em et~al.\/}(2012)Doostmohammadi, Stocker \&
  Ardekani]{doostmohammadi2012low}
{\sc \au{Doostmohammadi, A. }, \au{Stocker, R. } \& \au{Ardekani, A.~M. }}
  \yr{2012}  \at{Low-reynolds-number swimming at pycnoclines}.  \jt{Proceedings
  of the National Academy of Sciences}  \bvol{109}~(10),  \pg{3856--3861}.

\bibitem[Ern {\em et~al.\/}(2012)Ern, Risso, Fabre \& Magnaudet]{ern2012wake}
{\sc \au{Ern, P. }, \au{Risso, F. }, \au{Fabre, D. } \& \au{Magnaudet, J. }}
  \yr{2012}  \at{Wake-induced oscillatory paths of bodies freely rising or
  falling in fluids}.  \jt{Annual Review of Fluid Mechanics}  \bvol{44},
  \pg{97--121}.

\bibitem[Fabre {\em et~al.\/}(2008)Fabre, Auguste \&
  Magnaudet]{fabre2008bifurcations}
{\sc \au{Fabre, D. }, \au{Auguste, F. } \& \au{Magnaudet, J. }} \yr{2008}
  \at{Bifurcations and symmetry breaking in the wake of axisymmetric bodies}.
  \jt{Physics of Fluids}  \bvol{20}~(5),  \pg{051702}.

\bibitem[Feng {\em et~al.\/}(1994)Feng, Hu \& Joseph]{feng1994direct}
{\sc \au{Feng, J. }, \au{Hu, H.~H. } \& \au{Joseph, D.~D. }} \yr{1994}
  \at{Direct simulation of initial value problems for the motion of solid
  bodies in a newtonian fluid part 1. sedimentation}.  \jt{Journal of Fluid
  Mechanics}  \bvol{261},  \pg{95--134}.

\bibitem[Fernandes {\em et~al.\/}(2008)Fernandes, Ern, Risso \&
  Magnaudet]{fernandes2008dynamics}
{\sc \au{Fernandes, P.~C. }, \au{Ern, P. }, \au{Risso, F. } \& \au{Magnaudet,
  J. }} \yr{2008}  \at{Dynamics of axisymmetric bodies rising along a zigzag
  path}.  \jt{Journal of Fluid Mechanics}  \bvol{606},  \pg{209--223}.

\bibitem[Fernandes {\em et~al.\/}(2007)Fernandes, Risso, Ern \&
  Magnaudet]{fernandes2007oscillatory}
{\sc \au{Fernandes, P.~C. }, \au{Risso, F. }, \au{Ern, P. } \& \au{Magnaudet,
  J. }} \yr{2007}  \at{Oscillatory motion and wake instability of freely rising
  axisymmetric bodies}.  \jt{Journal of Fluid Mechanics}  \bvol{573},
  \pg{479--502}.

\bibitem[Field {\em et~al.\/}(1997)Field, Klaus, Moore \&
  Nori]{field1997chaotic}
{\sc \au{Field, S.~B. }, \au{Klaus, M. }, \au{Moore, M. } \& \au{Nori, F. }}
  \yr{1997}  \at{Chaotic dynamics of falling disks}.  \jt{Nature}
  \bvol{388}~(6639),  \pg{252--254}.

\bibitem[Gatignol {\em et~al.\/}(1983)]{gatignol1983faxen}
{\sc \au{Gatignol, R. } \& \au{others}} \yr{1983}  \at{The fax{\'e}n formulae
  for a rigid particle in an unsteady non-uniform stokes flow} .

\bibitem[Geyer {\em et~al.\/}(2008)Geyer, Scully \&
  Ralston]{geyer2008quantifying}
{\sc \au{Geyer, W.~R. }, \au{Scully, M.~E. } \& \au{Ralston, D.~K. }} \yr{2008}
   \at{Quantifying vertical mixing in estuaries}.  \jt{Environmental fluid
  mechanics}  \bvol{8}~(5-6),  \pg{495--509}.

\bibitem[Gibson {\em et~al.\/}(2007)Gibson, Atkinson \&
  Gordon]{gibson2007inherent}
{\sc \au{Gibson, R. }, \au{Atkinson, R. } \& \au{Gordon, J. }} \yr{2007}
  \at{Inherent optical properties of non-spherical marine-like particles from
  theory to observation}.  \jt{Oceanography and marine biology: an annual
  review}  \bvol{45},  \pg{1--38}.

\bibitem[Hanazaki {\em et~al.\/}(2009{\natexlab{{\em a\/}}})Hanazaki, Kashimoto
  \& Okamura]{hanazaki2009jets}
{\sc \au{Hanazaki, H. }, \au{Kashimoto, K. } \& \au{Okamura, T. }}
  \yr{2009{\natexlab{{\em a\/}}}}  \at{Jets generated by a sphere moving
  vertically in a stratified fluid}.  \jt{Journal of fluid mechanics}
  \bvol{638},  \pg{173--197}.

\bibitem[Hanazaki {\em et~al.\/}(2009{\natexlab{{\em b\/}}})Hanazaki, Konishi
  \& Okamura]{hanazaki2009schmidt}
{\sc \au{Hanazaki, H. }, \au{Konishi, K. } \& \au{Okamura, T. }}
  \yr{2009{\natexlab{{\em b\/}}}}  \at{Schmidt-number effects on the flow past
  a sphere moving vertically in a stratified diffusive fluid}.  \jt{Physics of
  Fluids}  \bvol{21}~(2),  \pg{026602}.

\bibitem[Henson {\em et~al.\/}(2011)Henson, Sanders, Madsen, Morris, Le~Moigne
  \& Quartly]{henson2011reduced}
{\sc \au{Henson, S.~A. }, \au{Sanders, R. }, \au{Madsen, E. }, \au{Morris,
  P.~J. }, \au{Le~Moigne, F. } \& \au{Quartly, G.~D. }} \yr{2011}  \at{A
  reduced estimate of the strength of the ocean's biological carbon pump}.
  \jt{Geophysical Research Letters}  \bvol{38}~(4).

\bibitem[Hilali {\em et~al.\/}(2022)Hilali, Pal, More, Saive \& Ardekani]{hilali2022sheared}
{\sc \au{Hilali, M.~M.}, \au{Pal, S.}, \au{More, R.~V.}, \au{Saive, R.}, \& \au{Ardekani, A.~M.}} \yr{2022} \at{Sheared thick-film electrode materials containing silver powders with nanoscale surface asperities improve solar cell performance}. \jt{Advanced Energy and Sustainability Research} \bvol{3}~(1), \pg{2100145}.

\bibitem[Jacobson \& Jacobson(2005)]{jacobson2005fundamentals}
{\sc \au{Jacobson, M.~Z. } \& \au{Jacobson, M.~Z. }} \yr{2005} {\em
  Fundamentals of atmospheric modeling\/}.  \publ{Cambridge university press}.

\bibitem[Jenny {\em et~al.\/}(2004)Jenny, Du{\v{s}}ek \&
  Bouchet]{jenny2004instabilities}
{\sc \au{Jenny, M. }, \au{Du{\v{s}}ek, J. } \& \au{Bouchet, G. }} \yr{2004}
  \at{Instabilities and transition of a sphere falling or ascending freely in a
  newtonian fluid}.  \jt{Journal of Fluid Mechanics}  \bvol{508},
  \pg{201--239}.

\bibitem[Kishore \& Gu(2011)]{kishore2011momentum}
{\sc \au{Kishore, N. } \& \au{Gu, S. }} \yr{2011}  \at{Momentum and heat
  transfer phenomena of spheroid particles at moderate reynolds and prandtl
  numbers}.  \jt{International Journal of Heat and Mass Transfer}
  \bvol{54}~(11-12),  \pg{2595--2601}.

\bibitem[Lofquist \& Purtell(1984)]{lofquist1984drag}
{\sc \au{Lofquist, K.~E. } \& \au{Purtell, L.~P. }} \yr{1984}  \at{Drag on a
  sphere moving horizontally through a stratified liquid}.  \jt{Journal of
  Fluid Mechanics}  \bvol{148},  \pg{271--284}.

\bibitem[MacIntyre {\em et~al.\/}(1995)MacIntyre, Alldredge \&
  Gotschalk]{macintyre1995accumulation}
{\sc \au{MacIntyre, S. }, \au{Alldredge, A.~L. } \& \au{Gotschalk, C.~C. }}
  \yr{1995}  \at{Accumulation of marines now at density discontinuities in the
  water column}.  \jt{Limnology and Oceanography}  \bvol{40}~(3),
  \pg{449--468}.

\bibitem[Magnaudet(1997)]{magnaudet1997forces}
{\sc \au{Magnaudet, J. }} \yr{1997} The forces acting on bubbles and rigid
  particles.  \bt{In {\em ASME Fluids Engineering Division Summer Meeting,
  FEDSM\/}}, ,  \vol{vol.~97},  \pg{pp. 22--26}.

\bibitem[Magnaudet \& Mercier(2019)]{Magnaudet2019}
{\sc \au{Magnaudet, J. } \& \au{Mercier, M.~J. }} \yr{2019}  \at{{Particles,
  Drops, and Bubbles Moving Across Sharp Interfaces and Stratified Layers}} .

\bibitem[Magnaudet \& Mougin(2007)]{magnaudet2007wake}
{\sc \au{Magnaudet, J. } \& \au{Mougin, G. }} \yr{2007}  \at{Wake instability
  of a fixed spheroidal bubble}.  \jt{Journal of Fluid Mechanics}  \bvol{572},
  \pg{311}.

\bibitem[Majlesara {\em et~al.\/}(2020)Majlesara, Abouali, Kamali, Ardekani \&
  Brandt]{majlesara2020numerical}
{\sc \au{Majlesara, M. }, \au{Abouali, O. }, \au{Kamali, R. }, \au{Ardekani,
  M.~N. } \& \au{Brandt, L. }} \yr{2020}  \at{Numerical study of hot and cold
  spheroidal particles in a viscous fluid}.  \jt{International Journal of Heat
  and Mass Transfer}  \bvol{149},  \pg{119206}.

\bibitem[Maxey \& Riley(1983)]{maxey1983equation}
{\sc \au{Maxey, M.~R. } \& \au{Riley, J.~J. }} \yr{1983}  \at{Equation of
  motion for a small rigid sphere in a nonuniform flow}.  \jt{The Physics of
  Fluids}  \bvol{26}~(4),  \pg{883--889}.

\bibitem[Mehaddi {\em et~al.\/}(2018)Mehaddi, Candelier \&
  Mehlig]{mehaddi2018inertial}
{\sc \au{Mehaddi, R. }, \au{Candelier, F. } \& \au{Mehlig, B. }} \yr{2018}
  \at{Inertial drag on a sphere settling in a stratified fluid}.  \jt{Journal
  of Fluid Mechanics}  \bvol{855},  \pg{1074--1087}.

\bibitem[Mercier {\em et~al.\/}(2020)Mercier, Wang, P{\'e}m{\'e}ja, Ern \&
  Ardekani]{mercier2020settling}
{\sc \au{Mercier, M. }, \au{Wang, S. }, \au{P{\'e}m{\'e}ja, J. }, \au{Ern, P. }
  \& \au{Ardekani, A. }} \yr{2020}  \at{Settling disks in a linearly stratified
  fluid}.  \jt{Journal of Fluid Mechanics}  \bvol{885}.

\bibitem[More \& Balasubramanian(2018)]{more2018mixing}
{\sc \au{More, R. } \& \au{Balasubramanian, S. }} \yr{2018}  \at{Mixing
  dynamics in double-diffusive convective stratified fluid layers}.  \jt{Curr.
  Sci}  \bvol{114},  \pg{1953--1960}.

\bibitem[More \& Ardekani(2020{\natexlab{{\em a\/}}})]{more2020interaction}
{\sc \au{More, R.~V. } \& \au{Ardekani, A.~M. }} \yr{2020{\natexlab{{\em
  a\/}}}}  \at{Hydrodynamic interactions between swimming microorganisms in a
  linearly density stratified fluid (under review)}.  \jt{Physical Review E} .

\bibitem[More \& Ardekani(2020{\natexlab{{\em b\/}}})]{more2020motion}
{\sc \au{More, R.~V. } \& \au{Ardekani, A.~M. }} \yr{2020{\natexlab{{\em
  b\/}}}}  \at{Motion of an inertial squirmer in a density stratified fluid}.
  \jt{Journal of Fluid Mechanics}  \bvol{905}.

\bibitem[Morris(1980)]{morris1980physiological}
{\sc \au{Morris, I. }} \yr{1980} Physiological ecology of phytoplankton.

\bibitem[Mrokowska(2018)]{mrokowska2018stratification}
{\sc \au{Mrokowska, M.~M. }} \yr{2018}  \at{Stratification-induced
  reorientation of disk settling through ambient density transition}.
  \jt{Scientific reports}  \bvol{8}~(1),  \pg{1--12}.

\bibitem[Mrokowska(2020{\natexlab{{\em a\/}}})]{mrokowska2020dynamics}
{\sc \au{Mrokowska, M.~M. }} \yr{2020{\natexlab{{\em a\/}}}}  \at{Dynamics of
  thin disk settling in two-layered fluid with density transition}.  \jt{Acta
  Geophysica}  \bvol{68}~(4),  \pg{1145--1160}.

\bibitem[Mrokowska(2020{\natexlab{{\em b\/}}})]{mrokowska2020influence}
{\sc \au{Mrokowska, M.~M. }} \yr{2020{\natexlab{{\em b\/}}}}  \at{Influence of
  pycnocline on settling behaviour of non-spherical particle and wake
  evolution}.  \jt{Scientific Reports}  \bvol{10}~(1),  \pg{1--14}.

\bibitem[Namkoong {\em et~al.\/}(2008)Namkoong, Yoo \&
  Choi]{namkoong2008numerical}
{\sc \au{Namkoong, K. }, \au{Yoo, J.~Y. } \& \au{Choi, H.~G. }} \yr{2008}
  \at{Numerical analysis of two-dimensional motion of a freely falling circular
  cylinder in an infinite fluid}.  \jt{Journal of Fluid Mechanics}  \bvol{604},
   \pg{33--53}.

\bibitem[Srdi{\'c}-Mitrovi{\'c} {\em et~al.\/}(1999)Srdi{\'c}-Mitrovi{\'c},
  Mohamed \& Fernando]{srdic1999gravitational}
{\sc \au{Srdi{\'c}-Mitrovi{\'c}, A. }, \au{Mohamed, N. } \& \au{Fernando, H. }}
  \yr{1999}  \at{Gravitational settling of particles through density
  interfaces}.  \jt{Journal of Fluid Mechanics}  \bvol{381},  \pg{175--198}.

\bibitem[Stone(2010)]{stone2010invisible}
{\sc \au{Stone, R. }} \yr{2010} The invisible hand behind a vast carbon
  reservoir.

\bibitem[Tchoufag {\em et~al.\/}(2014)Tchoufag, Fabre \&
  Magnaudet]{tchoufag2014global}
{\sc \au{Tchoufag, J. }, \au{Fabre, D. } \& \au{Magnaudet, J. }} \yr{2014}
  \at{Global linear stability analysis of the wake and path of buoyancy-driven
  disks and thin cylinders}.  \jt{Journal of fluid mechanics}  \bvol{740},
  \pg{278--311}.

\bibitem[Torres {\em et~al.\/}(2000)Torres, Hanazaki, Ochoa, Castillo \&
  Van~Woert]{torres2000flow}
{\sc \au{Torres, C. }, \au{Hanazaki, H. }, \au{Ochoa, J. }, \au{Castillo, J. }
  \& \au{Van~Woert, M. }} \yr{2000}  \at{Flow past a sphere moving vertically
  in a stratified diffusive fluid}.  \jt{Journal of Fluid Mechanics}
  \bvol{417},  \pg{211--236}.

\bibitem[Towns {\em et~al.\/}(2014)Towns, Cockerill, Dahan, Foster, Gaither,
  Grimshaw, Hazlewood, Lathrop, Lifka, Peterson {\em et~al.\/}]{towns2014xsede}
{\sc \au{Towns, J. }, \au{Cockerill, T. }, \au{Dahan, M. }, \au{Foster, I. },
  \au{Gaither, K. }, \au{Grimshaw, A. }, \au{Hazlewood, V. }, \au{Lathrop, S.
  }, \au{Lifka, D. }, \au{Peterson, G.~D. } \& \au{others}} \yr{2014}
  \at{Xsede: accelerating scientific discovery}.  \jt{Computing in science \&
  engineering}  \bvol{16}~(5),  \pg{62--74}.

\bibitem[Willmarth {\em et~al.\/}(1964)Willmarth, Hawk \&
  Harvey]{willmarth1964steady}
{\sc \au{Willmarth, W.~W. }, \au{Hawk, N.~E. } \& \au{Harvey, R.~L. }}
  \yr{1964}  \at{Steady and unsteady motions and wakes of freely falling
  disks}.  \jt{The physics of Fluids}  \bvol{7}~(2),  \pg{197--208}.

\bibitem[W{\"u}st {\em et~al.\/}(2017)W{\"u}st, Bittner, Yee, Mlynczak \&
  Russel~III]{wust2017variability}
{\sc \au{W{\"u}st, S. }, \au{Bittner, M. }, \au{Yee, J.-H. }, \au{Mlynczak,
  M.~G. } \& \au{Russel~III, J.~M. }} \yr{2017}  \at{Variability of the
  brunt-v{\"a}is{\"a}l{\"a} frequency at the oh* layer height} .

\bibitem[Yang \& Prosperetti(2007)]{yang2007linear}
{\sc \au{Yang, B. } \& \au{Prosperetti, A. }} \yr{2007}  \at{Linear stability
  of the flow past a spheroidal bubble}.  \jt{Journal of Fluid Mechanics}
  \bvol{582},  \pg{53}.

\bibitem[Yick {\em et~al.\/}(2009)Yick, Torres, Peacock \&
  Stocker]{yick2009enhanced}
{\sc \au{Yick, K.~Y. }, \au{Torres, C.~R. }, \au{Peacock, T. } \& \au{Stocker,
  R. }} \yr{2009}  \at{Enhanced drag of a sphere settling in a stratified fluid
  at small reynolds numbers}.  \jt{Journal of Fluid Mechanics}  \bvol{632},
  \pg{49--68}.

\end{thebibliography}

\end{document}